\documentclass[12pt, oneside]{article}   	
\usepackage{geometry}                		
\usepackage{afterpage}
\usepackage{float}

\PassOptionsToPackage{prologue,dvipsnames}{xcolor}
\usepackage{amssymb}
\usepackage{amsmath,bm,times}
\usepackage{verbatim}

\usepackage{caption}
\usepackage{subcaption} 

\usepackage{booktabs}  
\usepackage{longtable}
\usepackage{multirow}

\usepackage{mwe,tikz}
\usepackage{pgfplots}
\usepackage{pgfplotstable}

\usepgfplotslibrary{polar}

\usetikzlibrary{shapes,positioning,calc,arrows.meta,patterns}

\usepgfplotslibrary{statistics}

\newcommand*{\ReadOutElement}[4]{%
    \pgfplotstablegetelem{#2}{[index]#3}\of{#1}%
    \let#4\pgfplotsretval
}
	
\pgfplotstableread{./data.txt}\data
\pgfplotstableset{create on use/error/.style={
    create col/expr={\thisrow{uci}-\thisrow{mean}
    }
  }
}

\newcommand{\errplot}{%
  \begin{tikzpicture}[trim axis left,trim axis right]
    \begin{axis}[y=-\baselineskip,
        scale only axis,
        width             = 6.5cm,
        enlarge y limits  = {abs=0.5},
        axis y line*      = middle,
        y axis line style = dashed,
        ytick             = \empty,
        axis x line*      = bottom,
        scaled x ticks=false,
        xmin = -0.025,
        xmax=0.025,
         x tick label style={
        /pgf/number format/.cd,
        fixed,
        fixed zerofill,
        precision=2,
        /tikz/.cd
    }
      ]
      \addplot+[only marks][error bars/.cd,x dir=both, x explicit]
        table [x=mean,y expr=\coordindex,x error=error]{\data};
    \end{axis}
  \end{tikzpicture}%
}

\pgfplotsset{
    box plot/.style={
        /pgfplots/.cd,
        gray,
        only marks,
        mark=-,
        mark size=0.09em,
        /pgfplots/error bars/.cd,
        y dir=plus,
        y explicit
    },
    box plot box/.style={
        /pgfplots/error bars/draw error bar/.code 2 args={%
            \draw[fill, red]  ##1 -- ++(0.06em,0pt) |- ##2 -- ++(-0.06em,0pt) |- ##1 -- cycle;
        },
        /pgfplots/table/.cd,
        y index=2,
        y error expr={\thisrowno{3}-\thisrowno{2}},
        /pgfplots/box plot
    },
    box plot top whisker/.style={
        /pgfplots/error bars/draw error bar/.code 2 args={%
            \pgfkeysgetvalue{/pgfplots/error bars/error mark}%
            {\pgfplotserrorbarsmark}%
            \pgfkeysgetvalue{/pgfplots/error bars/error mark options}%
            {\pgfplotserrorbarsmarkopts}%
            \path ##1 -- ##2;
        },
        /pgfplots/table/.cd,
        y index=4,
        y error expr={\thisrowno{2}-\thisrowno{4}},
        /pgfplots/box plot
    },
    box plot bottom whisker/.style={
        /pgfplots/error bars/draw error bar/.code 2 args={%
            \pgfkeysgetvalue{/pgfplots/error bars/error mark}%
            {\pgfplotserrorbarsmark}%
            \pgfkeysgetvalue{/pgfplots/error bars/error mark options}%
            {\pgfplotserrorbarsmarkopts}%
            \path ##1 -- ##2;
        },
        /pgfplots/table/.cd,
        y index=5,
        y error expr={\thisrowno{3}-\thisrowno{5}},
        /pgfplots/box plot
    },
    box plot median/.style={
/pgfplots/box plot
    }
}

\usepackage[natbibapa]{apacite}
\providecommand{\keywords}[1]
{
  \small	
  \textbf{\textit{Keywords---}} #1
}

\usepackage[hyperfootnotes=false]{hyperref}
\hypersetup{
  colorlinks,
  citecolor=blue,
  linkcolor=red,
  urlcolor=blue}

\title{\Large Association between built environment characteristics and school run traffic congestion in Beijing, China}
\author{\normalsize Chaogui Kang*, Xiaxin Wu, Jialei Shi, Chao Yang*}
\date{\normalsize National Engineering Research Center of Geographic Information System,\\ China University of Geosciences, Wuhan, Hubei 430078, China\\\small * Correspondence to: kangchaogui@cug.edu.cn (C.K.), yangchao@cug.edu.cn (C.Y.)}							

\begin{document}
\maketitle

\begin{abstract}


School-escorted trips are a significant contributor to traffic congestion. Existing studies mainly compare road traffic during student pick-up/drop-off hours with off-peak times, often overlooking the fact that school-run traffic congestion is unevenly distributed across areas with different built environment characteristics. We examine the relationship between the built environment and school-run traffic congestion, using Beijing, China, as a case study. First, we use multi-source geospatial data to assess the built environment characteristics around schools across five dimensions: spatial concentration, transportation infrastructure, street topology, spatial richness, and scenescapes. Second, employing a generalized ordered logit model, we analyze how traffic congestion around schools varies during peak hours on school days, regular non-school days, and national college entrance exam days. Lastly, we identify the built environment factors contributing to school-run traffic congestion through multivariable linear regression and Shapley value explanations. Our findings reveal that: (1) School runs significantly exacerbate traffic congestion around schools, reducing the likelihood of free-flow by 8.34\% during school run times; (2) School-run traffic congestion is more severe in areas with multiple schools, bus stops, and scenescapes related to business and financial functions. These insights can inform the planning of new schools and urban upgrade strategies aimed at reducing traffic congestion.
   
\end{abstract}


\newpage
\section{Introduction}

School run traffic congestion has become prominent in cities as school-escorted trips account for an increasing proportion of the total trips during commute hours \citep{Yarlagadda2008}. Due to the higher utilization of private cars for student pick-up and drop-off, traffic congestion on roads in the school vicinity has remarkably undermined the efficiency of urban transport system and the comfort of school life \citep{An2021, Ding2023}. It is thus critical to understand the spatial distribution of school run traffic congestion and its association with the local urban built environment in order to mitigate such negative impacts.  

Existing studies have mainly addressed the relationship between traffic congestion and school runs from an aggregate perspective, by evaluating (1) whether the aggregate traffic congestion in the commuting hours on school-run days differs from that on non-school workdays \citep{Lu2017}, and/or (2) whether the aggregate traffic congestion on roads in the school vicinity differs from that on roads further away from schools \citep{Sun2021}. These attempts have provided rigorous evidences indicating that school runs (or more specifically, drive-to/from-school trips) disrupt road traffic significantly. However, such evidences fell short for the generation of operative interventions to mitigate the school run traffic congestion in that diaggregated spatial relationships between the school run traffic congestion and the ambient built environment were not provided.

To address this limitation, this article seeks to uncover the unevenly spatial distribution of traffic congestion in the school vicinity and, more importantly, how this uneven distribution is correlated with the local built environment based on multi-sourced geospatial data. More specifically, we will investigate the following research questions by taking Beijing, China as the case studied city :

\begin{itemize}
\item[Q1.] To what extent do drive-to/from-school trips contribute to the traffic congestion on roads in the school vicinity during student pick-up/drop-off hours?  
\item[Q2.] What kind of built environment characteristics in the school vicinity are plausible for lowering the risk of school run traffic congestion?
\end{itemize}

To answer Q1, we design a rigidly controlled experiment for evaluating the differences of trafffic congestion on roads in the school vicinity between school run days, normal non-school days, and national college entrance exam workdays.  The resultant findings provide additional evidence to the impact of school runs on road congestion around schools. To answer Q2, we measure the built environment characteristics of each school neighborhood in five distinct aspects and relate them to the risk of school run traffic congestion using quantitative and explainable models. The resultant findings uncover the contributing factors to traffic congestion around schools and provide insights for urban upgrade plans and school site allocations.  

The remainder of the paper is organized as follows. Section 2 summarizes previous studies on trafffic congestion around schools and the impact of urban built environment. Section 3 introduces the case studied city and the proposed regression models for addressing the aforementioned research questions in details. Section 4 presents the main results and findings. Section 5 discusses the relevant policy implications. Section 6 offers our conclusions.

\section{Related work}
\subsection{School run and traffic congestion}
\subsubsection{The increase of drive-to/from-school trips}

In recent years, the utilization of private cars for school-escorted trips has seen a marked rise \citep{Mcdonald2011}. In many western cities, as much as more than half of students take private cars to school \citep{VanRistell2013, Larouche2015, He2017}. Such situations are also emerging in large cities in developing countries such as China and India \citep{Singh2018}.  

Previous studies have mainly attributed such a rise to the facts that: (1) Schools are often located within the central areas in cities; (2) Housing prices are relatively higher surround schools; (3) Parents are forced to live far away from work places and schools \citep{Zheng2016, Sun2021}. Active school travels by walking, cycling, school bus and public transits are inplausible for countering the increased spatial seperations between home and school \citep{Muller2008, Ermagun2018, Ma2024}. Instead, parents often use private cars as the preferred travel mode for school-escorted trips due to its convinience, flexibility, and safety compared with walking, cycling, school buses and other public transits \citep{Zhang2017a,Dias2022}. 

\subsubsection{The impact of school run on traffic congestion}

Since student pick-up/drop-off hours coincide with daily commute hours, high traffic around schools often translate into school run traffic congestion. The majority of existing studies have compared the aggregate traffic congestion on school-run days (and/or hours) with that on non-shcool workdays (and/or hours) \citep{Sun2021}. Evidences have shown that the level of traffic congestion could be 5 to 20 percent lower on non-school holidays than nornal school days \citep{Lu2017}. Generally speaking, the main strategy of existing studies is to design an adequate controlled experiment that utilizes variations in traffic flows on roads by school days and school holidays, and/or by hours of school runs and other hours to identify the causal effect of school runs on the aggregate traffic congestion. However, such difference-in-difference (DID) settings are flawed in at least two aspects: (1) The difference in traffic between school days and school holidays is too coarse to concentrate on traffic congestion during student pick-up/drop-off hours that school run traffic congestion actually take places; (2) The difference in traffic between school attending/leaving hours and other time during a day neglects that the citywide travel demands are varying along with time, thus is not able to cancel this effect out. From a temporal perspective, it will be more convincing to analyze the difference in traffic between school attending/leaving hours on normal school run days and regular commute hours on non-school workdays outside of school holidays.  

Despite that previous studies have widely confirmed the causal effect of school run on aggregate traffic congestion, they neglected that school run traffic congestion was unevenly distributed across areas associated with different built environment characteristics. For example, in addition to the high concentration of drive-to/from-school trips, the availability of parking facilities and traffic guidances around schools also plays a critical role for alleviating the risk of school run traffic congestion. Therefore, from a disaggregated perspective, the impact of school runs on traffic congestion should be evaluated in finer spatial granularity by varying school, road, and neighborhood characteristics. A few empirical studies have already reported that traffic congestion is more severe around schools that are larger, better, public rather than private, in more expensive neighborhoods, and with no student accommodation \citep{Sun2021}. However, these studies almost distinguished schools by their own socioeconomic attributes such as school size, school quality and school accommodation. The more broader contextual characteristics around schools with regard to the overall quality of the built environment should be further investigated in order to conclude the causal effect of school run on traffic congestion in fidelity.  

\subsection{Urban built environment and traffic congestion}
\subsubsection{The measurement of built environment characteristics}

Built environment characteristics sit at the heart of many urban analyses and applications \citep{Rothman2014, Silver2016, Nabipour2022, Boakye2023}. Traditional studies heavily rely on socioeconomic data including land uses and points-of-interest (POIs) to characterize the urban built environment based on the well-established 5Ds (Density, Diversity, Design, Distance to transit, and Destination accessibility) system \citep{Yue2017, AnD2019, AnR2022}. However, these traditional measurements are incapable to perceive the built environment comprehensively. Fortunately, with the accumulation of ubiquitous street-view images (SVIs) in past decades, we are able to quantify built environment characteristics in a much broader spectrum through their visual appearance \citep{Fan2023a}.

To faciliate the extraction of urban features from SVIs, a couple of computer vision algorithms and benchmark datasets have been introduced. For example, Places365-VGG \citep{Zhou2018} and PSPnet \citep{Zhao2017} have been widely adopted for urban scene-centric recognition problems comprising hundreds of unique street scene categories. Those extracted scene categories provide a comphensive collection of features for characterizing the built environment from diverse perspectives and scales, ranging from objective perceptual attributes such as openness, greenness, enclosure, walkability, and imageability to subjective perceptual attributes such as safety, vibrancy, tediousness, wealthiness, depression, and beauty \citep{Dubey2016, Qiu2022}. Those works also provide popular scene-centric benchmarks including Place Pulse \citep{Salesses2012}, Cityscapes \citep{Cordts2016}, Places \citep{Zhou2018}, and ADE20K \citep{Zhou2017}.

\subsubsection{The impact of built environment on traffic congestion}

With the assistance of both traditional and newly-emerging data sources, a rich body of literature has scrunitized the relationship between built environment characteristics and traffic congestion conditions. Earlier studies have primarly evaluated the impact of land uses, road features, and public transport facilities on the road traffic congestion \citep{Pan2020}. For example, it has been reported that the densities of bus stops, hospitals, parking entrance/exit, residential estates, and office buildings significantly affect the average travel speed of road traffic \citep{Nian2021}. Besides, the relationship between the built environment and traffic congestion has been found to vary in space and time, which depends on the supply of public transit in different land uses and activity hours \citep{Bao2023}. In general, the majority of previous studies followed the aforementioned 5Ds measurement system and explained traffic congestion patterns based on POI and land use distributions.

A few studies have applied SVIs to examine the relationship between built environment characteristics and traffic congestion \citep{Qin2020}. However, they often utilized the black-boxed machine learning models for achieving the task, thus were not able to explicitly quantify the impact of ad-hoc built environment feature on the risk level of road congestion. In addition, specific deep learning models were also proposed to incorporate scene features extracted from SVIs into graph convolutional neural networks for estimate traffic speed \citep{Jiao2023}, taxi trip volume \citep{Zhang2019a} and other road traffic statistics. Desipte that explainable tools were utilized in those machine learning models for assessing the discriminative information in SVIs, the quantitative relationships between built environment characteristics and traffic congestion indices were not explicitly established.    

\subsection{Research gaps}

In summary, we argue that there are two important research gaps in the existing literature as below:

\begin{itemize}
\item[(1)] The existing DID alike settings by differentiating either traffic on school-run days and school-holiday workdays or traffic on school attending/leaving hours and other time in a day are not sufficient for revealing the causal effect of school run on traffic congestion. Therefore, from a temporal perspective, we have to establish a more convincing DID setting that enable us to directly analyze the difference in traffic between school attending/leaving hours on normal school run days and regular commute hours on non-school workdays outside of (summer and winter) school holidays.  
\item[(2)] The established models of utilizing built environment characteristics for the understanding of traffic congestion patterns are black-boxed and feature-specific. The comprehensive features derived from SVIs are not fully considered as complementary to the 5Ds system of urban built environment measurements. Therefore, from a spatial perspective, we have to characterize the built environment around schools in more details by taking advantage of SVIs and establish regression models to explicitly decouple the association between built environment characteristics and school run traffic congestion. 
\end{itemize}

\section{Methods}
\subsection{Study area}

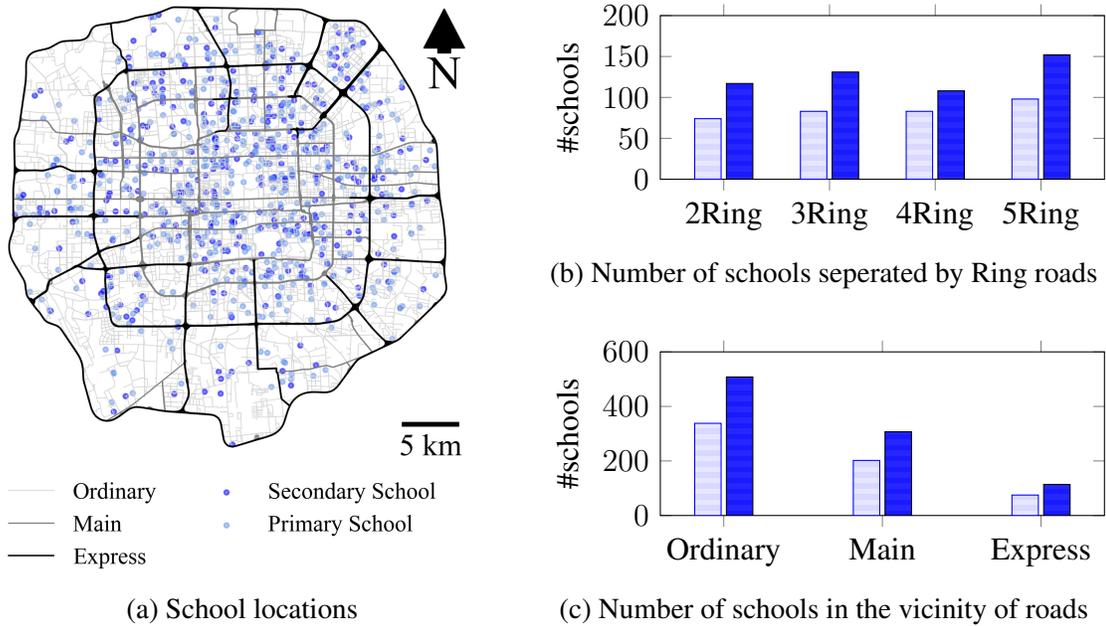
\begin{figure}[htbp]
\sbox0{\begin{subfigure}[b]{\dimexpr 0.5\textwidth-0.5\columnsep}
  \includegraphics[width=\linewidth]{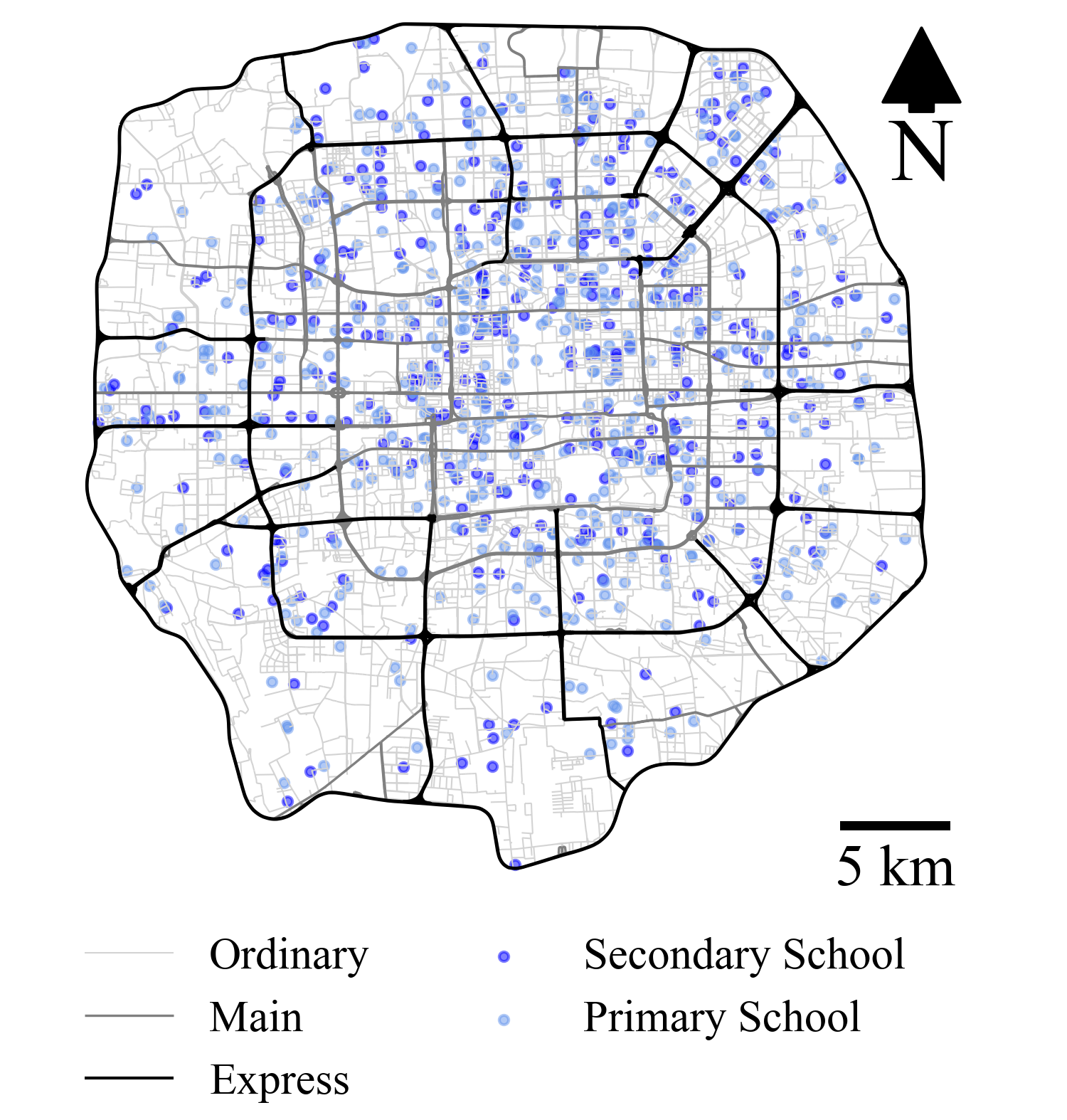}
  \caption{School locations}
\end{subfigure}}%
\usebox0\hfill\begin{minipage}[b][\ht0][s]{\wd0}
  \begin{subfigure}{\linewidth}
            \begin{tikzpicture}
        	\begin{axis}[
	   scale only axis,
            ybar,
            width=0.8\textwidth,
            height=0.295\textwidth,
            enlarge x limits=0.2,
            ymin=0, ymax=200,
            xtick=data,
            ylabel={\#schools},
            symbolic x coords={2Ring, 3Ring, 4Ring, 5Ring}
       	 ]
            \addplot [draw=blue,
        pattern=horizontal lines light blue,
    ] coordinates {
        (2Ring,74) (3Ring,83) (4Ring,83) (5Ring,98)
    };
    \addplot [draw=black,
        pattern=horizontal lines dark blue,
    ] coordinates {
        (2Ring,117) (3Ring,131) (4Ring,108) (5Ring,152)
    };
        \end{axis}
    	\end{tikzpicture}
    \caption{Number of schools seperated by Ring roads}
  \end{subfigure}\par
  \vfill
  \begin{subfigure}[b]{\linewidth}
             \begin{tikzpicture}
        	\begin{axis}[
	   scale only axis,
            ybar,
            width=0.8\textwidth,
            height=0.295\textwidth,
            enlarge x limits=0.2,
            ymin=0, ymax=600,
            xtick=data,
            ylabel={\#schools},
            symbolic x coords={Ordinary, Main, Express}
       	 ]
            \addplot [draw=blue,
        pattern=horizontal lines light blue,
    ] coordinates {
        (Ordinary,338) (Main,202) (Express,75)
    };
    \addplot [draw=black,
        pattern=horizontal lines dark blue,
    ] coordinates {
        (Ordinary,508) (Main,307) (Express,114)
    };
        \end{axis}
    	\end{tikzpicture}
    \caption{Number of schools in the vicinity of roads}
  \end{subfigure}
\end{minipage}
\caption{The spatial distribution of primary and secondary schools within the case studied area. To count the number of schools in the vicinity of roads by category, we use a buffer area in the range of 500 meters away from the road.}
        \label{fig1:study_area}
\end{figure}

As the capital city of China, Beijing has been suffering severe traffic congestion around schools during daily commuting hours. It has been previously reported that the population and the (primary and secondary) schools in Beijing were mainly concentrated in the central area, or more specifically within the 3rd Ring road \citep{Feng2009}. Moreover, about 30\% of primary school students lived 10 kilometers or more far away from their schools and the proportion of school-escorted trips taken by private cars reached nearly 50\% for certain primary schools \citep{Yu2011}. Due to these facts and its significance, we thus choose the core urban area (i.e., the area within the 5th Ring road, as shown in Fig.~\ref{fig1:study_area}) of Beijing as the case studied area.

Within the case studied area, there are in total 846 primary and secondary schools (according to POIs provided by Baidu Map, \url{https://map.baidu.com}). For each school, we define the spatial scope of its school neighborhood by a (buffering) bandwidth of 500 meters, which is in line with previous studies that have indicated that school run traffic congestion is less significant beyond 400 to 500 meters away from the school. Based on the extracted school neighborhoods, we collect comprehensive information about the traffic congestion situation and the built environment around schools to understand the spatiotemporal pattern of school run traffic congetion and the influencing factors in terms of built environment characteristics. Details on the data source and the preprocessing process are described below.

\subsection{Data preparation}

\subsubsection{Spatiotemporal distribution of traffic congestion around schools}

To analyze traffic congestion around schools, we first obtain the road network from the National Geomatics Center of China (NGCC) and collect the real-time traffic index on each roads within the case studied area during a carefully-screened week from June 5, 2023 to June 11, 2023 via the API provided by Baidu Map. This screening period of dates include two normal school-run days (i.e., June 5-6, 2023; Monday-Tuesday), four national college entrance examination days (i.e., June 7-10, 2023; Wednesday-Saturday; Note that no exams are scheduled on Thursday morning) when primary and secondary schools are closed, and a normal non-school weekend (i.e., June 11, 2023; Sunday). Therefore, it enables us to conduct a DID analysis by comparing the difference in traffic during the regular commute hours between normal school-run workdays and non-school workdays outside of (summer and winter) school holidays.  

In each day, we further concentrate on three representative time slots: (1) The morning hours from 6:30 am to 10:30 am, which covers the school attending period from 7:30 am to 8:00 am; (2) The noon time at 12:30 pm; and (3) The afternoon hours from 4:30 pm to 6:30 pm, which covers the school leaving period from 4:30 pm to 5:30 pm. During these time slots, the congestion index ranging from “smooth”, “slow”, “congested” to “severely congested” is recorded hourly for each road. As shown in Fig.~\ref{fig2:congestion}(a), during the observed hours, traffic congestion happens more often on roads within the school neighborhood than roads outside the school neighborhood. Moreover, the proportion of congested roads during the school attending period on the normal school-run day (e.g., Monday and Tuesday) is much higher than that during the morning commute period on the non-school workday (i.e., Thursday), which confirms that school runs and traffic congestion are closely correlated with each other.

\afterpage{%
  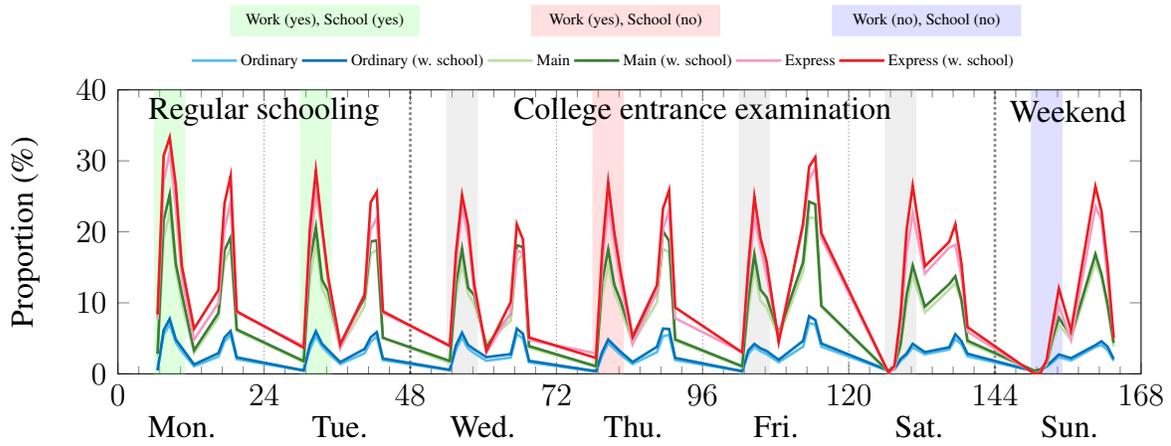
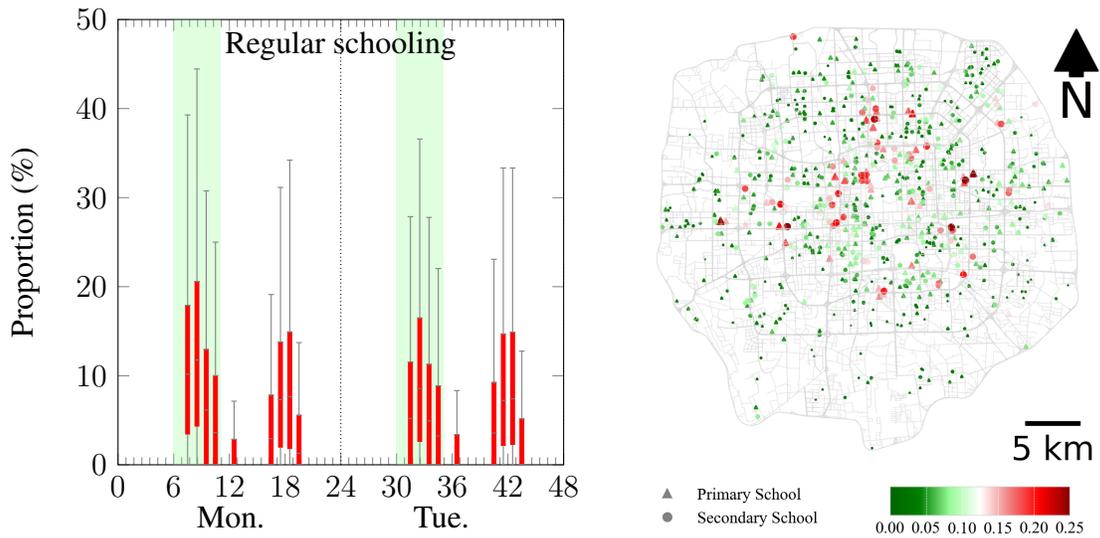
\begin{figure}[H]
     \centering
     \begin{subfigure}[b]{\textwidth}
         \centering
         \begin{tikzpicture}
        \begin{axis}[
	   scale only axis,
            width=0.9\textwidth,
            height=0.25\textwidth,
            xmin=0, xmax=168,
            ymin=0, ymax=40,
            xtick={0, 24, 48, 72, 96, 120, 144, 168},
            ylabel={Proportion (\%)},
            minor x tick num= 6,
	legend style={at={(axis cs:10.5,42)}, nodes={scale=0.5, transform shape}, anchor=south west, draw=none, legend columns=6}
	]
	    \draw[densely dotted, opacity=0.5] (24, 0) -- (24, 400);
            \draw[densely dotted, line width=1, opacity=0.5] (48, 0) -- (48, 400);
            \draw[densely dotted, opacity=0.5] (72, 0) -- (72, 400);
            \draw[densely dotted, opacity=0.5] (96, 0) -- (96, 400);
            \draw[densely dotted, opacity=0.5] (120, 0) -- (120, 400);
            \draw[densely dotted, line width=1, opacity=0.5] (144, 0) -- (144, 400);
            
            \filldraw[green!50,opacity=0.25] (6,0) rectangle ++(5,400);
            \filldraw[green!50,opacity=0.25] (30,0) rectangle ++(5,400);
            \filldraw[gray!50,opacity=0.25] (54,0) rectangle ++(5,400);
            \filldraw[red!50,opacity=0.25] (78,0) rectangle ++(5,400);
            \filldraw[gray!50,opacity=0.25] (102,0) rectangle ++(5,400);
            \filldraw[gray!50,opacity=0.25] (126,0) rectangle ++(5,400);
            \filldraw[blue!50,opacity=0.25] (150,0) rectangle ++(5,400);
            
	\addplot[Cerulean!60, mark=none, empty line=none, line width=1] table[x=t, y=B] {timeseries.dat};
            \addlegendentry{Ordinary}
            \addplot[RoyalBlue, mark=none, empty line=none, line width=1] table[x=t, y=A] {timeseries.dat};
            \addlegendentry{Ordinary (w. school)}  
            \addplot[YellowGreen!60, mark=none, empty line=none, line width=1] table[x=t, y=D] {timeseries.dat};
            \addlegendentry{Main}     
            \addplot[OliveGreen, mark=none, empty line=none, line width=1] table[x=t,y=C] {timeseries.dat};
            \addlegendentry{Main (w. school)}
            \addplot[VioletRed!60, mark=none, empty line=none, line width=1] table[x=t,y=F] {timeseries.dat};
            \addlegendentry{Express}
            \addplot[Red, mark=none, empty line=none, line width=1] table[x=t, y=E] {timeseries.dat};
            \addlegendentry{Express (w. school)}
            \draw (24, 370) node {Regular schooling};
            \draw (96, 370) node {College entrance examination};
            \draw (156, 370) node {Weekend};
        \end{axis}
        \draw (0.85, -0.7) node {Mon.};
        \draw (2.95, -0.7) node {Tue.};
        \draw (4.85, -0.7) node {Wed.};
        \draw (6.85, -0.7) node {Thu.};
        \draw (8.75, -0.7) node {Fri.};
        \draw (10.65, -0.7) node {Sat.};
        \draw (12.65, -0.7) node {Sun.};
        
          \filldraw[green!50,opacity=0.25] (1.5,4.5) rectangle ++(2.5,0.4) node[midway, color=black, opacity= 1.0,  scale=0.5, transform shape] {Work (yes), School (yes)};
            \filldraw[red!50,opacity=0.25] (5.5,4.5) rectangle ++(2.5,0.4) node[midway, color=black, opacity= 1.0,  scale=0.5, transform shape] {Work (yes), School (no)};
          \filldraw[blue!50,opacity=0.25] (9.5,4.5) rectangle ++(2.5,0.4) node[midway, color=black, opacity= 1.0, scale=0.5, transform shape] {Work (no), School (no)};
          
    \end{tikzpicture}
         \caption{Proportion of congested roads on school-run days and non-school days}
         \label{fig:five over x}
     \end{subfigure}
     \break
          \begin{subfigure}[b]{0.49\textwidth}
         \centering
         \begin{tikzpicture}
        \begin{axis}[
	   scale only axis,
	   boxplot/draw direction=y,
            width=0.8\textwidth,
            height=0.8\textwidth,
            xmin=0, xmax=48,
            ymin=0, ymax=50,
            xtick={0, 6, 12, 18, 24, 30, 36, 42, 48},
            ylabel={Proportion (\%)},
            minor x tick num= 6,
	legend style={at={(axis cs:10.5,62)}, nodes={scale=0.5, transform shape}, anchor=south west, draw=none, legend columns=6}
	]
	            \draw[densely dotted] (24, 0) -- (24, 600);
            \draw[densely dotted] (48, 0) -- (48, 600);
            \draw[densely dotted] (72, 0) -- (72, 600);
            \draw[densely dotted] (96, 0) -- (96, 600);
            \draw[densely dotted] (120, 0) -- (120, 600);
            \draw[densely dotted] (144, 0) -- (144, 600);
            
            \filldraw[green!50,opacity=0.25] (6,0) rectangle ++(5,600);
            \filldraw[green!50,opacity=0.25] (30,0) rectangle ++(5,600);
            \filldraw[gray!50,opacity=0.25] (54,0) rectangle ++(5,600);
            \filldraw[red!50,opacity=0.25] (78,0) rectangle ++(5,600);
            \filldraw[gray!50,opacity=0.25] (102,0) rectangle ++(5,600);
            \filldraw[gray!50,opacity=0.25] (126,0) rectangle ++(5,600);
            \filldraw[blue!50,opacity=0.25] (150,0) rectangle ++(5,600);

	\addplot [box plot median] table {schooltime.dat};
    \addplot [box plot box] table {schooltime.dat};
    \addplot [box plot top whisker] table {schooltime.dat};
    \addplot [box plot bottom whisker] table {schooltime.dat};
	
            \draw (24, 470) node {Regular schooling};
        \end{axis}
        \draw (1.5, -0.7) node {Mon.};
        \draw (4.3, -0.7) node {Tue.};          
    \end{tikzpicture}
         \caption{Difference of the proportions of congested roads by school neighborhoods}
         \label{fig:five over x}
     \end{subfigure}    
     \hspace{0.75em}
     \begin{subfigure}[b]{0.475\textwidth}
     	\centering
         \includegraphics[width=0.9\textwidth, trim={0 0.2cm 0 0.5cm},clip]{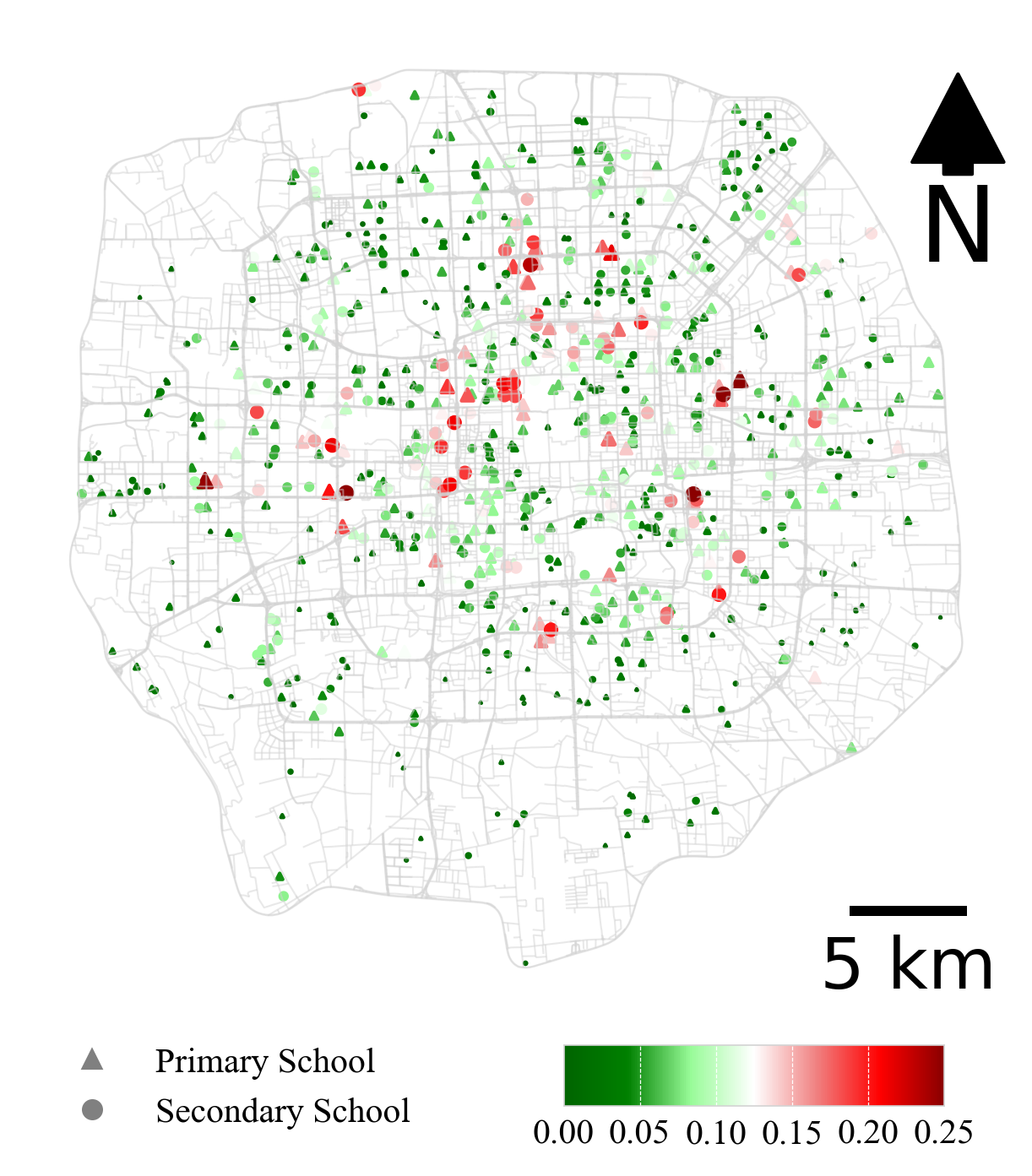}
         \caption{Spatial distribution of the probability of traffic congestion by school neighborhoods}
     \end{subfigure}
        \caption{The distribution of school run traffic congestion in the case studied area}
        \label{fig2:congestion}
\end{figure}
}

In addition, as shown in Fig.~\ref{fig2:congestion}(b), the proportion of congested roads varies markedly in different school neighborhoods. It suggests that school run traffic congestion should be understood according to built environment features associated with each school neighborhood. Therefore, we compute the average frequency of traffic congestion on all the roads within each school neighborhood as the indicator of the probability of traffic congestion in each school neighborhood. According to the spatial distribution of the derived indicators for all the schools (see Fig.~\ref{fig2:congestion}(c)), certain schools suffer from traffic congestion more frequently when compared with others. 

\subsubsection{Physical features extracted from traditional data}

To characterize the built environement for each school neighborhood, we further categorize the roads within the case studied area into three groups as ordinary road, main road, and express road. The spatial distributions of roads in the three categories are illustrated in Fig.~\ref{fig1:study_area}(a). Along with the configuration of roads, transport-related POIs (including subway stations, bus stops, parking lots), land uses (including green spaces, administration and public services, commercial and business facilities, educational areas, industrial areas, while residential areas are excluded because they are largely represented by building footprints and so to to avoid multicollinearity), and building footprints (including shapes, heights) around schools are also collected to meaure the physical characteristics of each school neighborhood.

Based on these datasets, we measure the physical characteristics of each school neighborhood in the case studies area from 4 distinct dimensions: (1) Spatial concentration, which quantifies the closeness between schools and the flux of population flow around schools. Note that the hourly population flux for each school neighborhood is derived from the mobile phone data provided by China Unicom.; (2) Transport facility, which accounts for the distributions of bus stops, subway stations, and parking lots around schools; (3) Road topology, which includes betweenness centrality, spatial integration, angular choice, and density of intersections derived from the underlying road network; (4) Spatial richness, which measures the mixture of land uses and building features. The definitions of those measurements are described in Table \ref{tab1:physical}.  

\begin{longtable}{p{0.2\textwidth}p{0.2\textwidth}p{0.15\textwidth}p{0.35\textwidth}}
\caption{Variables with regard to physical characteristics of the built environment} \\
\toprule
Category & Name & Symbol & Description \\
\midrule
\endfirsthead

\multicolumn{4}{l}%
{\tablename\ \thetable\ -- \textit{Continued from previous page}} \\
\toprule
Category & Name & Symbol & Description \\
\midrule
\endhead

\midrule \multicolumn{4}{r}{\textit{Continued on next page}} \\
\endfoot

\bottomrule
\endlastfoot

\multirow{11}{*}{Spatial concentration}  & No. of School & $school\_mix$ & Presence of neighboring schools; 0 for no, 1 for yes \\
& Angle from City East & $angle$&  Angles measured counterclockwise from east, ranging from 0 to 360 degrees \\
 & Distance to City Center & $distance$ &  The radial distance in the polar coordinate system from Tiananmen Square (in km)\\
   & Population Flow & $population$ &  Whether the count of mobile signalings is greater than 10,000; 0 for no, 1 for yes \\
\midrule
\multirow{6}{*}{Transport facility} & No. of Bus Stop & $bus\_stop$ &  Whether there are more than 5 bus stops; 0 for no, 1 for yes \\
 & Access to Subway& $subway$ &  Presence of subway station; 0 for no, 1 for yes \\
 & No. of Parking Lot &  $parking\_lot$ & Availability of more than 50 parking lots; 0 for no, 1 for yes \\
\midrule
& Betweenness Centrality &  $betweeness$ & Average road betweeness centrality in the school neighborhood \\
\multirow{7}{*}{Road topology}  & Spatial Integration & $integration$ & Mean value of road integration in the school neighborhood; see \cite{Van2021} \\
& Spatial Choice & $choice$ & Mean value of road choice in the school neighborhood; also see \cite{Van2021} \\
& Intersection Density &  $intersecton$ & No. of road intersection in the school neighborhood \\
\midrule
\multirow{13}{*}{Spatial richness} & Mix of New / Old Building & $building\_age$ &  Whether the difference between the percentages of old and new buildings is less than 10\%; 0 for no, 1 for yes \\
 & Average Building Height & $building\_height$ &  Whether the average height of buildings is larger than 6 stories; 0 for no, 1 for yes \\
& Mix of High / Low Building & $building\_mix$ & Whetherthe difference between the percentages of high and low buildings is less than 30\%; 0 for no, 1 for yes \\
& Mix of Land Use  & $landuse\_mix$  & Whether there are 5 land use categories; 0 for no, 1 for yes
\label{tab1:physical}
\end{longtable}

\subsubsection{Scene-centric features extracted from SVIs}

Recall that SVIs provide visual clues for measuring the built environment in a more subjective and holistic manner. We therefore collect the panoramic SVIs at every 50m intervals along the obtained road network via the API provided by Baidu Map (\url{https://lbsyun.baidu.com/}). The widely-used Places365-CNNs model is applied to extract scene-centric features from the collected SVIs. The model derives 365 predefined scene labels and their corresponding probabilities from each SVI, out of which the 159 outdoor, man-made scene labels are adopted for further analysis.  

\afterpage{%
  \begin{figure}[H]
     \centering
     \begin{subfigure}[b]{0.3\textwidth}
         \centering
         \includegraphics[width=\textwidth]{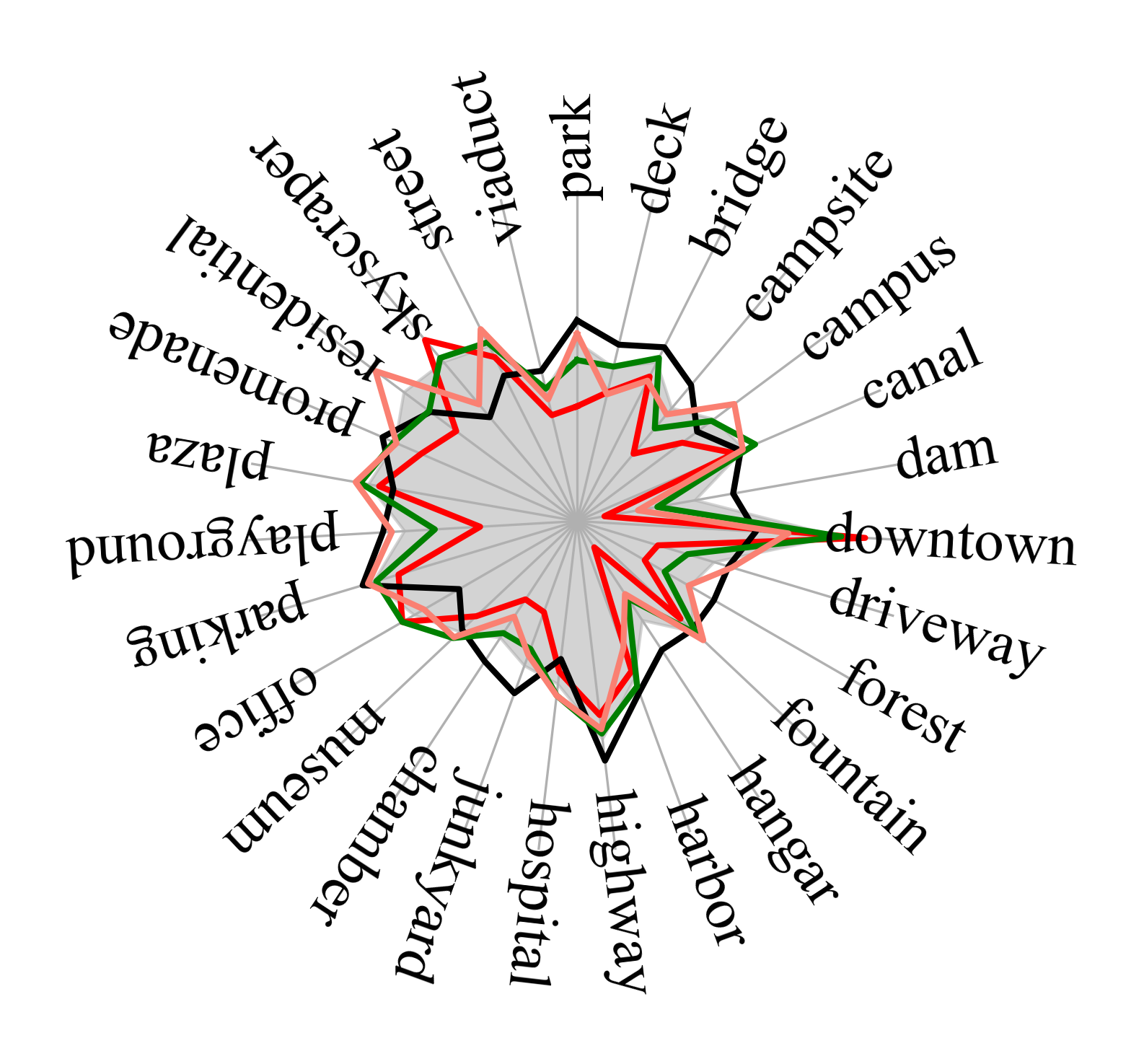}
         \caption{Ordinary roads}
         \label{fig:y equals x}
     \end{subfigure}
     \hfill
     \begin{subfigure}[b]{0.3\textwidth}
         \centering
         \includegraphics[width=\textwidth]{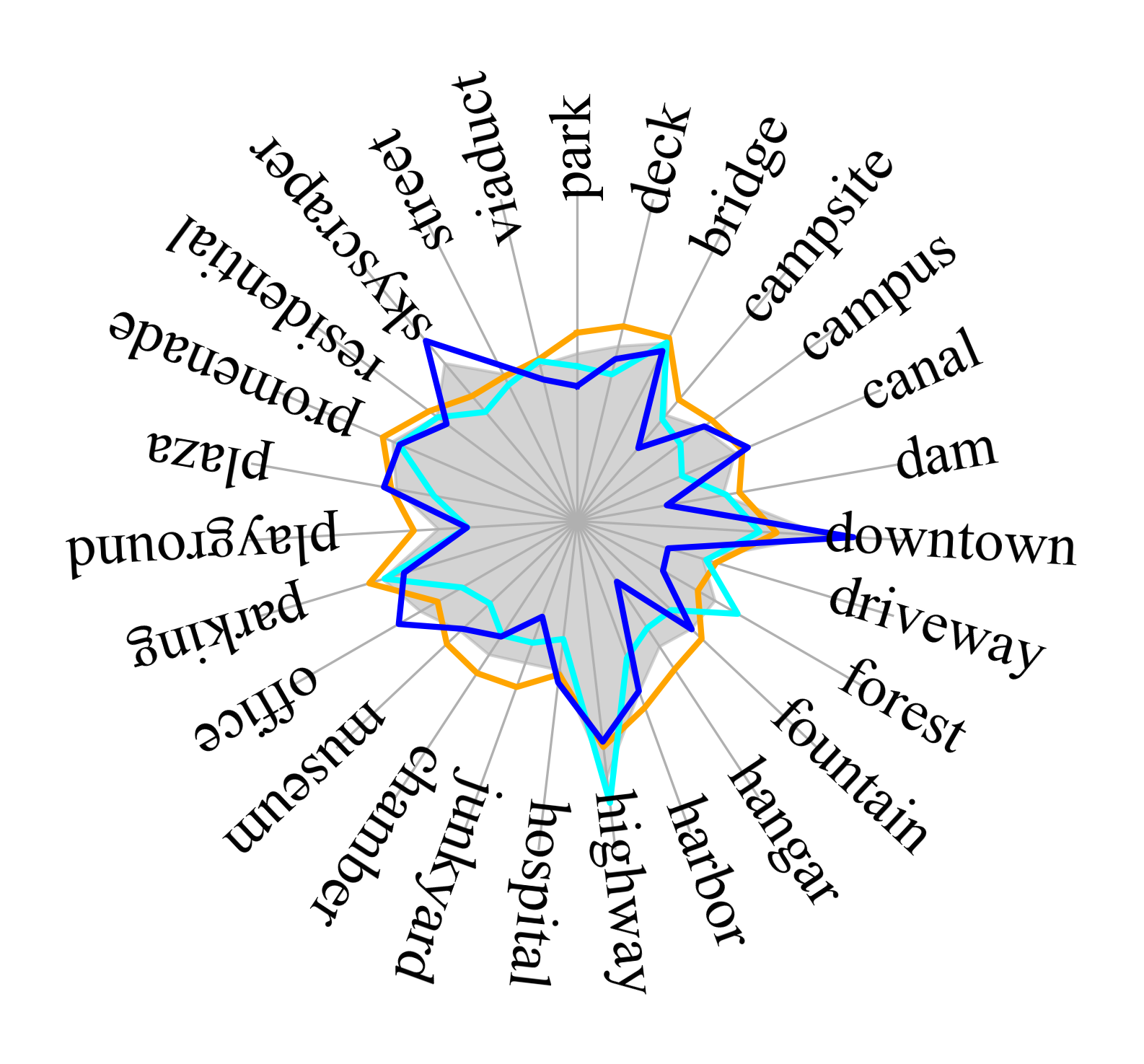}
         \caption{Main roads}
         \label{fig:three sin x}
     \end{subfigure}
     \hfill
     \begin{subfigure}[b]{0.3\textwidth}
         \centering
         \includegraphics[width=\textwidth]{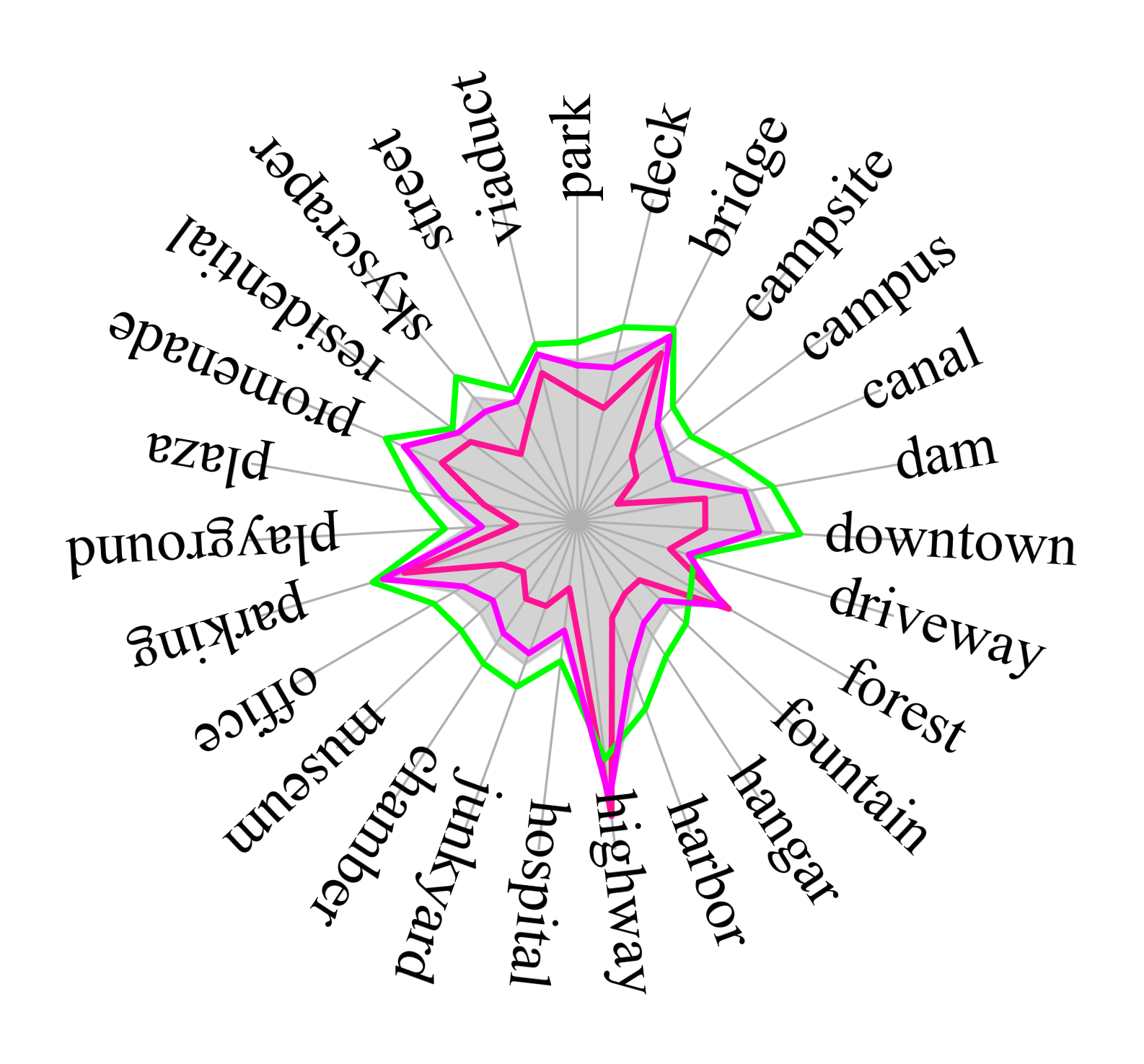}
         \caption{Express roads}
         \label{fig:five over x}
     \end{subfigure}
     \break
          \begin{subfigure}[b]{\textwidth}
         \centering
         \begin{tikzpicture}[      
        every node/.style={anchor=south west,inner sep=0pt},
        x=1mm, y=1mm,
      ]   
     \node (fig1) at (0,0)
       {\includegraphics[scale=0.7]{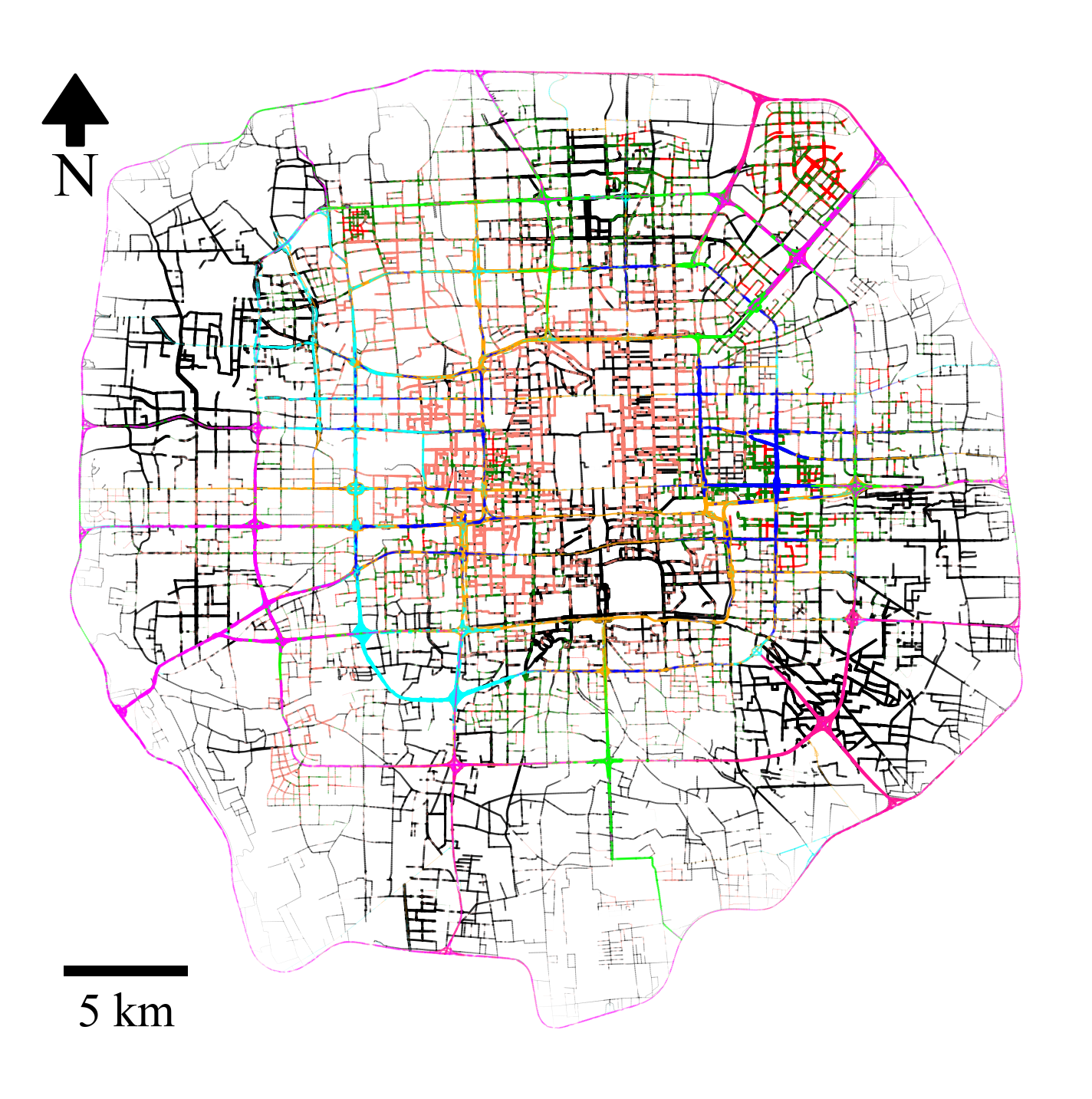}};
     \draw[draw=red, fill=red] (85,75) rectangle ++(4,7) node[above, xshift=-0.2cm,yshift=-0.5cm, color=white] {1};
     \node (fig2) at (90,75)
       {\includegraphics[scale=0.04]{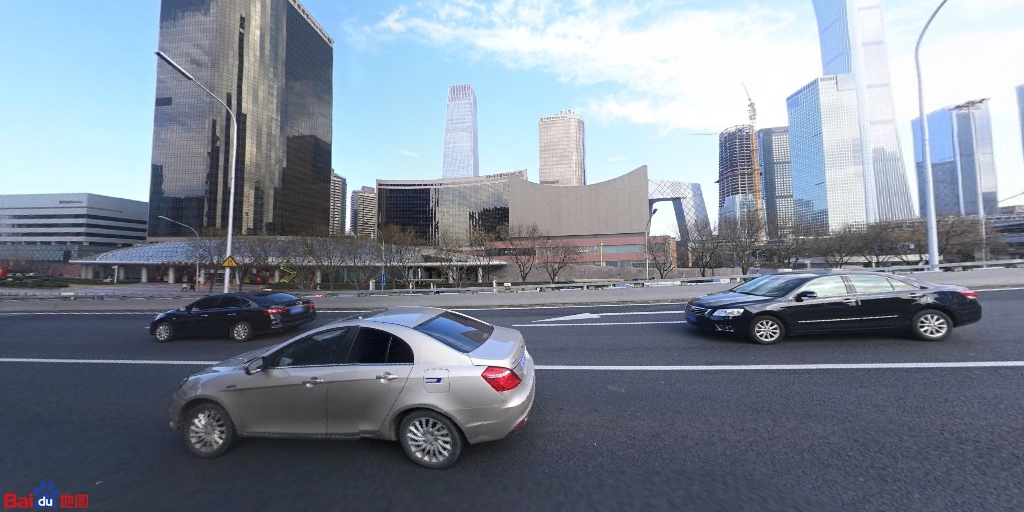}};  
       \node (fig2) at (105,75)
       {\includegraphics[scale=0.04]{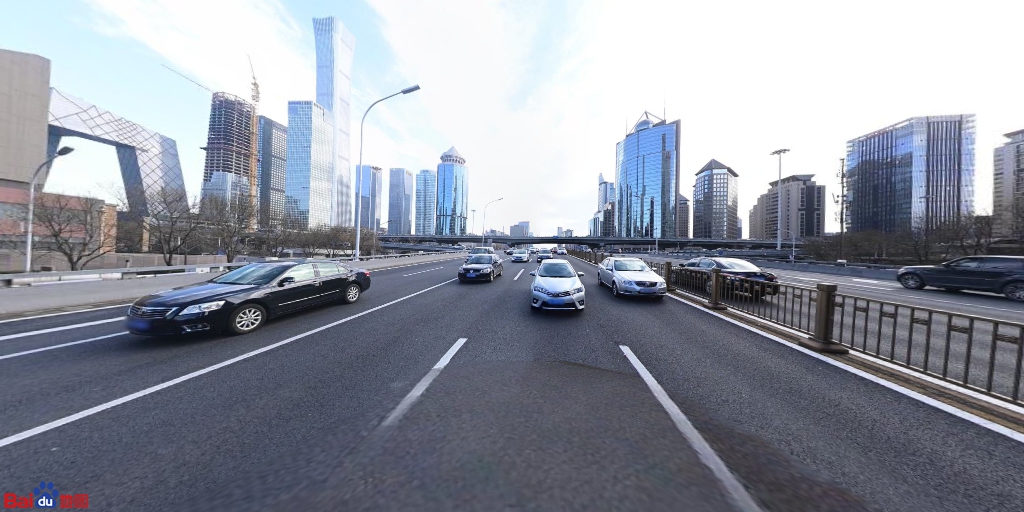}};  
       \node (fig2) at (120,75)
       {\includegraphics[scale=0.04]{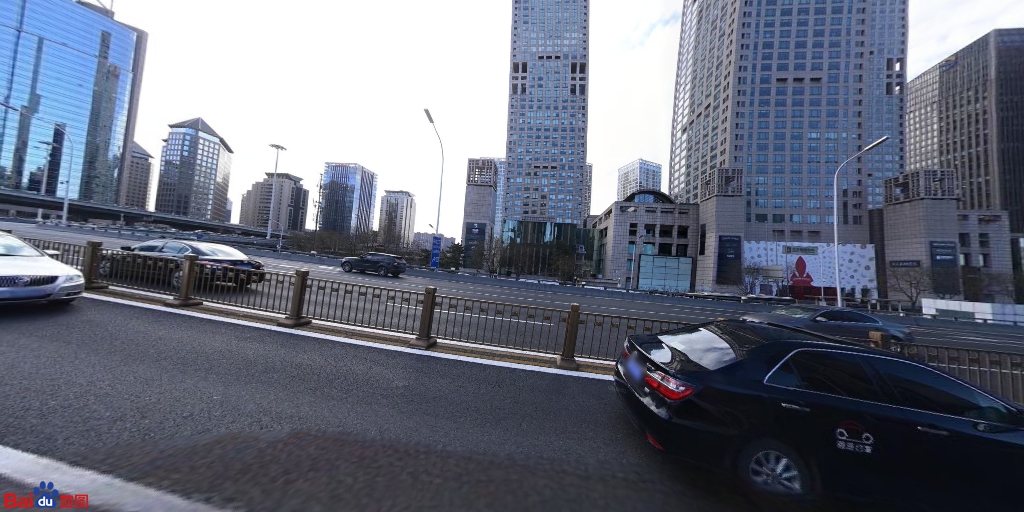}};  
       \node (fig2) at (135,75)
       {\includegraphics[scale=0.04]{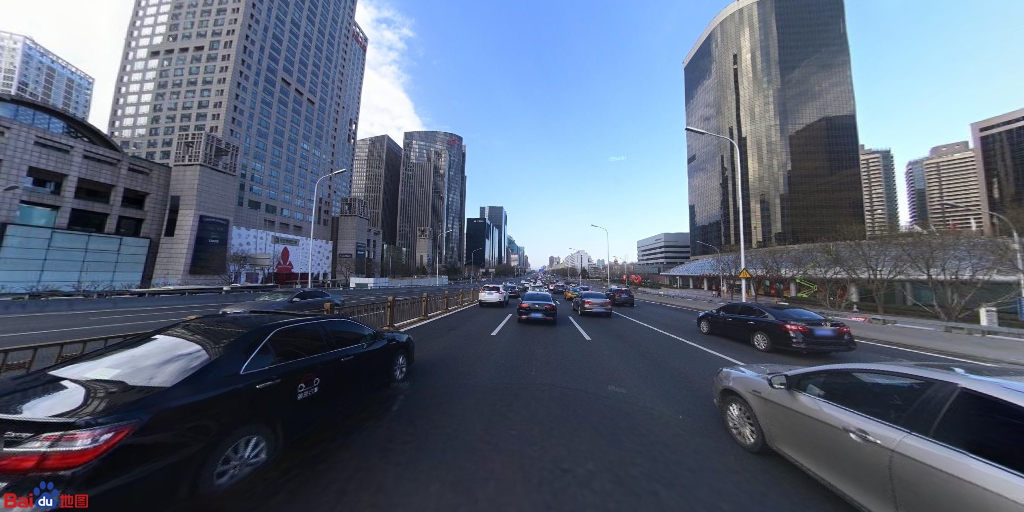}};  
       \draw[draw=black, fill=black] (85,67) rectangle ++(4,7) node[above, xshift=-0.2cm,yshift=-0.5cm, color=white] {2};
            \node (fig2) at (90,67)
       {\includegraphics[scale=0.04]{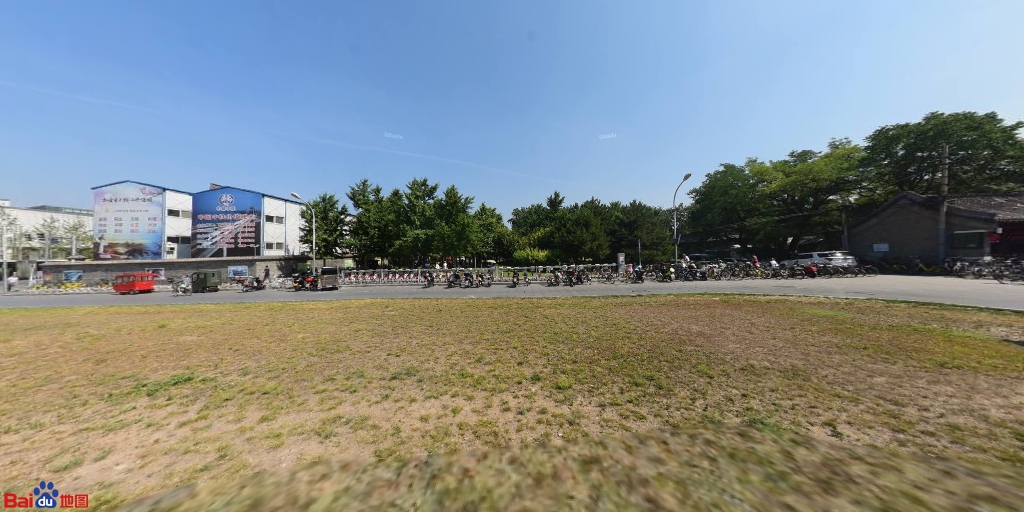}};  
       \node (fig2) at (105,67)
       {\includegraphics[scale=0.04]{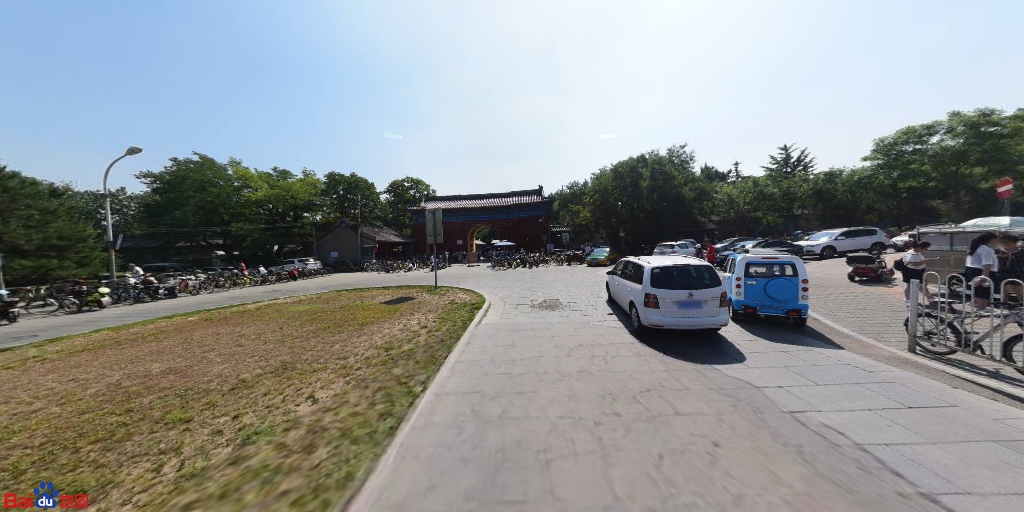}};  
       \node (fig2) at (120,67)
       {\includegraphics[scale=0.04]{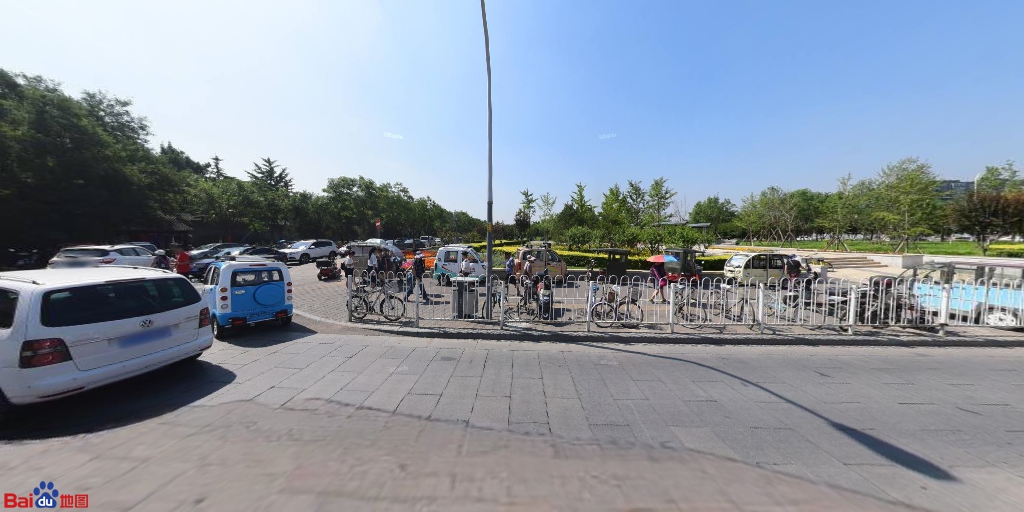}};  
       \node (fig2) at (135,67)
       {\includegraphics[scale=0.04]{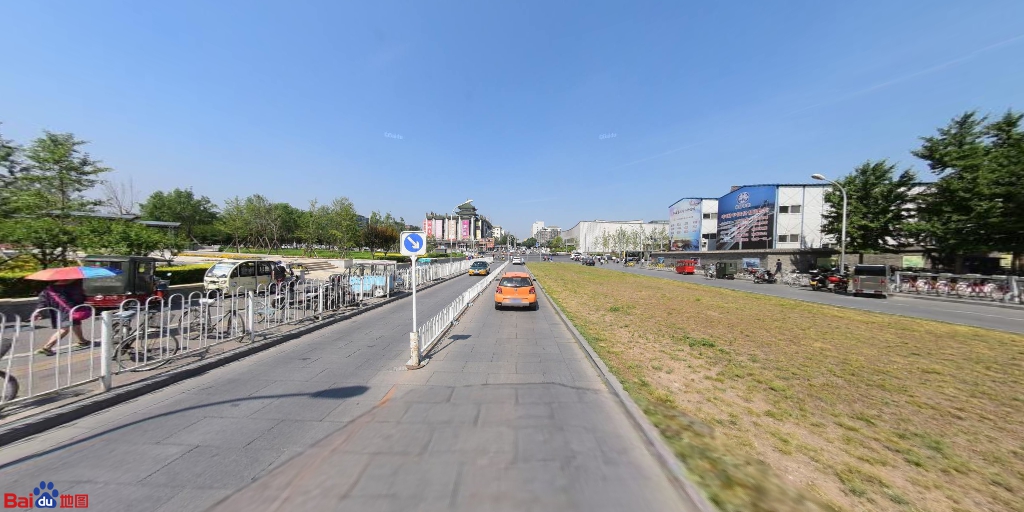}};  
        \draw[draw=black!50!green, fill=black!50!green] (85,59) rectangle ++(4,7) node[above, xshift=-0.2cm,yshift=-0.5cm, color=white] {3};
            \node (fig2) at (90,59)
       {\includegraphics[scale=0.04]{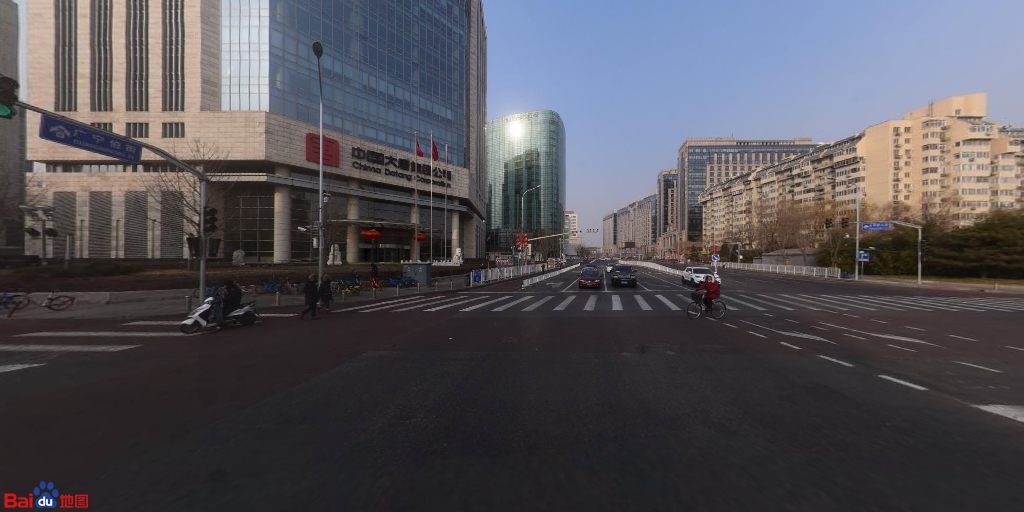}};  
       \node (fig2) at (105,59)
       {\includegraphics[scale=0.04]{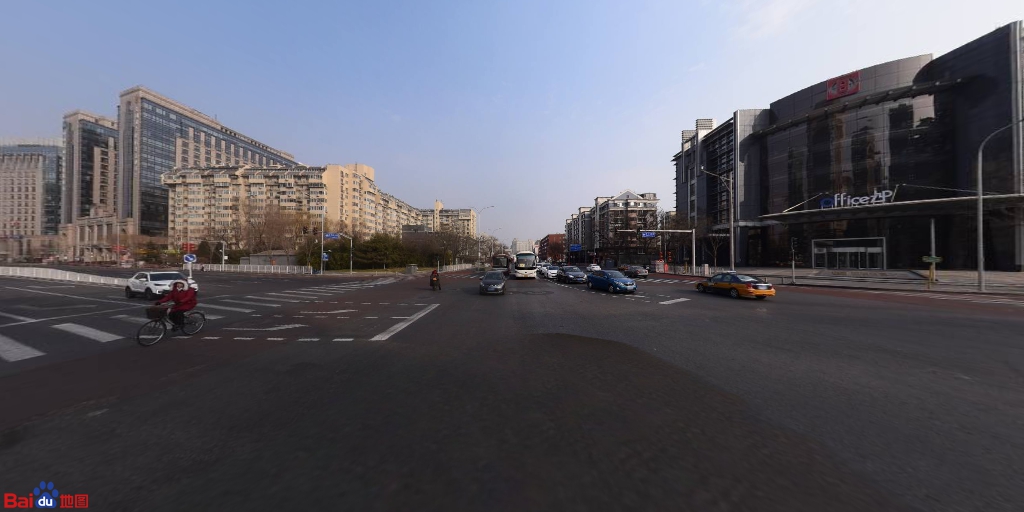}};  
       \node (fig2) at (120,59)
       {\includegraphics[scale=0.04]{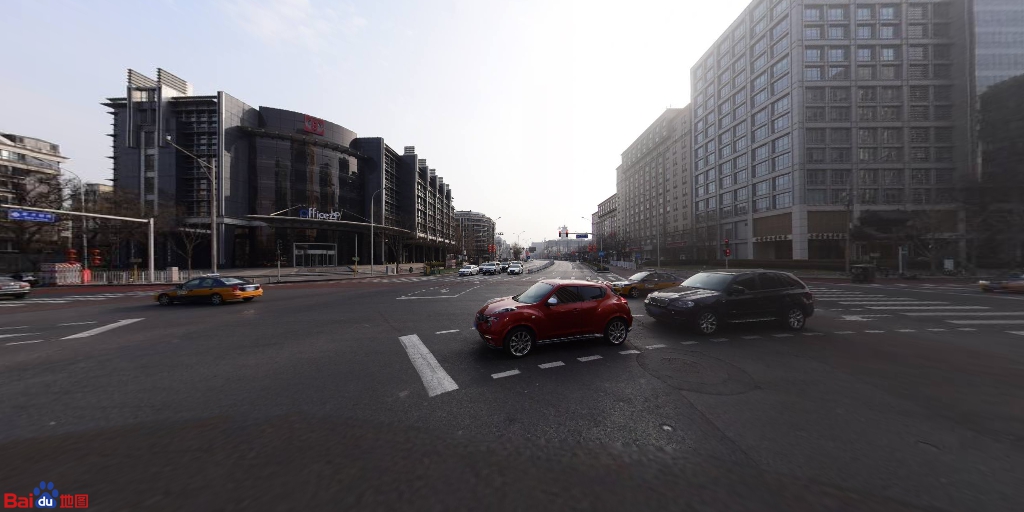}};  
       \node (fig2) at (135,59)
       {\includegraphics[scale=0.04]{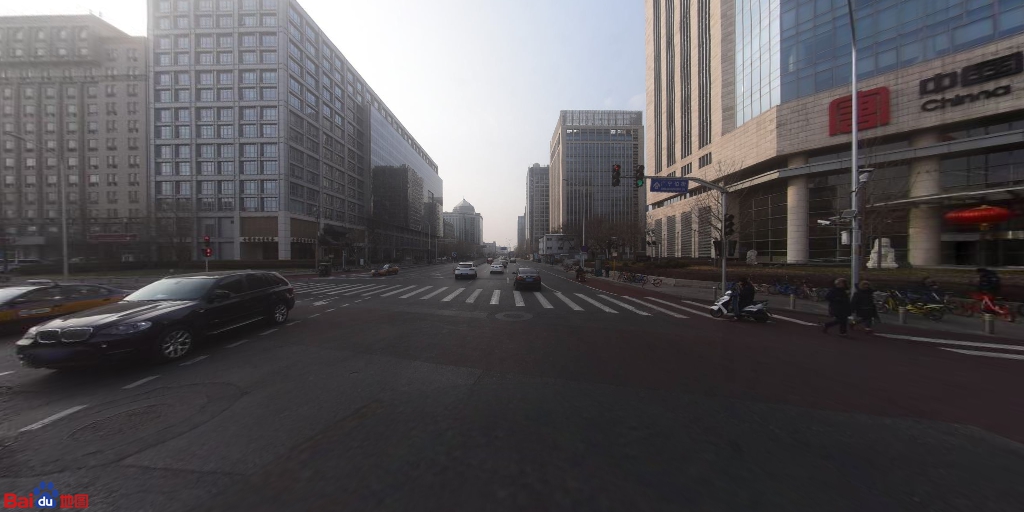}};  
        \draw[draw=black!5!pink, fill=black!5!pink] (85,51) rectangle ++(4,7) node[above, xshift=-0.2cm,yshift=-0.5cm, color=white] {4};
            \node (fig2) at (90,51)
       {\includegraphics[scale=0.04]{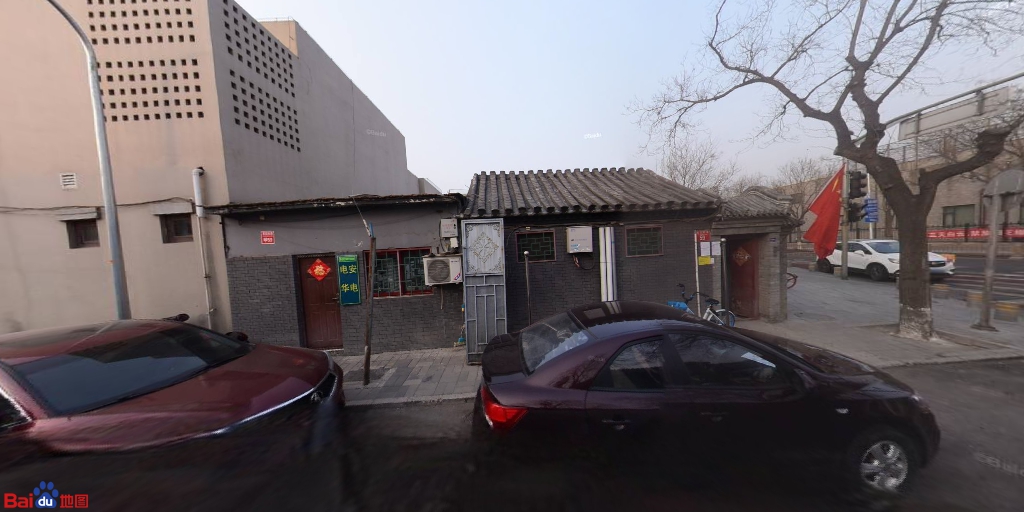}};  
       \node (fig2) at (105,51)
       {\includegraphics[scale=0.04]{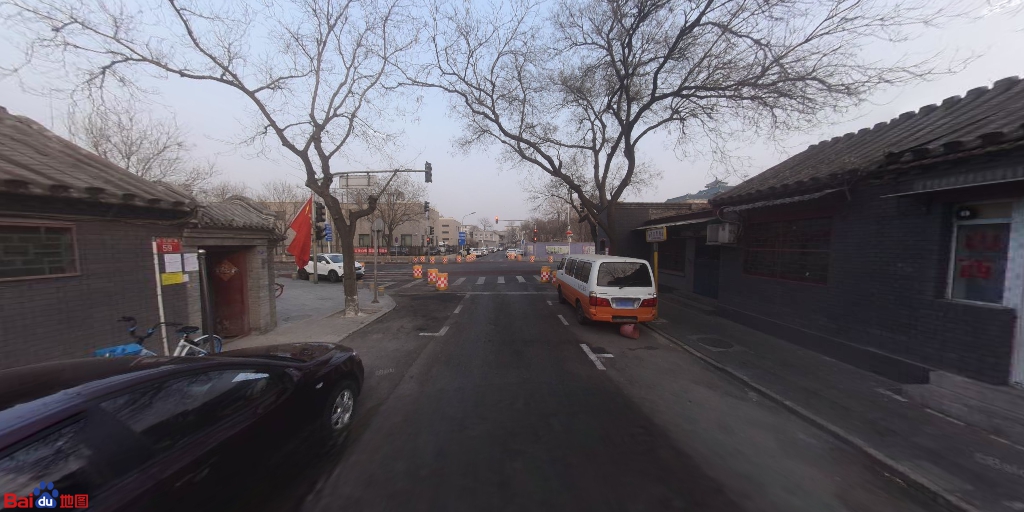}};  
       \node (fig2) at (120,51)
       {\includegraphics[scale=0.04]{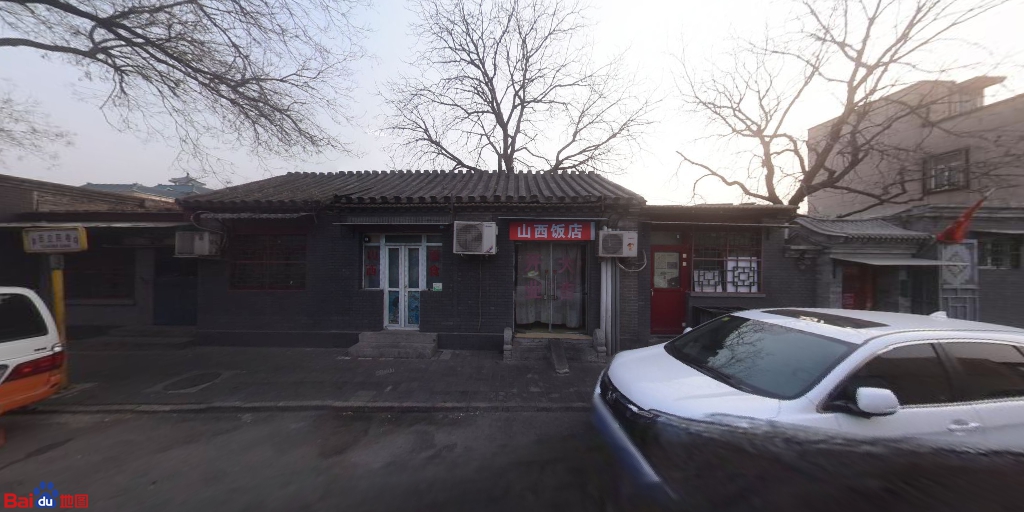}};  
       \node (fig2) at (135,51)
       {\includegraphics[scale=0.04]{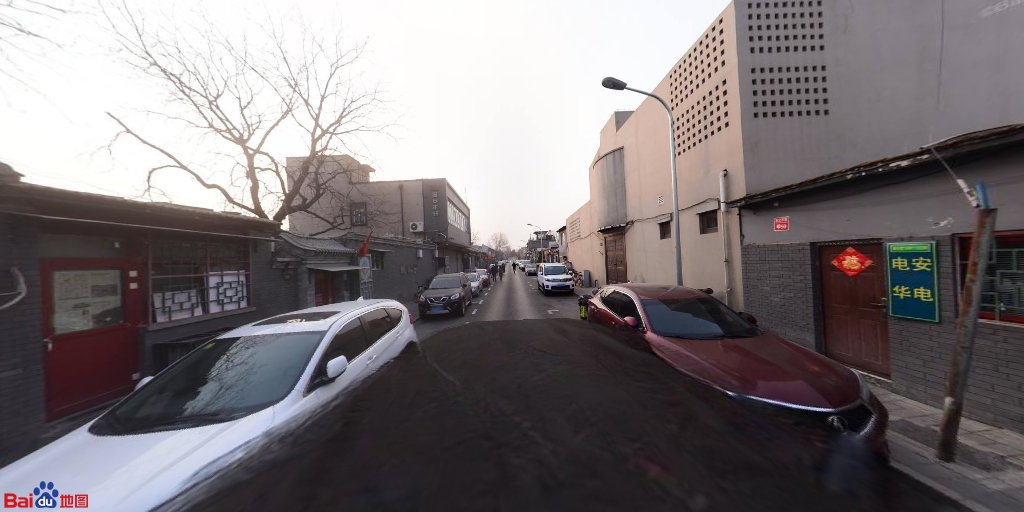}};  
        \draw[draw=orange, fill=orange] (85,43) rectangle ++(4,7) node[above, xshift=-0.2cm,yshift=-0.5cm, color=white] {5};
            \node (fig2) at (90,43)
       {\includegraphics[scale=0.04]{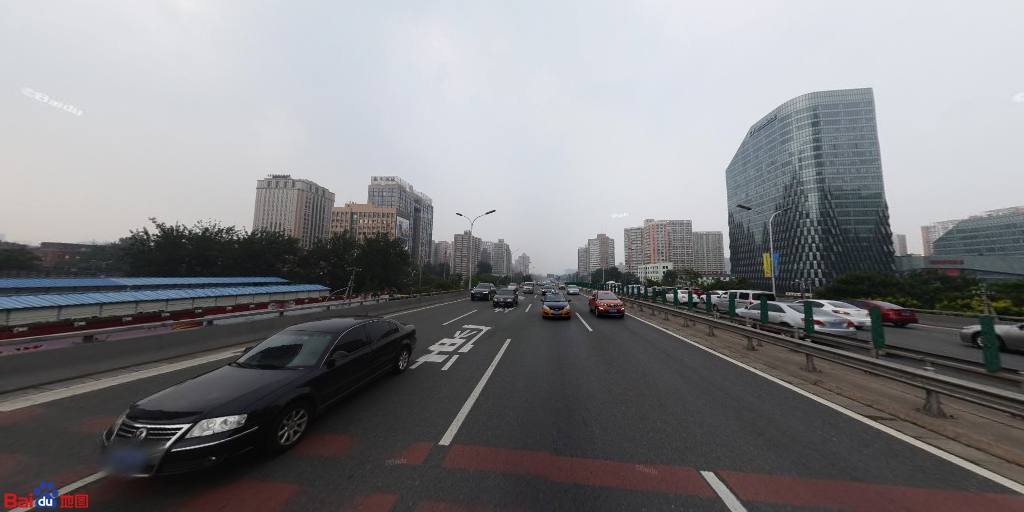}};  
       \node (fig2) at (105,43)
       {\includegraphics[scale=0.04]{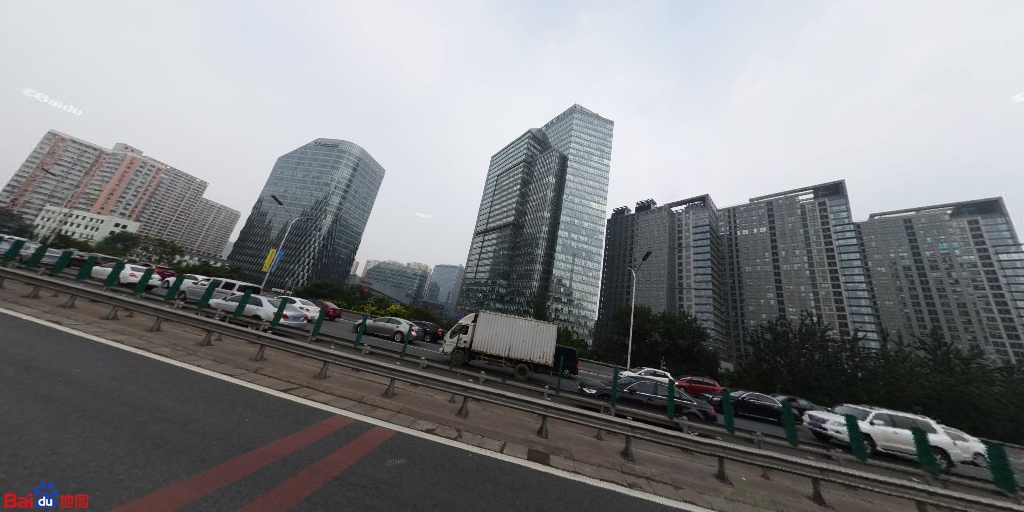}};  
       \node (fig2) at (120,43)
       {\includegraphics[scale=0.04]{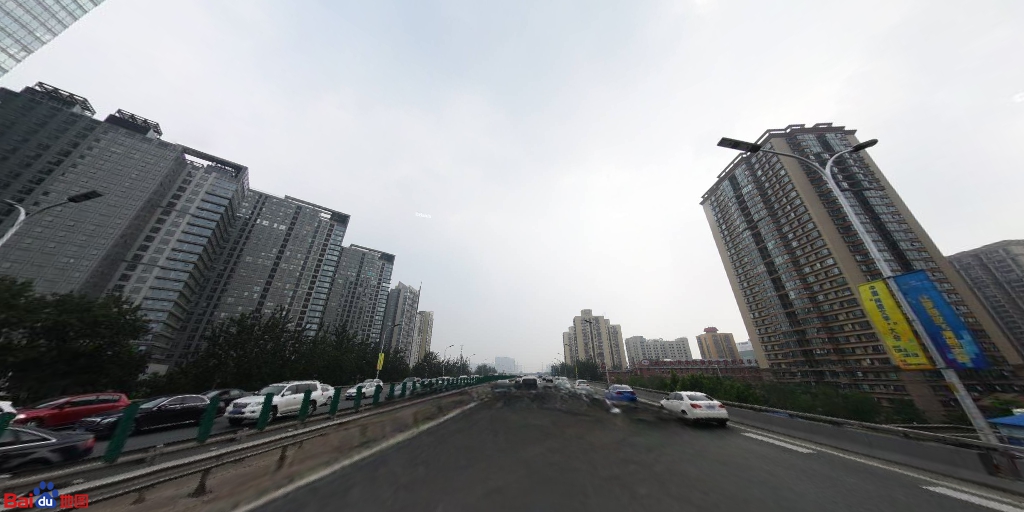}};  
       \node (fig2) at (135,43)
       {\includegraphics[scale=0.04]{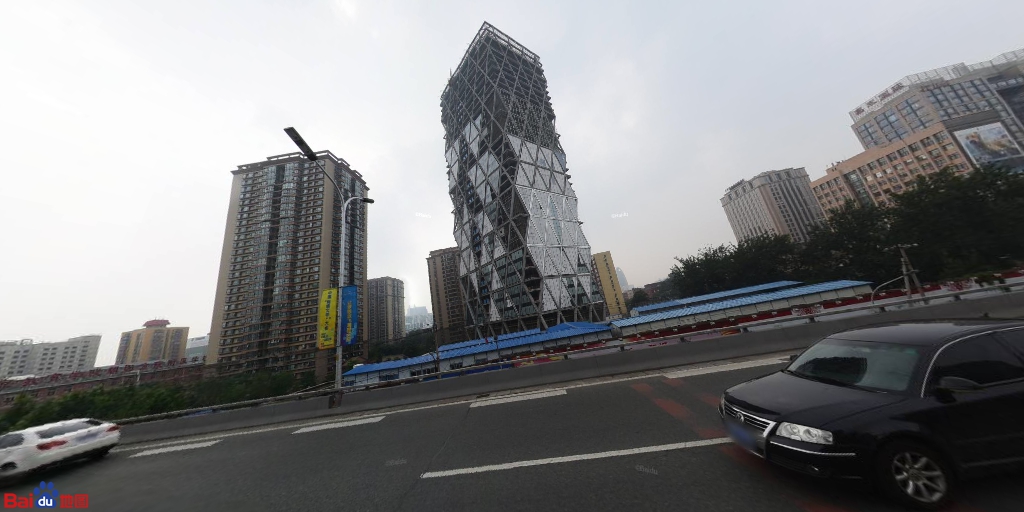}};  
        \draw[draw=cyan!50, fill=cyan!50] (85,35) rectangle ++(4,7) node[above, xshift=-0.2cm,yshift=-0.5cm, color=white] {6};
            \node (fig2) at (90,35)
       {\includegraphics[scale=0.04]{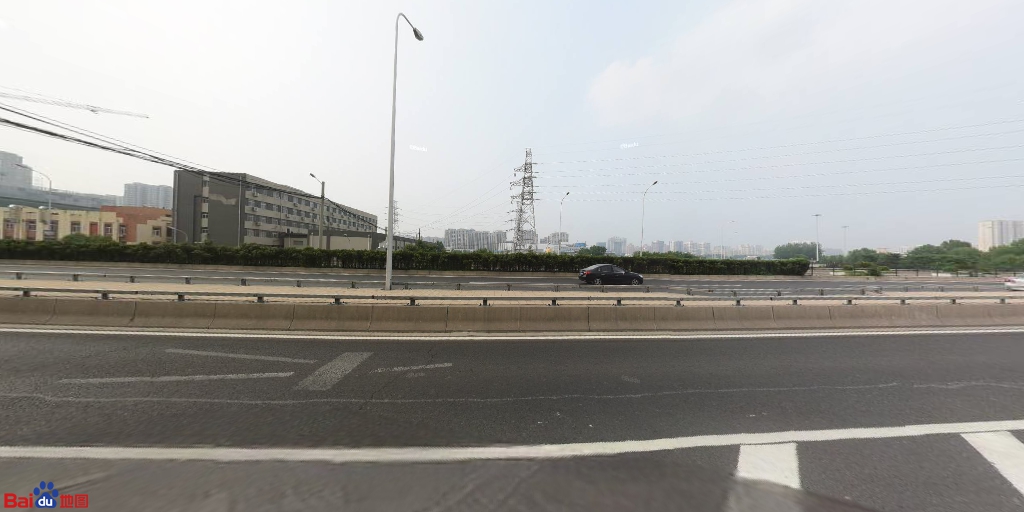}};  
       \node (fig2) at (105,35)
       {\includegraphics[scale=0.04]{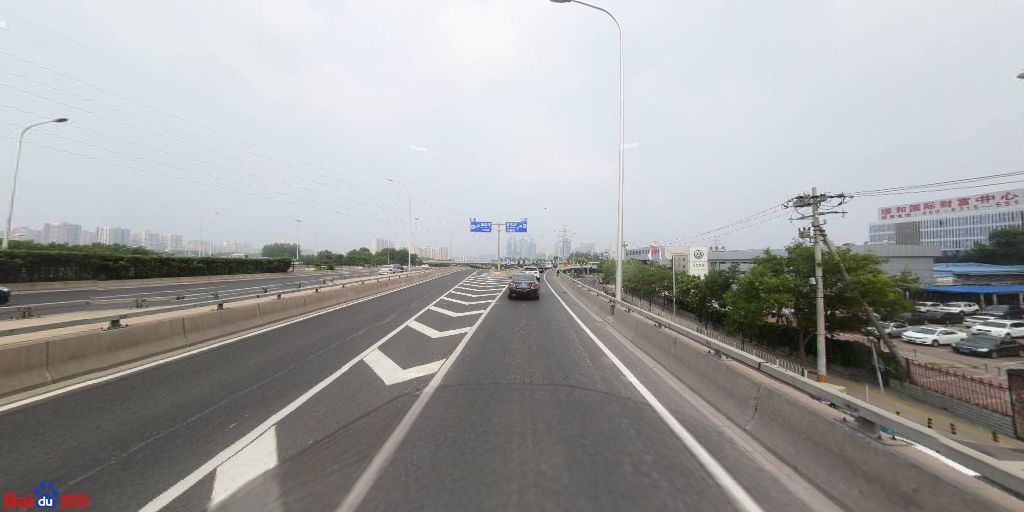}};  
       \node (fig2) at (120,35)
       {\includegraphics[scale=0.04]{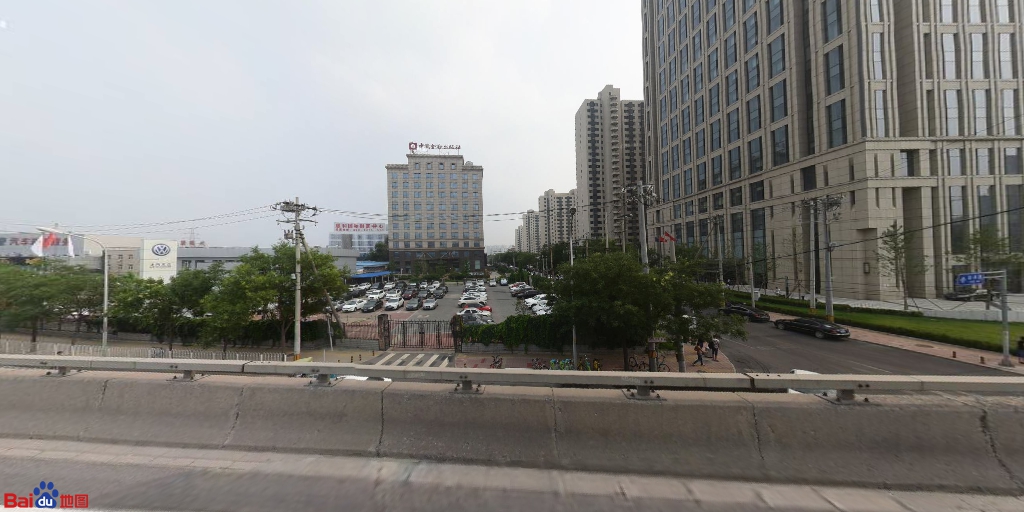}};  
       \node (fig2) at (135,35)
       {\includegraphics[scale=0.04]{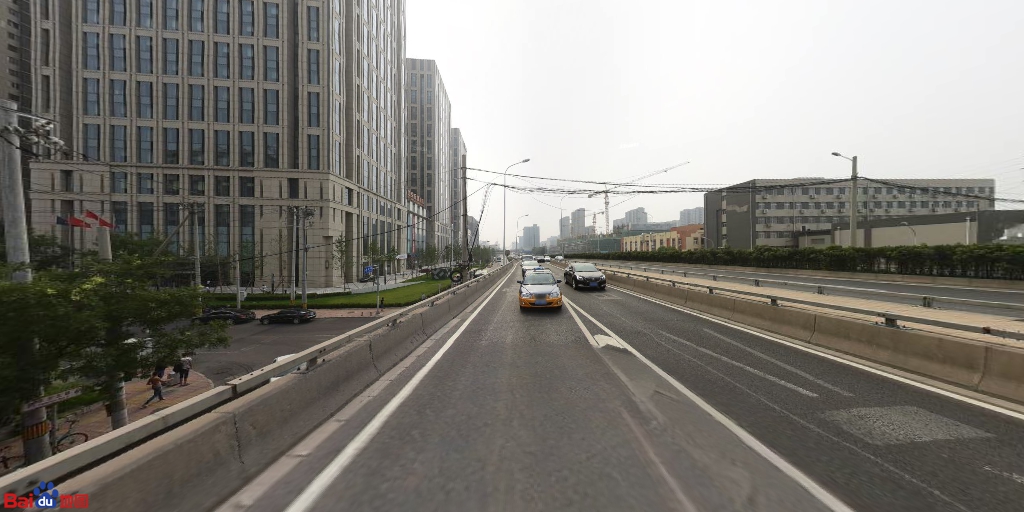}};  
        \draw[draw=blue, fill=blue] (85,27) rectangle ++(4,7) node[above, xshift=-0.2cm,yshift=-0.5cm, color=white] {7};
            \node (fig2) at (90,27)
       {\includegraphics[scale=0.04]{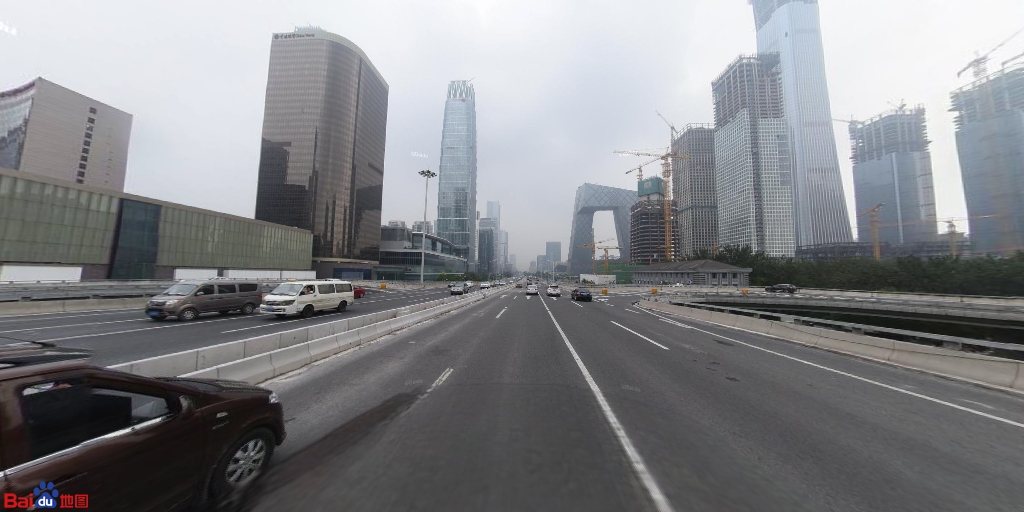}};  
       \node (fig2) at (105,27)
       {\includegraphics[scale=0.04]{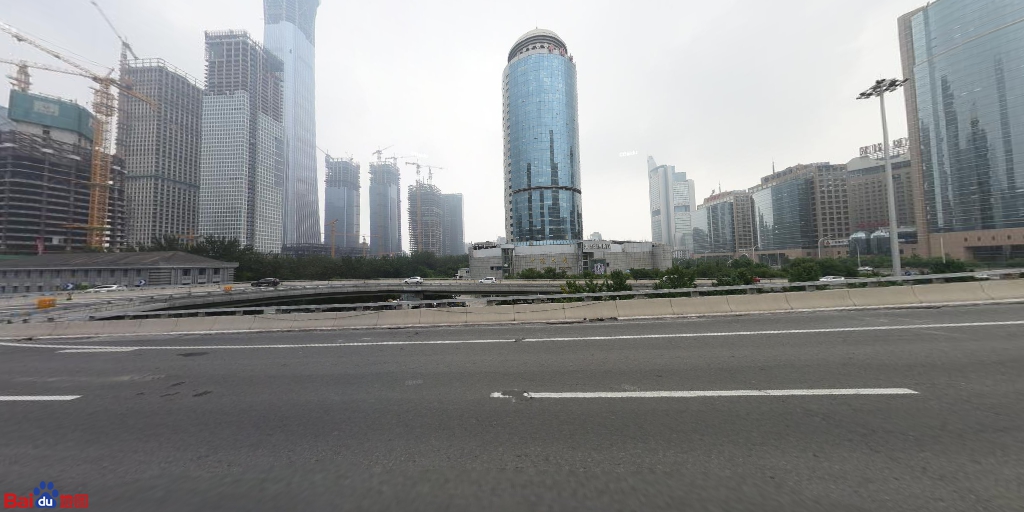}};  
       \node (fig2) at (120,27)
       {\includegraphics[scale=0.04]{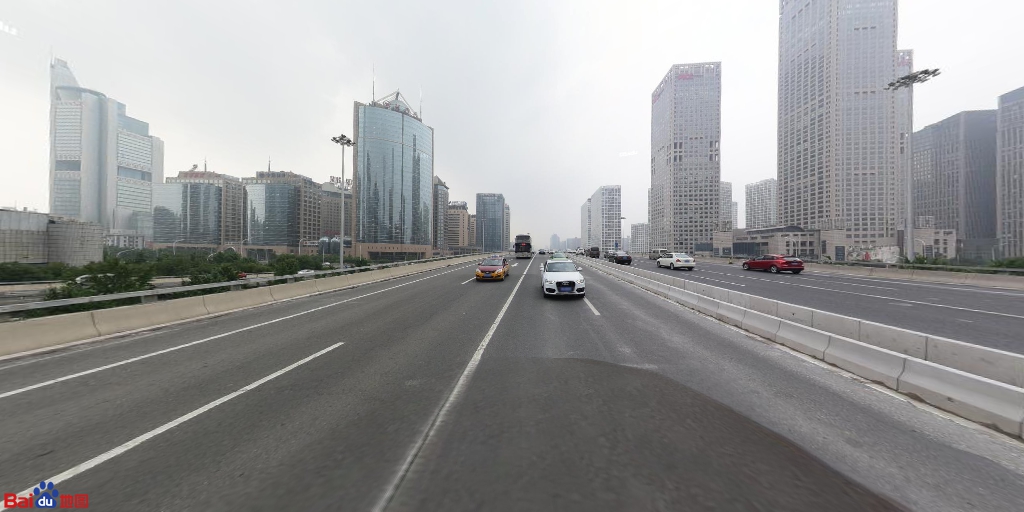}};  
       \node (fig2) at (135,27)
       {\includegraphics[scale=0.04]{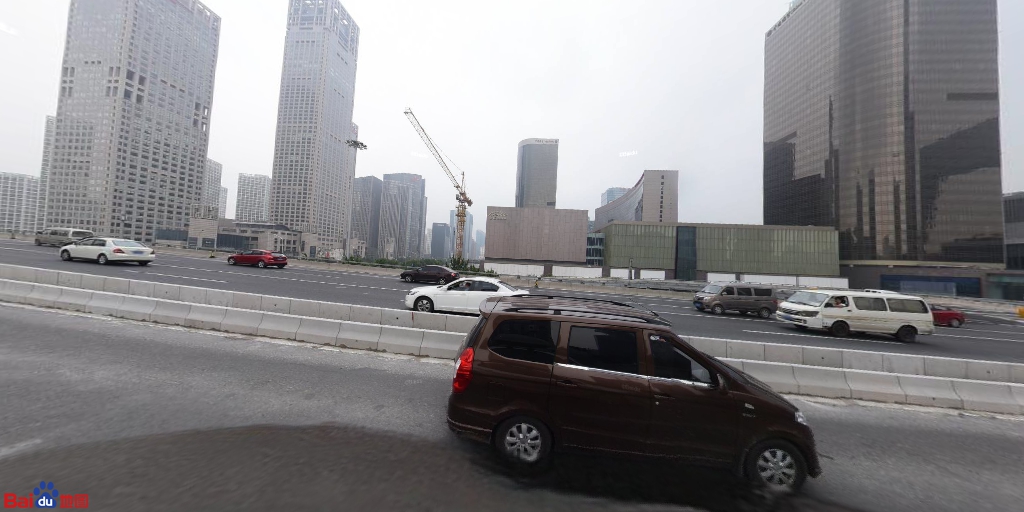}};  
        \draw[draw=magenta, fill=magenta] (85,19) rectangle ++(4,7) node[above, xshift=-0.2cm,yshift=-0.5cm, color=white] {8};
            \node (fig2) at (90,19)
       {\includegraphics[scale=0.04]{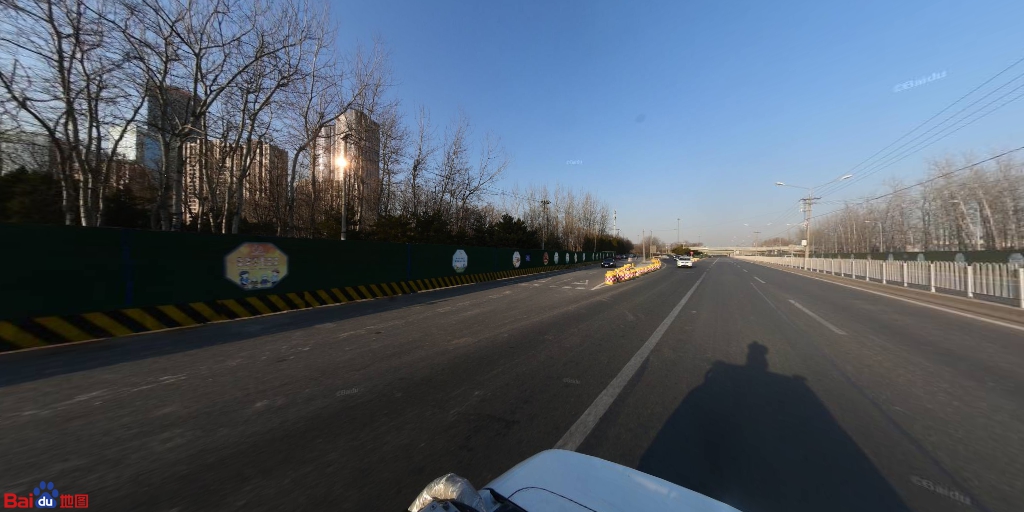}};  
       \node (fig2) at (105,19)
       {\includegraphics[scale=0.04]{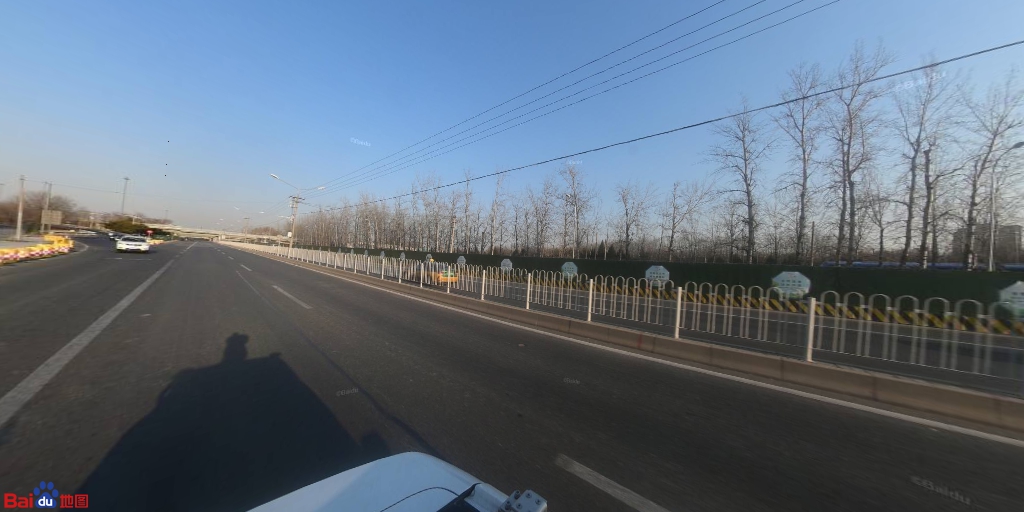}};  
       \node (fig2) at (120,19)
       {\includegraphics[scale=0.04]{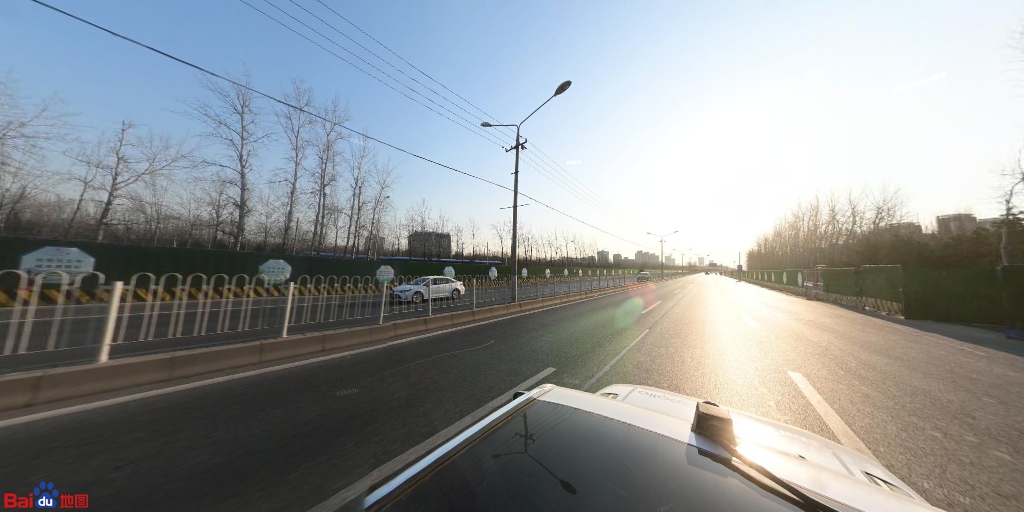}};  
       \node (fig2) at (135,19)
       {\includegraphics[scale=0.04]{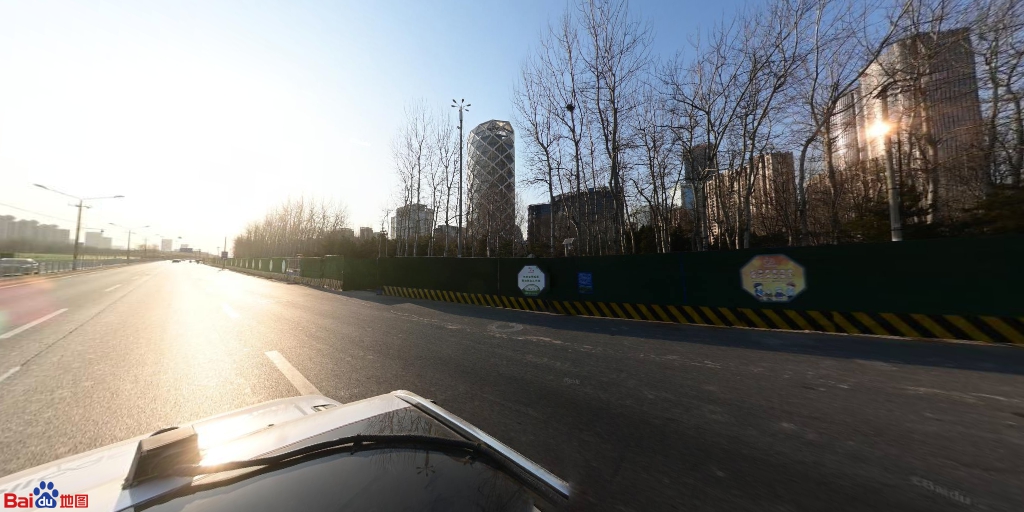}};  
        \draw[draw=green, fill=green] (85,11) rectangle ++(4,7) node[above, xshift=-0.2cm,yshift=-0.5cm, color=white] {9};
            \node (fig2) at (90,11)
       {\includegraphics[scale=0.04]{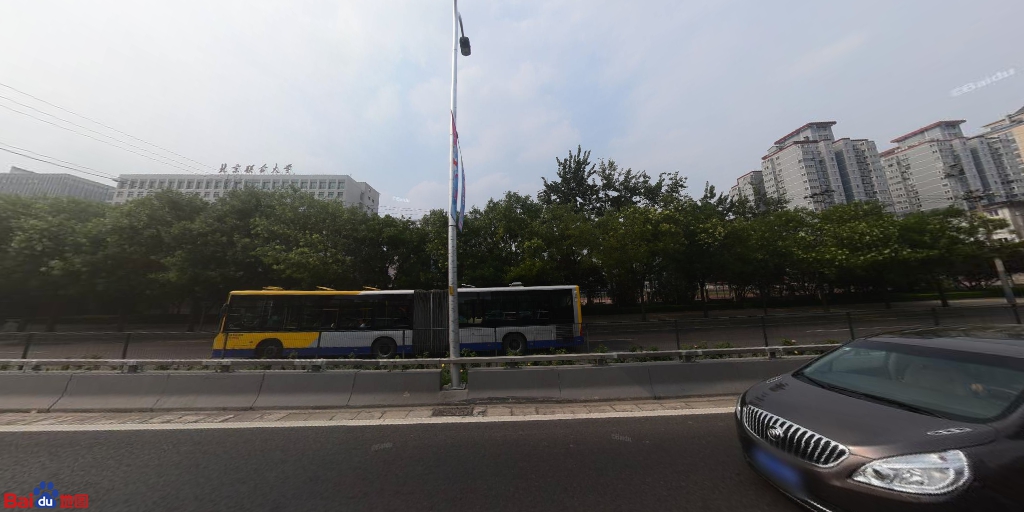}};  
       \node (fig2) at (105,11)
       {\includegraphics[scale=0.04]{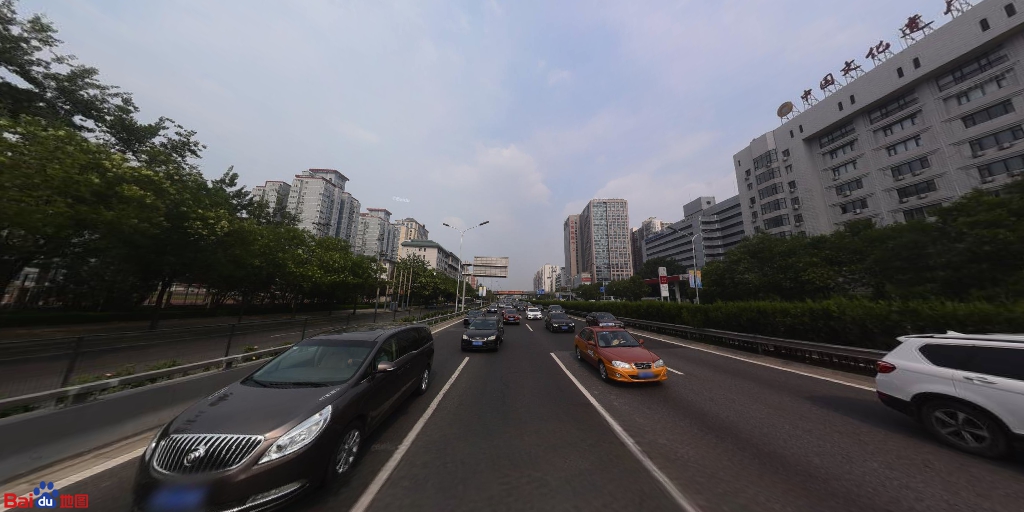}};  
       \node (fig2) at (120,11)
       {\includegraphics[scale=0.04]{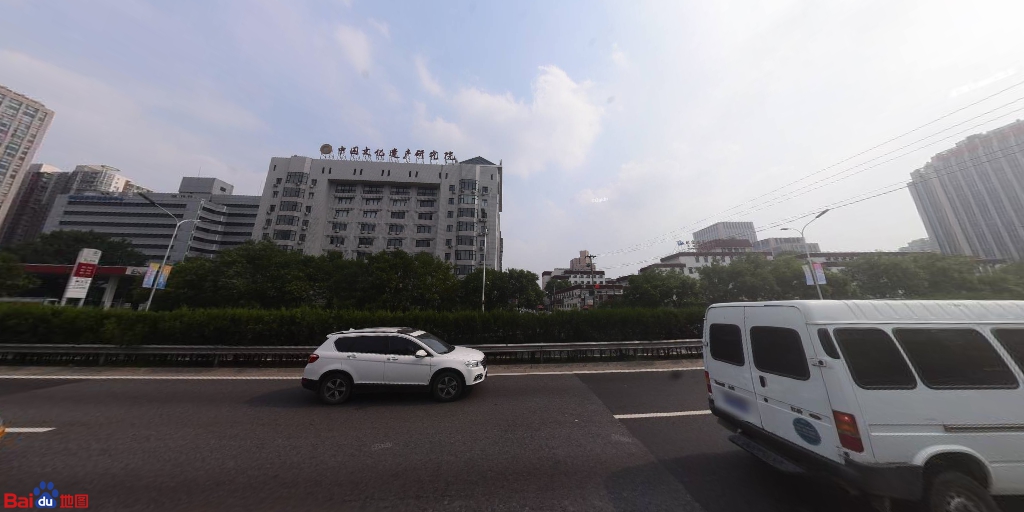}};  
       \node (fig2) at (135,11)
       {\includegraphics[scale=0.04]{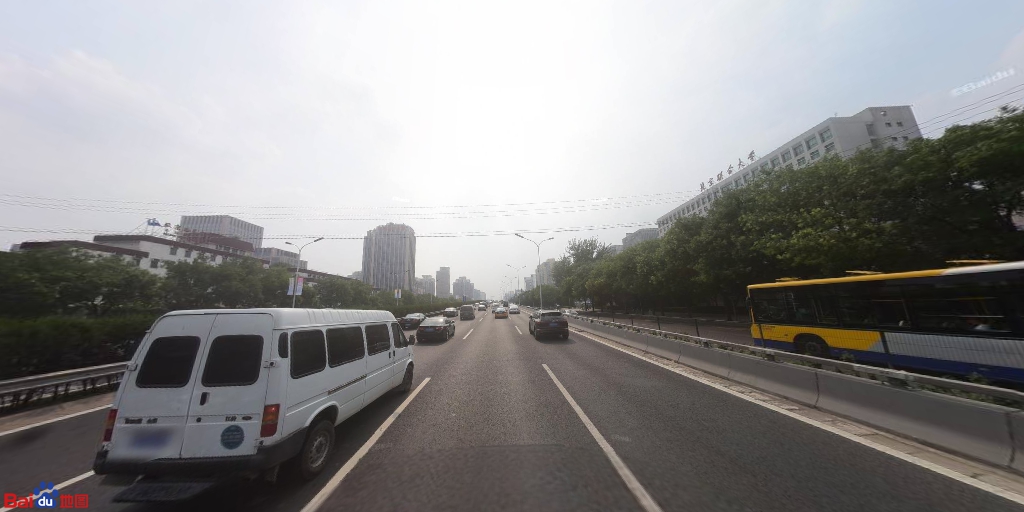}};  
        \draw[draw=violet!50, fill=violet!50] (85,3) rectangle ++(4,7) node[above, xshift=-0.2cm,yshift=-0.5cm, color=white] {10};
            \node (fig2) at (90,3)
       {\includegraphics[scale=0.04]{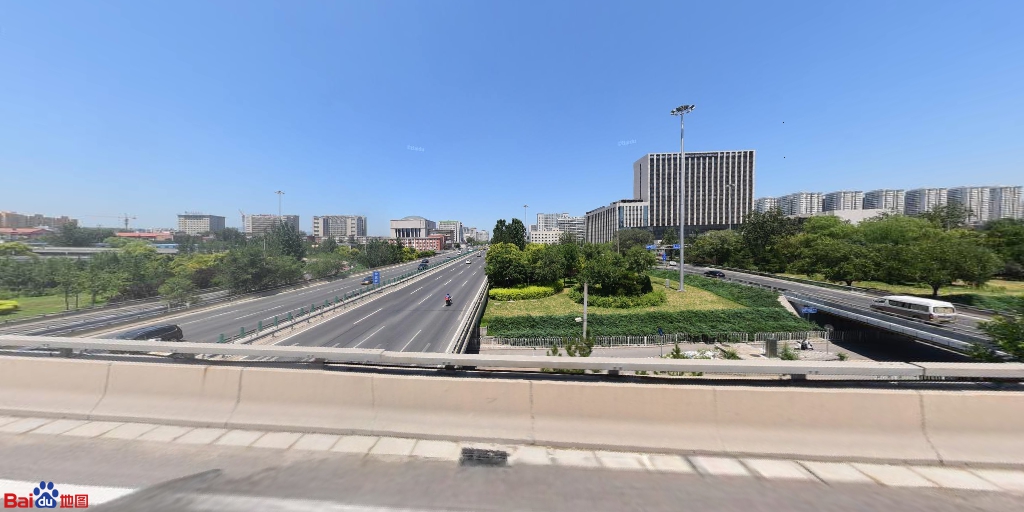}};  
       \node (fig2) at (105,3)
       {\includegraphics[scale=0.04]{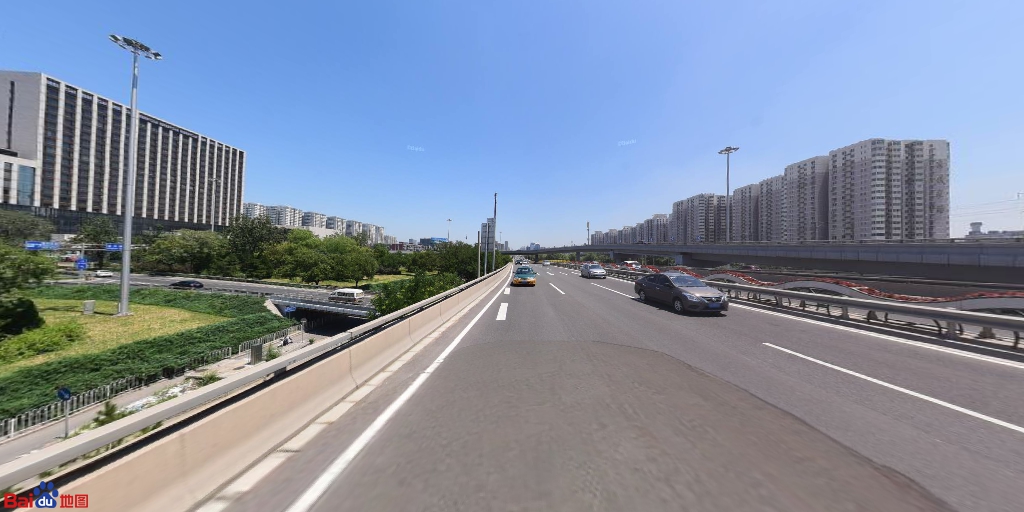}};  
       \node (fig2) at (120,3)
       {\includegraphics[scale=0.04]{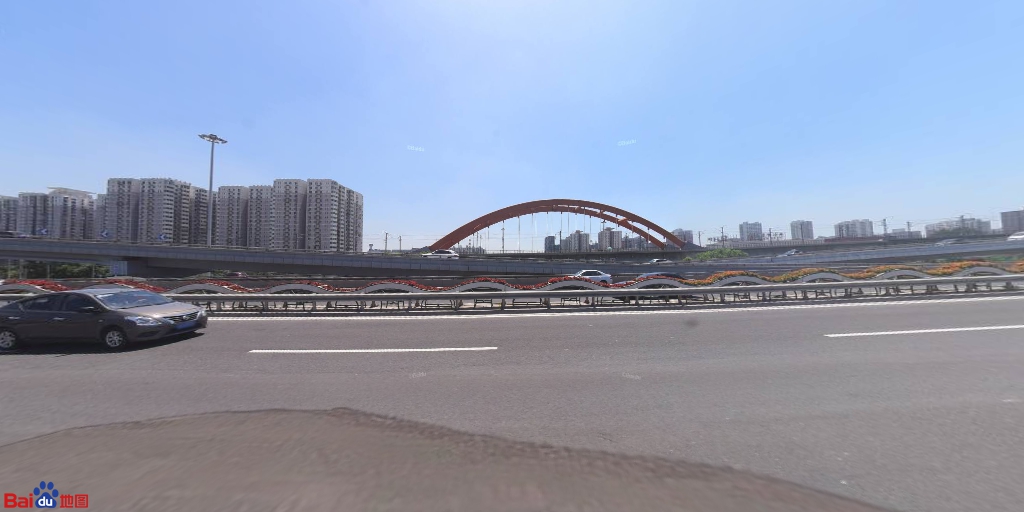}};  
       \node (fig2) at (135,3)
       {\includegraphics[scale=0.04]{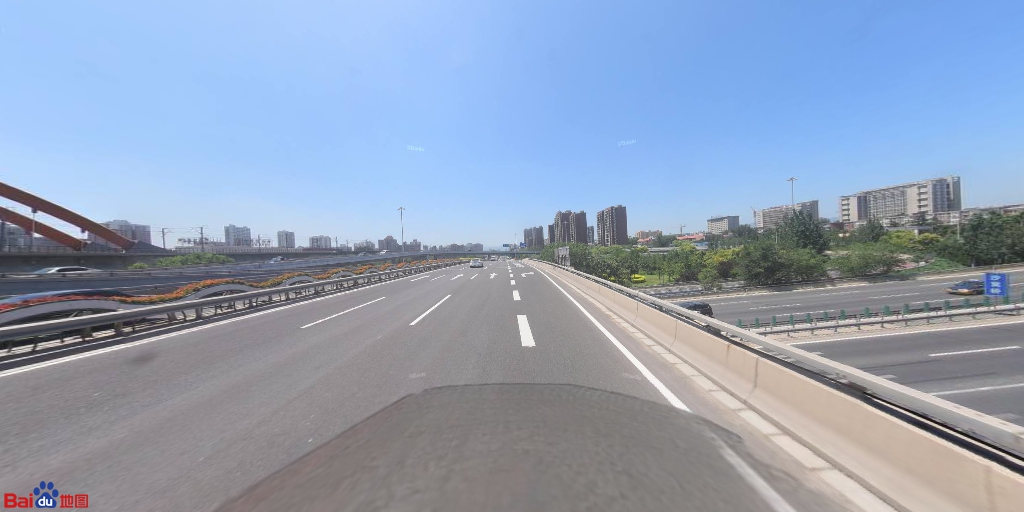}};  
\end{tikzpicture}
         \caption{Typical scenescapes and the corresponding street-view images}
         \label{fig:five over x}
     \end{subfigure}
        \caption{The distribution of typical scenescapes in the case studied area. (a-c) The top-15 scene labels associated with each typical scenescape; (d) Representive SVIs are selected based on visual interpretation.}
        \label{fig3:scene}
\end{figure}
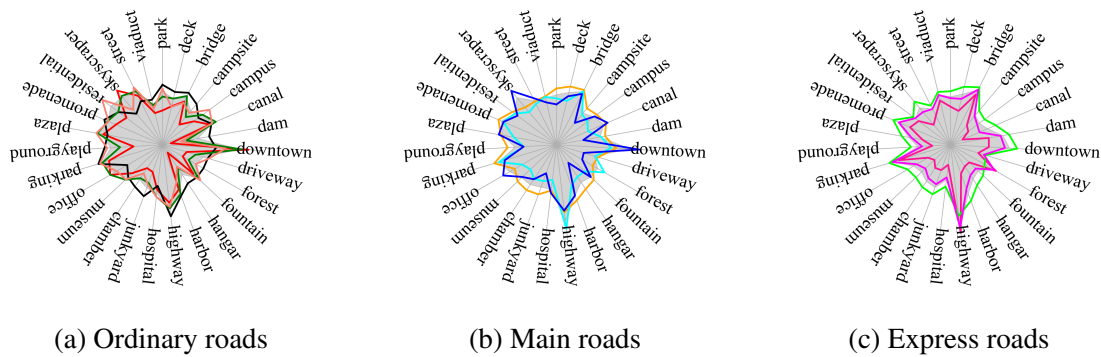
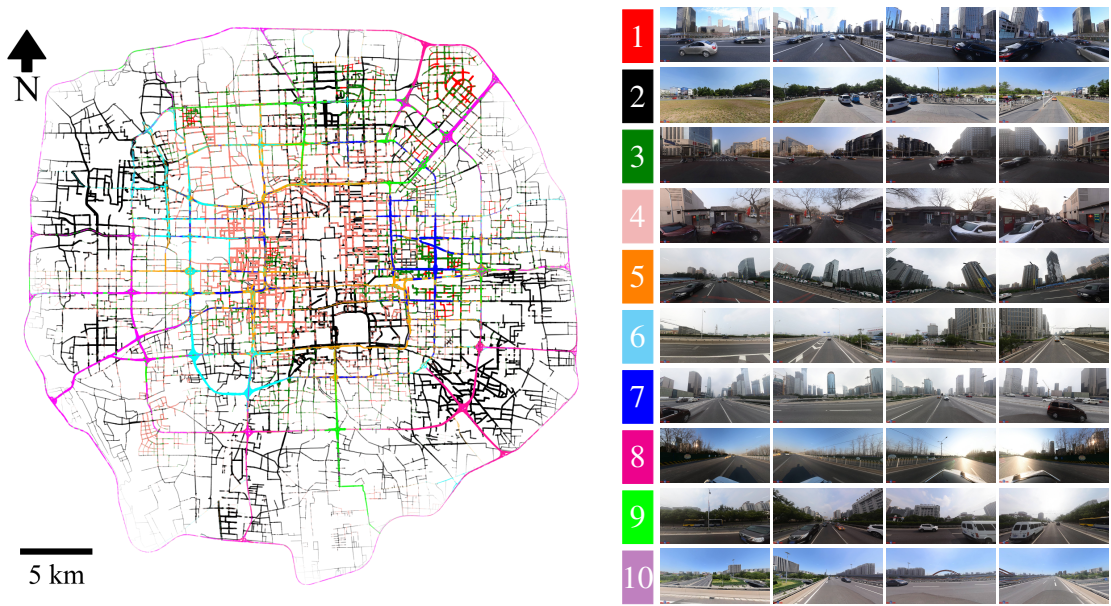
}

To obtain the scene-centric features of the built envionment in each school neighborhood, we aggregate the distribution of the scene labels using the K-means clustering algorithm. In details, four typical scenescapes are extracted for SVIs of the ordinary road (i.e., Scenescape 1-4); Three typical scene scenarios are extracted for SVIs of the main road (i.e., Scenescape 3-7); Three typical scene scenarios are extracted for SVIs of the express road (i.e., Scenescape 8-10). The top-15 scene labels associated with each typical scenescape are illustrated in Fig.~\ref{fig3:scene}. By acknowledging expert knowledge, we summarize the derived scenescapes in Table \ref{tab2:scene} and take them as the measurements of scene-centric characteristics of the built environment around schools. The spatial distribution of the extracted scenescapes is shown in Fig.~\ref{fig3:scene}(d).   

\begin{table}[htbp]
   \centering
   \caption{Variables with regard to scene-centric characteristics of the built environment.}
   \begin{tabular}{@{} lccr @{}} 
      \toprule
      Category    &  Index & Name & Symbol  \\
      \midrule
      \multirow{10}{*}{Scenescape}      & 1 & Urban, financial, business & $UFB$   \\
                & 2 & Urban, park, leisure     & $UPL$     \\
             & 3 & Business, residential  & $BR$    \\
             & 4 & Hutong, residential  & $HR$    \\
       & 5 & Finanical, business, highway   & $FBH$     \\
       & 6 & Residential, highway & $RH$  \\
       & 7 & Downtown, business, highway & $DBH$  \\
       & 8 & Suburban, park, highway & $SPH$  \\
       & 9 & Educational, cultural, highway & $ECH$  \\
       & 10 & Business, residential, highway & $BRH$  \\
      \bottomrule
   \end{tabular}
   \label{tab2:scene}
\end{table}

\subsection{Regression models}

In order to answer the aforementioned research questions, two regression models are established. The first one (Model 1) is a generalized ordered Logit regression model, which enables us to determine whether school runs put a significant impact on traffic congestion around schools (Q1). The second one (Model 2) is a multiple linear regression model, which enables us to decouple the relationship between built environment characteristics and the risk of traffic congestion in different school neighborhoods (Q2). Descriptive information of the two proposed models is given in Fig.~\ref{fig4:model}.

\afterpage{%
  \begin{figure}[H]
   \centering

\pgfdeclarelayer{background}
\pgfdeclarelayer{foreground}
\pgfsetlayers{background,main,foreground}

\tikzstyle{sensor}=[draw, fill=blue!20, text width=5em, 
    text centered, minimum height=2.5em]
\tikzstyle{ann} = [above, text width=5em]
\tikzstyle{golrs} = [sensor, text width=6em, fill=red!20, 
    minimum height=12em, rounded corners]
    
\tikzstyle{dvs}=[fill=green!20, text width=5em, 
    text centered, minimum height=2.5em, rounded corners, anchor=center]
\tikzstyle{idvs}=[fill=orange!20, text width=5em, 
    text centered, minimum height=2.5em, rounded corners, anchor=center]  
\tikzstyle{cidvs}=[fill=gray!20, text width=5em, 
    text centered, minimum height=2.5em, rounded corners, anchor=center]      
\tikzstyle{scns}=[fill=blue!20, text width=5em, 
    text centered, minimum height=2.5em, rounded corners, anchor=center]        
\tikzstyle{ops1} = [dvs, text width=6em, text centered, fill=red!20, 
    minimum height=4em, rounded corners, anchor=center]
\tikzstyle{ops2} = [dvs, text width=6em, text centered, fill=red!20, 
    minimum height=4em, rounded corners, anchor=center]
\tikzstyle{rqs} = [text width=12em, text centered, fill=cyan!20, 
    minimum height=6em, rounded corners, anchor=center]    
        
\def\blockdist{2.3}
\def\edgedist{2.5}

\tikzset{global scale/.style={
    scale=##1,
    every node/.append style={scale=##1}
  }
}
\begin{tikzpicture}[global scale = 0.9]
    \node (golr) [ops1] {Generalized ordered logit regression};
    
    \path (golr)+(0,+1.5*\blockdist) node (cindex) [dvs] {Congestion index};
    
    \path (cindex)+(3.75, 0) node (m1fa) [idvs] {Workday};
    \path (m1fa)+(0, -0.5*\blockdist) node (m1fb) [idvs] {School run};
    \path (m1fb)+(0, -0.625*\blockdist) node (m1fc) [idvs] {College entrance examination};
    \path (m1fc)+(0.05, -0.55*\blockdist) node (m1f) [text centered] {\textbf{Model 1}};
    
    \path (m1fa)+(2.8, 0) node (mcfa) [cidvs] {School location};
    \path (mcfa)+(0, -0.5*\blockdist) node (mcfb) [cidvs] {Road structure};
    \path (mcfb)+(0, -0.5*\blockdist) node (mcfc) [cidvs] {Transport facilities};
    
    \path (mcfa)+(2.8, 0) node (m2fa) [cidvs] {Buildings};
    \path (m2fa)+(0, -0.5*\blockdist) node (m2fb) [cidvs] {Land uses};
    \path (m2fb)+(0, -0.625*\blockdist) node (m2fc) [cidvs] {Mobile signaling counts};
    \path (m2fc)+(0.0, -0.55*\blockdist) node (m2f) [text centered] {\textbf{Model 2}};
        
    \path (m2fa)+(3.75, 0) node (cratio) [dvs] {Congestion ratio};
    
    \path (cratio)+(0,-1.5*\blockdist) node (mlr) [ops2] {Multiple linear regression};
    
    \path (cindex)+(0.9, -1.2) node (street) {\scriptsize Road};
     \path (cindex)+(0.9, -1.5) node (street) {\scriptsize segment};
    \draw[-{Latex[length=3mm]}] ([xshift=-0.2em, yshift=-1.5em] m1fb.west) -| (golr.north);
     \path (cratio)+(-0.9, -1.2) node (street) {\scriptsize School};
     \path (cratio)+(-0.9, -1.5) node (street) {\scriptsize neighborhood};
    \draw[-{Latex[length=3mm]}] ([xshift=0.2em, yshift=-1.5em] m2fb.east) -| (mlr.north);

    \path (mcfc)+(0, -0.52*\blockdist)  node (sce) [scns] {Scene categories};
    
     \path (sce)+(0, -0.95*\blockdist) node (svi) [scns] {Street-view imagery};
     \path (svi)+(-2.8, 0)  node (svia) [scns] {Place365-CNN};
     \path (svi)+(+2.8, 0)  node (svib) [scns] {K-means};
     
      \draw [-{Latex[length=3mm]}] (svia) edge (svi);
      \draw [-{Latex[length=3mm]}] (svib) edge (svi);
    \draw [shorten <=0.125cm,-{Latex[length=3mm]}] (svi) edge (sce);
    
    \path (golr)+(1.3,-2.2*\blockdist) node (rq1) [rqs] {Do drive-to-school travel activities contribute to traffic congestion in school neighborhoods?};
    
    \path (mlr)+(-1.3,-2.2*\blockdist) node (rq2) [rqs] {Do built environment characteristics contribute to traffic congestion in school neighborhoods?};
    
    \draw [-{Latex[length=3mm]}] ([yshift=0.5em] rq1.east) to[bend left] ([yshift=0.5em] rq2.west);
     \draw [-{Latex[length=3mm]}] ([yshift=-0.5em] rq2.west) to[bend left] ([yshift=-0.5em] rq1.east);
    
    \node (CI) [above of=cindex] {\textbf{Dependent variable}};
    \node (CR) [above of=cratio] {\textbf{Dependent variable}};
    \node (IV) [above of=mcfa] {\textbf{Independent variables}};
    
    \path (svi)+(0,-2.8) node (RQ) {\textbf{Research Questions}};
    
    \node (M1) [below =0.15 of golr] {\textbf{Model 1}};
    \node (M2) [below =0.15 of mlr] {\textbf{Model 2}};
    
    \draw [shorten <=0.25cm,-{Latex[length=3mm]}] (cindex) edge (golr);
     \draw [shorten <=0.15cm,-{Latex[length=3mm]}] (M1.south) to ([xshift=0cm]M1.south |- rq1.north);
     
      \path (golr)+(0.9, -2.5) node (street) {\scriptsize Marginal};
      \path (golr)+(0.9, -2.8) node (street) {\scriptsize effect};
     
     \path (mlr)+(-0.9, -2.5) node (street) {\scriptsize Spatial};
      \path (mlr)+(-0.9, -2.8) node (street) {\scriptsize quality};
      
           \path (mlr)+(0.9, -2.5) node (street) {\scriptsize SHAP};
      \path (mlr)+(0.9, -2.8) node (street) {\scriptsize interpreter};
     
    \draw [shorten <=0.25cm,-{Latex[length=3mm]}] (cratio) edge (mlr);
     \draw [shorten <=0.15cm,-{Latex[length=3mm]}] (M2.south) to ([xshift=0cm]M2.south |- rq2.north);
     
    \begin{pgfonlayer}{background}
        \path (CI.west |- golr.north)+(-0.05,0.3) node (a) {};
        \path (M1.south -| CI.east)+(+0.05,-0.2) node (b) {};
        \path[fill=red!10,rounded corners, draw=black!50, dashed]
            (a) rectangle (b);
        
        \path (CR.west |- mlr.north)+(-0.05,0.3) node (a) {};
        \path (M2.south -| CR.east)+(+0.05,-0.2) node (b) {};
        \path[fill=red!10,rounded corners, draw=black!50, dashed]
            (a) rectangle (b);    
            
        \path (CI.north west)+(-0.05,0.2) node (a) {};
        \path (cindex.south -| CI.east)+(+0.05,-0.3) node (b) {};
        \path[fill=green!10,rounded corners, draw=black!50, dashed]
            (a) rectangle (b);
            
        \path (CR.north west)+(-0.05,0.2) node (a) {};
        \path (cindex.south -| CR.east)+(+0.05,-0.3) node (b) {};
        \path[fill=green!10,rounded corners, draw=black!50, dashed]
            (a) rectangle (b);
            
	\path (m1fa.west |- IV.north)+(-0.2,0.2) node (a) {};
        \path (sce.south -| m2fa.east)+(+0.2,-0.3) node (b) {};
        \path[fill=gray!10,rounded corners, draw=black!50, dashed]
            (a) rectangle (b);

	\path (svia.north west)+(-0.2,0.2) node (a) {};
        \path (svib.south east)+(+0.2,-0.2) node (b) {};
        \path[fill=blue!10,rounded corners, draw=black!50, dashed]
            (a) rectangle (b);
            
         \path (m1fa.north west)+(-0.1,0.15) node (a) {};
        \path (sce.south west)+(-0.275,-0.15) node (b) {};
        \path[fill=yellow!10, rounded corners, draw=red!50, dashed, fill opacity=0.5]
            (a) rectangle (b);
            
         \path (m2fa.north east)+(0.1,0.15) node (a) {};
        \path (sce.south west)+(-0.125,-0.15) node (b) {};
        \path[fill=green!10, rounded corners, draw=black!50, dashed, fill opacity=0.5]
            (a) rectangle (b);    
         
        \path (CI.west |- rq1.north)+(-0.05,0.3) node (a) {};
        \path (rq2.south -| CR.east)+(+0.05,-0.3) node (b) {};
        \path[fill=cyan!10,rounded corners, draw=black!50, dashed]
            (a) rectangle (b);      
    \end{pgfonlayer}
\end{tikzpicture}
   \caption{The proposed regression models for school run traffic congestion analysis}
   \label{fig4:model}
\end{figure}
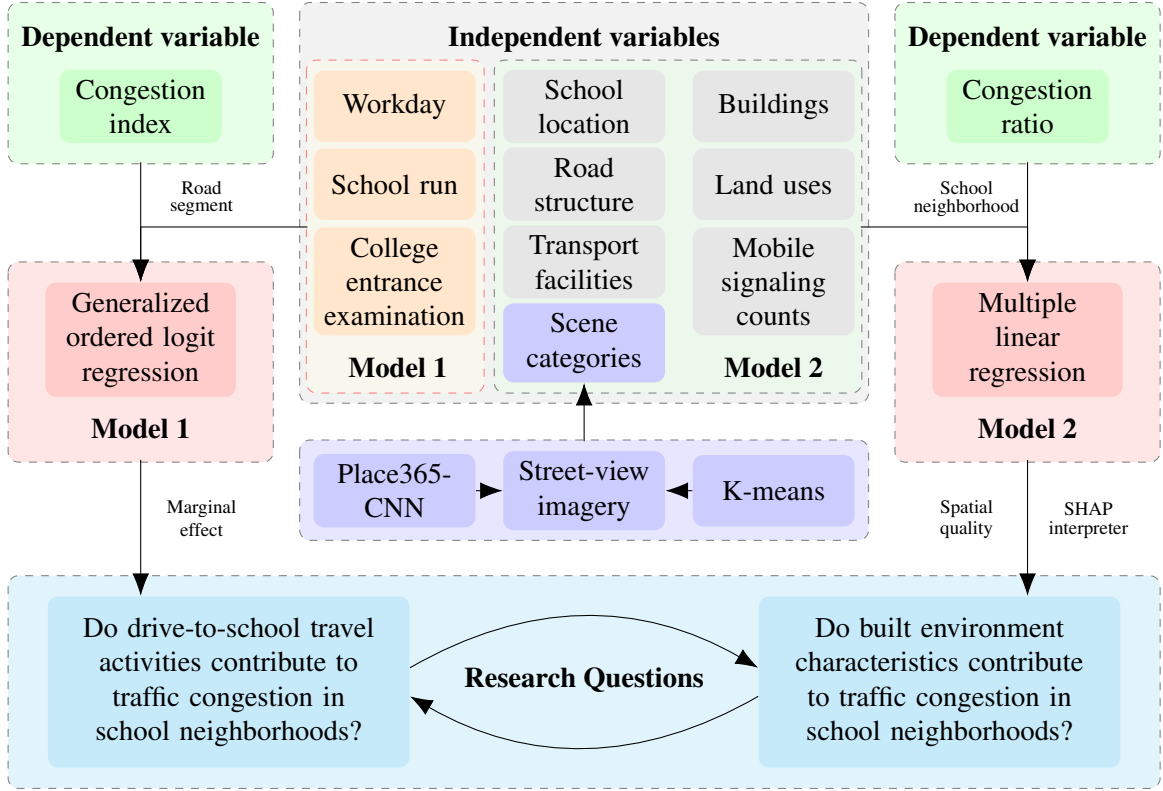
}

\subsubsection{Deciphering the impact of school run on traffic congestion}

Considering that the obtained traffic congestion index on each road is an ordinal, categorical variable (i.e., ``smooth", ``slow", ``congested", or ``severely congested"), we establish a generalized ordered Logit model which has been proven to be plausible for quantifying the effect of explanatory variables on such ordinal outcome while be less restrictive to the proportional odds/parallel lines assumptions \citep{Williams2016}. 

Mathematically, given an ordered dependent variable $Y$ with $M$ categories, the generalized ordered Logit model can be written as 
\begin{align} 
   P(Y > j) = g(X\beta_{j})=\frac{\exp(\alpha_{j} + X\beta_{j})}{1+[\exp(\alpha_{j} + X\beta_{j})]}, j=1,2,\cdots,M-1
\end{align}
where $X$ is the explanatory (or independent) variable; $g(*)$ is the regression function; $\alpha_{j}$ is the model intercept; and $\beta_{j}$ is the model coefficient to be estimated. Intuitively, the model yields the probability of observing the outcome $Y$ in each category $j=1, 2,\cdots, M$ as
\begin{equation}
  P(Y =j)  =
    \begin{cases}
      1 - P(Y >1)= 1- g(X\beta_{1}) & \text{for $j=1$}\\
      P(Y >j-1)-P(Y >j)=g(X\beta_{j-1}) - g(X\beta_{j}) & \text{for $j=2, \cdots, M-1$}\\
      P(Y >M-1)= g(X\beta_{M-1}) & \text{for $j =M$}
    \end{cases}       
\end{equation}

In our case study, since the outcome variable $Y$ (i.e., the traffic congestion index) has four possible values (i.e., $M=4$), the proposed model will estimate three sets of coefficients $\beta$ by performing binary logistic regressions for (1) ``smooth" against ``slow, congested, or severly congestion"; (2) ``smooth or slow" against "congested or severely congested"; and (3) ``smooth, slow, or congested" against ``severely congested". After the beta coefficients are estimated, the effect of each explanatory variable $X_{k}$ on each traffic congestion category $Y_{i}$ can be quantified by the marginal effects as
\begin{equation}
  \frac{\partial P(Y=j)}{\partial X_{k}}  =
    \begin{cases}
      -g(X\beta_{1})(1-g(X\beta_{1}))\beta_{1k} & \text{for $j=1$}\\
      g(X\beta_{j-1})(1-g(X\beta_{j-1}))\beta_{(j-1)k}-g(X\beta_{j})(1-g(X\beta_{j}))\beta_{jk} & \text{for $j=2, \cdots, M-1$}\\
      g(X\beta_{M-1})(1-g(X\beta_{M-1}))\beta_{(M-1)k} & \text{for $j =M$}
    \end{cases}       
\end{equation}
where the estimated coefficient $\beta_{jk}$ explicitly accounts for the effect of the given explanatory (or independent) variable $X_{k}$ on the $j$th category of the outcome (or dependent) variable $Y$.

In the case studied area, we take the traffic congestion index on roads in the vicinity of schools (i.e., $\leq 500$ meters) at the pre-selected hourly time slots as the dependent variable $Y$, and three dummy variables indicating (1) whether the observed time slot is on the normal workday; (2) whether the observed time slot is within the school attending/leaving hours on the normal school-run day (which is meantime the normal workday); and (3) whether the observed time slot is within the national college entrance examination hours on the workday (which is meantime the non-school day). More detailed descriptions for the dependent variable and the independent variables are given in Table~\ref{tab3:gologit}. By taking the three explanatory variables into Model 1, we not only account for the interaction effect of daily commute, school run time and distance to school, but also set a more reliable DID situation than existing studies to distinguish the impact of student pick-up/drop-off activities on traffic congestion around schools.  

\begin{table}[htbp]
   \centering
   \caption{Description of the variables for Model 1}
   \begin{tabular}{@{} p{2.5cm}p{3.5cm}p{1.5cm}p{6cm} @{}} 
      \toprule
      \textbf{Category}    & \textbf{Name} & \textbf{Symbol} & \textbf{Description}\\
      \midrule
      Dependent \break (road-wise) & Traffic congestion \break index & $traffic$ & 1 for ``smooth"; 2 for ``slow"; 3 for ``congested"; 4 for ``severely congested"\\
      \cmidrule(r){2-4} 
      \multirow{7}{*}{Independent}      & Working & $work$ & 1 for any time slot during Monday and Friday; 0 for others \\
       \multirow{5}{*}{(dummy)}  & School attending \break / leaving     &  $school$ & 1 for student pick-up and drop-off hours on Monday and Tuesday; 0 for others\\
      & National college \break entrance exam  & $exam$ & 1 for the National College Entrance Examination hours during Wednesday and Friday excluding Thursday morining; 0 for others\\
      \bottomrule
   \end{tabular}
   \label{tab3:gologit}
\end{table}

\subsubsection{Deciphering the impact of built environment characteristics on school run traffic congestion}

Noticing that the frequency of traffic congestion vary in different school neighborhoods (see Fig.~\ref{fig2:congestion}(c)), we further establish a multiple linear regression model (Model 2) to uncover the association between built environment characteristics and the frequency of traffic congestion in each school neighborhood. In specifc, we take the derived physical and scene-centric features of the built environment as the explanatory variables, and the average frequency of traffic congestion on all the roads within each given school neighborhood as the outcome variable (see Table \ref{tab4:mlr}). All the independent variables are normalized by the Z-score method.

\begin{table}[htbp]
   \centering
   \caption{Description of the variables for Model 2}
   \begin{tabular}{@{} p{2.5cm}p{4cm}p{1.5cm}p{6cm} @{}} 
      \toprule
      \textbf{Category}    & \textbf{Name} & \textbf{Symbol} & \textbf{Description}\\
      \midrule
      Dependent \break (school-wise) & Congestion frequency & $jam$ & The average frequency of observing  traffic congestion on each road in each school neighborhood during the school attending/leaving period on Monday and Tuesday\\
      \cmidrule(r){2-4} 
      \multirow{5}{*}{Independent}      & Spatial concentration & \multirow{5}{*}{$\cdots$}    & \multirow{5}{*}{Please refer to Table 1 and 2.}  \\
       & Transport facility    &   & \\
      & Road topology  &  & \\
      & Spatial richness  &  & \\
      & Scenescape  &  & \\            
      \bottomrule
   \end{tabular}
   \label{tab4:mlr}
\end{table}

In addition to the estimated coefficients for each built environment characteristic, we compute the Shapley values  to enhance model interpretability \citep{Lundberg2017}. The Shapley additive explanations enable us to assign each built environment feature an importance value for the occurence of traffic congestion in each school neighborhood. Mathematically, the Shapley value is calculated based on the average marginal contribution of all possible feature permutations as defined below:
\begin{align} 
   \phi_{i} = \sum_{S \subseteq F \backslash \{i\}} \frac{|S|!(|F|-|S|-1)!}{|F|!} [f_{S \cup \{i\}}(x_{S \cup \{i\}})-f_{S}(x_{S})]
\end{align}
where $S$ is all feature subsets; $F$ is the set of all features; $f_{S \cup \{i\}}$ is the model trained with the given feature; $f_{S}$ is the model trained with the given feature withheld; and $x_{S}$ represents the values of the input features in the set $S$.

Based on the Shapley values of different independent variables for each individual school, we can determine the contribution of each built environment feature to the (average) congestion probability of each school neighborhood, as well as the cumulative contribution of each built environment feature in all school neighborhoods, which represents the overall impact of each built environment feature on school run traffic congestion:
\begin{align} 
   g(z^{'}) = \phi_{0} + \sum_{i=1}^{M} \phi_{i} z_{i}^{'}
\end{align}
where $z_{'} \in \{0,1\}^{M}$, and $M$ is the number of simplified input features.

By using Model 2 and its associated SHAP explanations, we quantitatively measure the effect of built environment characteristics on the risk of school run traffic congestion using the model coefficients, the Shapley values, and the SHAP interaction values. With these values, we are also able to explicitly evaluate the overall spatial quality of the built environment around schools.

\section{Results}

\subsection{The impact of school run on trafffic congestion around schools}

For the 10 observed timestamps, a total of 1,401,259 traffic congestion indices are recorded in the vicinity ($\leq 500$ m) of the 846 schools. Among them, 1,314,210 records are ``smooth"; 41,193 records are ``slow"; 39,078 records are ``congested"; and 6,778 records are ``severely congested". By feeding these data into Model 1, we calculate the marginal effects for the generalized ordered Logit regression equation and uncover the impact of each independent variable on the different levels of the ordinal dependent variable (see Fig.~\ref{fig5:time}). It is also noteworthy that the resultant generalized ordered logit model has passed the multicollinearity test and all the estimated coefficients $\beta_{jk}$ for the marginal effects are significant (with the $p$-value $<0.01$). 

\afterpage{%
  \begin{figure}[H]
   \centering
%
\definecolor{bblue}{HTML}{4F81BD}
\definecolor{rred}{HTML}{C0504D}
\definecolor{ggreen}{HTML}{9BBB59}
\definecolor{ppurple}{HTML}{9F4C7C}

\pgfplotsset{
/pgfplots/my legend/.style={
legend image code/.code={
\draw[thick,black](-0.05cm,0cm) -- (0.3cm,0cm);%
   }
  }
}
          \begin{subfigure}[b]{\textwidth}
         \centering
\begin{tikzpicture}
    \begin{axis}[
        width  = 0.85*\textwidth,
        height = 8cm,
        xtick style={/pgfplots/major tick length=-1.5mm,draw=black},
        ybar=2*\pgflinewidth,
        bar width=16pt,
        ymajorgrids = false,
        ylabel = {Coefficient of marginal effect (\%)},
        symbolic x coords={$work$,$school$,$exam$},
        xtick = data,
        nodes near coords={
   \pgfmathprintnumber[fixed , fixed zerofill, precision=2] 
   {\pgfplotspointmeta}
   },
        every node near coord/.append style={font=\tiny},
   	nodes near coords align={vertical},
        scaled y ticks = false,
        enlarge x limits=0.3,
        ymin=-7,
        ymax=7,
        legend cell align=left,
        legend columns=4,
        legend style={
                at={(0.975,0.85)},
                anchor=south east,
                column sep=1ex,
                draw = gray
        },
        extra y ticks = 0,
        extra y tick labels={},
        extra y tick style={grid=major,major grid style={dashed, draw=gray}}
    ]
        \addplot[style={ggreen,fill=ggreen,mark=none}]
            coordinates {($work$, -3.83) ($school$,-5.05) ($exam$,-1.45)};

        \addplot[style={bblue,fill=bblue,mark=none}]
             coordinates {($work$,0.91) ($school$,1.99) ($exam$,0.87)};

        \addplot[style={rred,fill=rred,mark=none}]
             coordinates {($work$,2.37) ($school$,2.55) ($exam$, 0.54)};

        \addplot[style={ppurple,fill=ppurple,mark=none}]
             coordinates {($work$,0.54) ($school$,0.51) ($exam$, 0.05)};

        \legend{Smooth,Slow,Congested,Severely congested}
    \end{axis}
\end{tikzpicture}
\caption{Comparison between workday hours, school run hours, and exam hours}
\end{subfigure}
\break
          \begin{subfigure}[b]{\textwidth}
         \centering
\begin{tikzpicture}
    \begin{axis}[
        width  = 0.85*\textwidth,
        height = 8cm,
        xtick style={/pgfplots/major tick length=-1.5mm,draw=black},
        ybar=2*\pgflinewidth,
        bar width=25pt,
        ymajorgrids = false,
        ylabel = {Coefficient of marginal effect (\%)},
        symbolic x coords={$work\_commute\_no\_school$, $work\_commute\_school$},
        xtick = data,
        nodes near coords,
        every node near coord/.append style={font=\tiny},
   nodes near coords align={vertical},
        scaled y ticks = false,
        enlarge x limits=0.5,
        ymin=-10,
        ymax=10,
        legend cell align=left,
        legend columns=4,
        legend style={
                at={(0.975,0.85)},
                anchor=south east,
                column sep=1ex,
                draw = gray
        },
        extra y ticks = 0,
        extra y tick labels={},
        extra y tick style={grid=major,major grid style={dashed, draw=gray}}
    ]
        \addplot[style={ggreen,fill=ggreen,mark=none}]
            coordinates {($work\_commute\_school$, -8.34) ($work\_commute\_no\_school$,-3.61) };

        \addplot[style={bblue,fill=bblue,mark=none}]
             coordinates {($work\_commute\_school$, 3.08) ($work\_commute\_no\_school$, 1.30) };

        \addplot[style={rred,fill=rred,mark=none}]
             coordinates {($work\_commute\_school$, 4.20) ($work\_commute\_no\_school$,1.94) };

        \addplot[style={ppurple,fill=ppurple,mark=none}]
            coordinates {($work\_commute\_school$, 1.05) ($work\_commute\_no\_school$,0.37) };

        \legend{Smooth,Slow,Congested,Severely congested}
    \end{axis}
\end{tikzpicture}
\caption{Comparison between commute hours with and without school-escorted trips}
\end{subfigure}
   \caption{The marginal effect ($\beta_{jk}$) of temporal variables in Model 1}
   \label{fig5:time}
\end{figure}
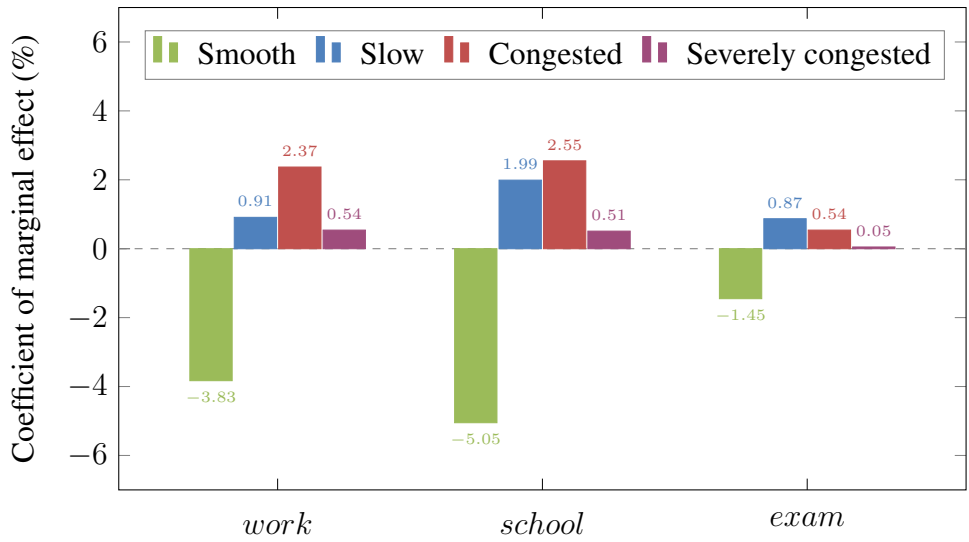
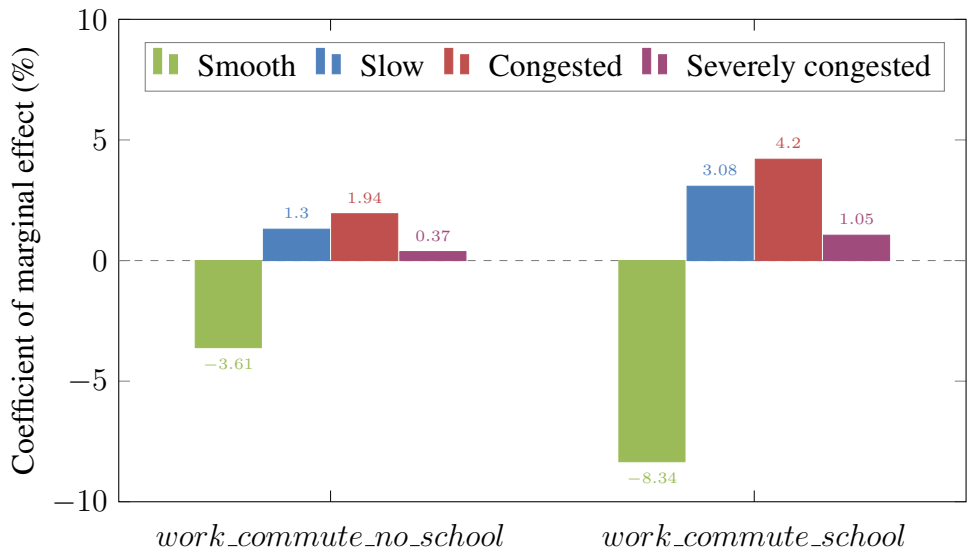
}

As shown in Fig.~\ref{fig5:time}(a), the marginal effects of comparing traffic congestion between workday hours, school run hours, and exam (non-school) hours indicate that: (1) Traffic condition is worse on workdays than weekends, during which the probability of roads around schools in the ``smooth" state is $3.83\%$ lower while the probability of ``slow", ``congested" and ``severely congested" states are $0.91\%$, $2.37\%$ and $0.54\%$ higher, respectively. It confirms that human mobility is more active on workdays, which puts more stresses on the road traffic; (2) During the school attending/leaving hours on the workday, road traffic is further disrupted. The probability of roads around schools in the ``smooth" state decreases by $5.05\%$, whereas the probability of ``slow", ``congested" and ``severely congested" states increase by $1.99\%$, $2.55\%$ and $0.51\%$, respectively. Since the impact of regular commute trips are not accounted by the marginal effect of school runs, it implies that school runs increase the occurence of traffic congestion around them in the case studied area; (3) During the national college entrance exam hours on the workday, since the school-escorted trips within most of the school neighborhoods (excluding schools assigned as the sites for the national college entrance exam) are filtered out, the probability of roads around schools in ``congested" and ``severely congested" states are not disinctly difference when compared with the normal workday (i.e., changes of probability for ``slow", ``congested" and ``severely congested" states are $0.87\%$ vs. $0.54\%$ and $0.05\%$, respectively). This slight difference provides additional evidences for the casual effect of school runs on the traffic congestion around schools. 

More importantly, Fig.~\ref{fig5:time}(b) demonstrates the marginal effect of comparing traffic congestion during the morning commute hours on Monday/Tuesday with traffic congestion during the morning commute hours on Thursday within the case studied time period. Recall that our DID setting involves Thursday morning commute hours when school-escorted trips are excluded. Therefore, the marginal effect of this time period only accounts for the impact of regular commute trips on traffic congestion (i.e., -3.61\%, 1.3\%, 1.94\% and 0.37\% for ``smooth", ``slow", ``congested" and ``severely congested" traffic, respectively). In other words, its marginal effect can be taken as the baseline for decoupling the impact of school-escorted trips on school run traffic congestions. Since both regular commute trips and school-escorted trips account for traffic congestion around schools on Monday/Tuesday morning, the impact of this time period on traffic congestion is the strongest. Therefore, the marginal effects of Monday/Tuesday morining commute hours explicitly measure the impact of adding school-escorted trips into Thursday morning commute hours on traffic congestion; that is, the impact of school run (or more specifically, school-escorted trips) on traffic congestion during the school attending/leaving period. In the case studied area, we can conclude that school runs reduces the probability of ``smooth" traffic by 8.34\%, while increase the probabilities of ``slow", ``congested", and ``severely congested" traffic states by 3.08\%, 4.2\%, and 1.05\%, respectively.    

\subsection{The impact of built environment characteristics on traffic congestion around schools}

By observing the traffic congestion situations at 7:30am, 8:30am, 4:30pm and 5:30pm on Monday and Tuesday, we build the multiple linear regression model as aforementioned. The variance inflation factor (VIF) for all the built environment characteristics is lower than the commonly defined threshold 10, implying that the multicollinearity between the built environment features could be neglected. The resultant model yields an moderate adjusted $R^{2}=0.4524$ and identifies 13 explanatory variables (out of all the 25 built environment characteristics) with a significance less than $10\%$ (i.e., $p$-value$<0.1$), including angle to city east, distance to city center, school mix, bus stops, betweenness centrality, new/old buildings, average building height, and scenescapes 1, 5, 6, 7, 9, 10 (see Fig.~\ref{fig6:mlr} and \ref{fig6:shap} for details).    

\afterpage{%
  \begin{figure}[H]
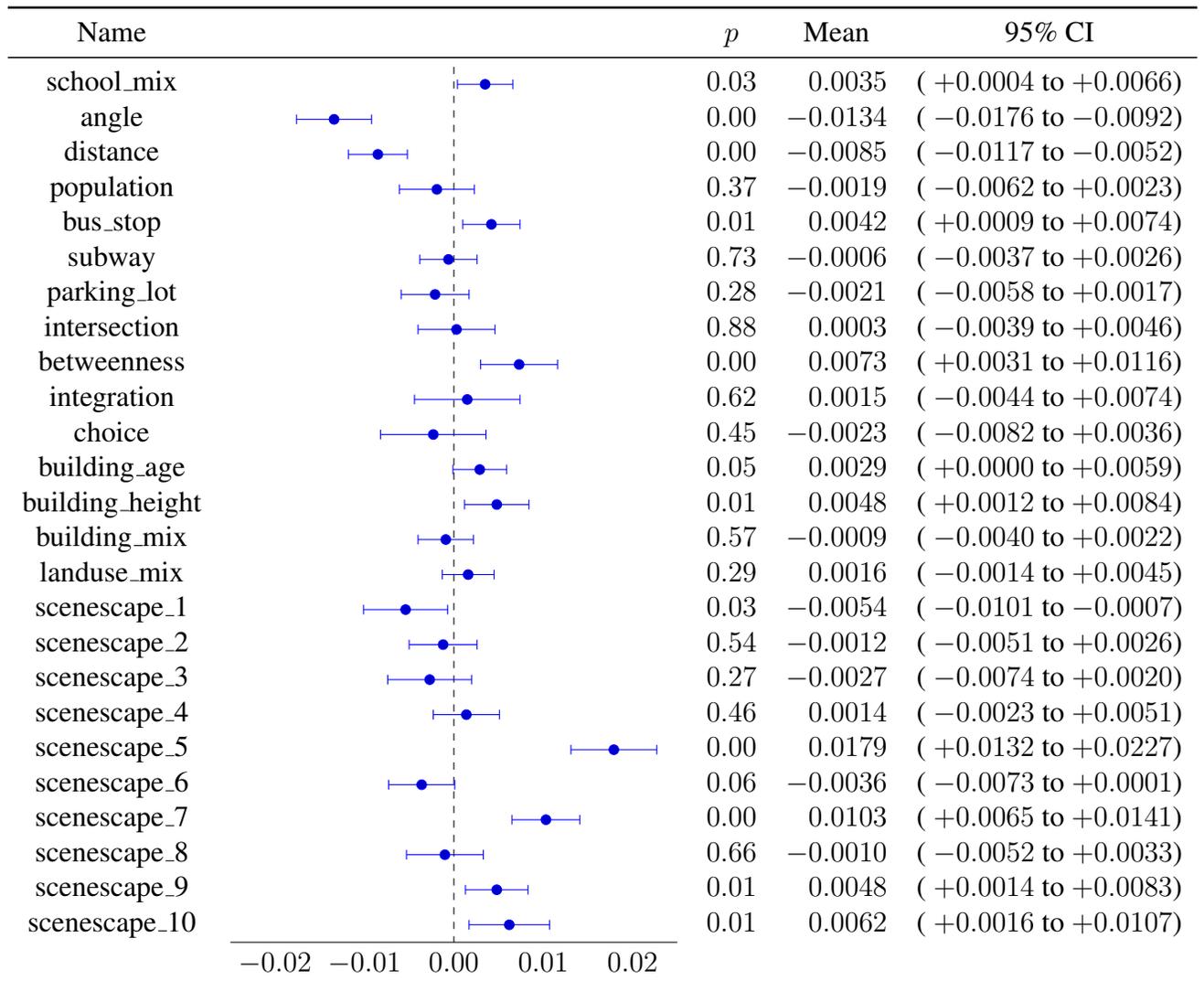

\noindent\makebox[\textwidth]{
\pgfplotstablegetrowsof{\data}
\let\numberofrows=\pgfplotsretval
\pgfplotstabletypeset
  [
    columns={name,error,p,mean,ci},
    every head row/.style = {before row=\toprule, after row=\midrule},
    every last row/.style = {after row=[3ex]\bottomrule},
    columns/name/.style = {string type, column name=Name},
    columns/error/.style = {
      column name = {},
      assign cell content/.code = {
      \ifnum\pgfplotstablerow=0
        \pgfkeyssetvalue{/pgfplots/table/@cell content}
        {\multirow{\numberofrows}{6.5cm}{\errplot}}%
      \else
        \pgfkeyssetvalue{/pgfplots/table/@cell content}{}%
      \fi
      }
    },
    columns/mean/.style = {column name = Mean, fixed ,fixed zerofill, precision=4, dec sep align},
    columns/p/.style    = {column name = $p$, fixed, fixed zerofill, dec sep align},
    columns/ci/.style   = {string type, column name = 95\% CI},
    create on use/ci/.style={
    create col/assign/.code={\edef\value{(
      \noexpand\pgfmathprintnumber[showpos,fixed,fixed zerofill, precision=4]{\thisrow{lci}}
      to \noexpand\pgfmathprintnumber[showpos,fixed,fixed zerofill, precision=4]{\thisrow{uci}})}
      \pgfkeyslet{/pgfplots/table/create col/next content}\value
      }
    }
  ]
{\data}
}
\caption{The relationship between built environment characteristics and traffic congestion around schools derived from Model 2}
        \label{fig6:mlr}
\end{figure}
}

Among all the built environment variables, the ones that are positively correlated with traffic congestion at the 1\% significance level include betweenness centrality, average building height, scenescape 5, scenescape 7, scenescape 9, and scenescape 10. For each 1-unit increase in these variables, the average congestion probability increases by 0.735\%, 0.482\%, 1.793\%, 1.029\%, 0.481\%, and 0.615\%, respectively. Since betweenness centrality of the road network often represents the likelihood of a road being selected in the route choice \citep{Gao2013}, a higher average betweenness centrality in the areas surrounding schools means that the roads are more likely to be designated as essential routes for the daily travel. Therefore, traffic volumes on those roads are often higher and can easily lead to traffic congestion. In a sense, the correlation between average building height and traffic congestion also adhere to this rationale in that high buidings often provide a higher capacity to accommodate urban population, which also implies higher travel demands and road traffic in the area. Similarly, scenescapes 5, 7, 9 and 10 are in general associated with financial, commercial and educational activities, thus traffic volumes are higher. This finding can be also verified by the effect of the mixture of new and old buildings (with a 1-unit increase leading to a 0.293\% increase in the average congestion probability at the 10\% significance level), which has been well-received as an essential indicator of urban vibrancy \citep{Jacobs1961}.  

At the 5\% significance level, the variables positively correlated include the number of neighboring schools and bus stops, for which each 1-unit increase translates into an increase of the average congestion probability increases by 0.349\% and 0.418\%, respectively. One reason for this relationship is that, if the distance between two schools is not far enough, a large number of private cars temporarily park along the roadside during student pick-up/drop-off hours, which easily cause traffic congestion status \citep{Rothman2017}. When there are a large number of bus stops close to schools, the presence of buses often signifcantly affects the driving speed of other vehicles on the road (particularly during the daily commuting hours), thus disrupting the overall traffic around schools and causing traffic congestion \citep{Song2019}. This finding also highlights the importance of traffic management around schools during student pick-up/drop-off hours.

In contrast, the variables negatively correlated with traffic congestion include angle and distance to the city center (at the 1\% significance level),  scenescape 1 (at the 5\% significance level), and scenescape 6 (at the 10\% significance level). For each 1-unit increase in the angle and distance, the average congestion probability decreases by 0.845\% and 1.338\%, respectively. In other words, schools located closer to the city center (i.e., Tiananmen Square) and in the northeast area of the case studied area (i.e., Wang Jing) have a higher likelihood of experiencing congestion in the surrounding areas. For scenescape 1, each 1-unit increase leads to a 0.541\% decrease in average congestion probability, while for scenescape 6, a 1-unit increase leads to a 0.357\% decrease in the average congestion probability. Interestingly, despite that scenescape 1 is dominated by financial and commercial scenes, the visual appearances of the roads within the scenescape is clean and well organized, which avoid traffic congestion situations. Scenescape 6 is dominated by residential estates, implying that the seperation between home and school is less far in the area. Besides, areas associated with this scenescape are often close to express roads, allowing fast traffic going through the school neighborhood to avoid congestion.

\afterpage{%
  \begin{figure}[H]
     \begin{subfigure}[b]{0.485\textwidth}
         \centering
         \includegraphics[width=\textwidth]{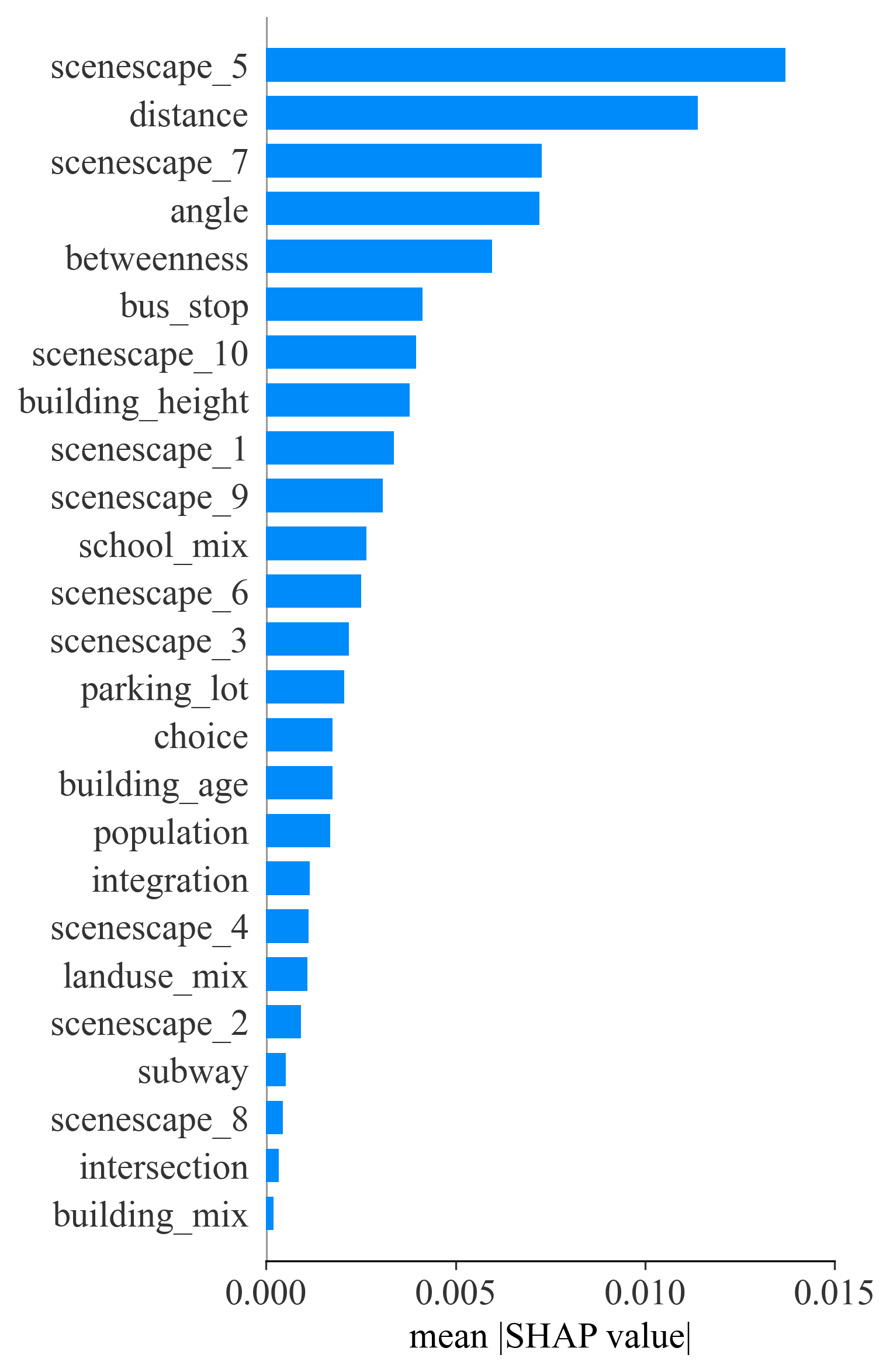}
         \caption{The average feature importance}
         \label{fig:five over x}
     \end{subfigure}
     \hfill
     \begin{subfigure}[b]{0.485\textwidth}
         \centering
         \includegraphics[width=\textwidth]{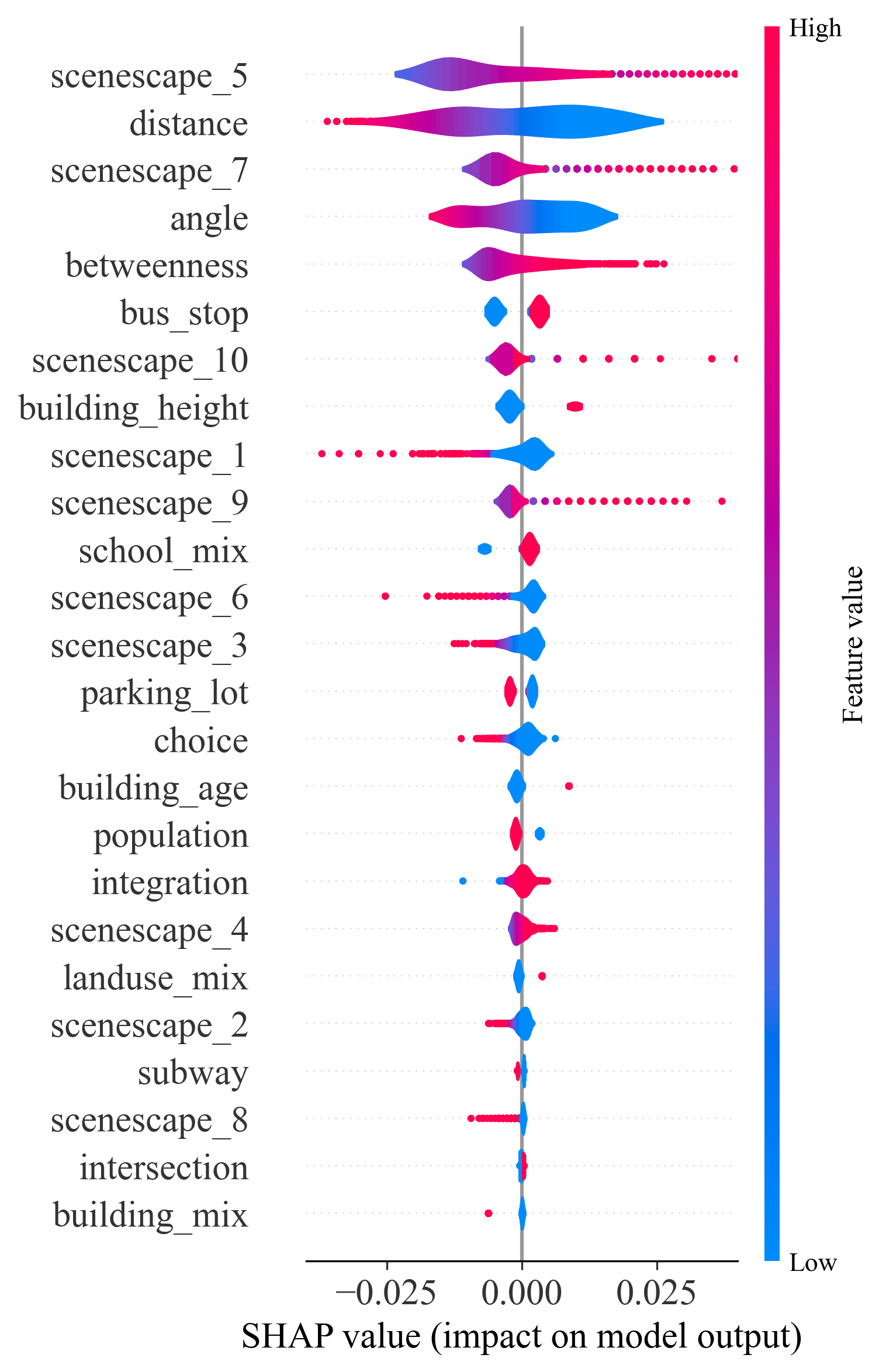}
         \caption{The SHAP summary plot}
         \label{fig:five over x}
     \end{subfigure}
\caption{The importance of built environment features based on SHAP explanations}
        \label{fig6:shap}
\end{figure}
}

The Shapley additive explanations (see Fig.~\ref{fig6:shap}(a) and \ref{fig6:shap}(b)) further confim above findings. In essence, the SHAP feature importance refers to the degree to which each feature contributes to improving the model's overall predictive ability. The higher the importance of a explanatory feature (i.e., built environment characteristics), the greater its impact on the dependent variable (i.e., the congestion probability in areas surrounding primary and secondary schools). According to the SHAP feature importance, we find that the top-ranked features including scenescape 5, distance to city center, scenescape 7, angle to city east, betweenness centrality, bus stops, scenescape 10, average building height, scenescape 1, scenescape 9, school mix and scenescape 6 are highly consistent with the variables that are significant in the multiple linear regression model.  

\section{Policy implications}

\subsection{Evaluating the quality of the built environment around schools}

The above analyses have established a valid regression model for the relationship between built environment characteristics and traffic congestion around schools. Intuitively, the established regression model could provide a useful tool for computing the score of the overall quality of the built environment around schools. Therefore, according to the derived coefficients $\beta_{i}$ and their corresponding significance values $p_{i}$, we select a subset explanatory variables $X$ from all the built environement characteristics by the threshod $p_{i}<0.1$ and normalize their coefficients as $\beta_{i}^{'} = \frac{\beta_{i}}{\max(|\beta_{i}|)}$ to obtain a linear function $f(X)$ for scoring the school neighborhood as below:
\begin{align} 
\label{eq:score}
   env\_score = f(X) = \alpha - \sum_{i}^{N} \beta_{i}^{'} X_{i}
\end{align}
where the intercept $\alpha$ is set as 14 to avoid negative scores; $N$ is the number of selected significant variables; and $\sum_{i}^{N}\beta_{i}^{'}X_{i}$ can also be taken as the score of the risk of traffic congestion. 

\afterpage{%
  \begin{figure}[H]
     \centering
          \begin{subfigure}[b]{0.5\textwidth}
         \centering
                             \pgfplotsset{width=\textwidth,height=7.35cm,}
         \begin{tikzpicture}
\begin{axis}[%
xmin=0,xmax=20,
ymin=0,ymax=40,
xlabel={$jam\_score$},
ylabel={Congestion (\%)},
scatter/classes={%
    a={mark=o,
    draw=blue}}
    ]
\addplot[
scatter,
only marks,%
mark=o,
mark size=0.05em,
    draw=blue,
    fill=blue,
    opacity=0.5,
    scatter src=explicit symbolic
    ]%
	table[x=x, y=y] {scores.dat};
\addplot [red, dashed, mark=none] 
        table[y={create col/linear regression={y=y}}]{scores.dat};	
	\node[] at (150,350) {$R^{2}=0.461$};
\end{axis}
\end{tikzpicture}
         \caption{Correlation between the estimated score and the observed probability of traffic congestion around schools}
         \label{fig:three sin x}
     \end{subfigure}
     \hfill
     \begin{subfigure}[b]{0.45\textwidth}
         \centering
         \includegraphics[width=\textwidth, trim={0 0.2cm 0 0.5cm},clip]{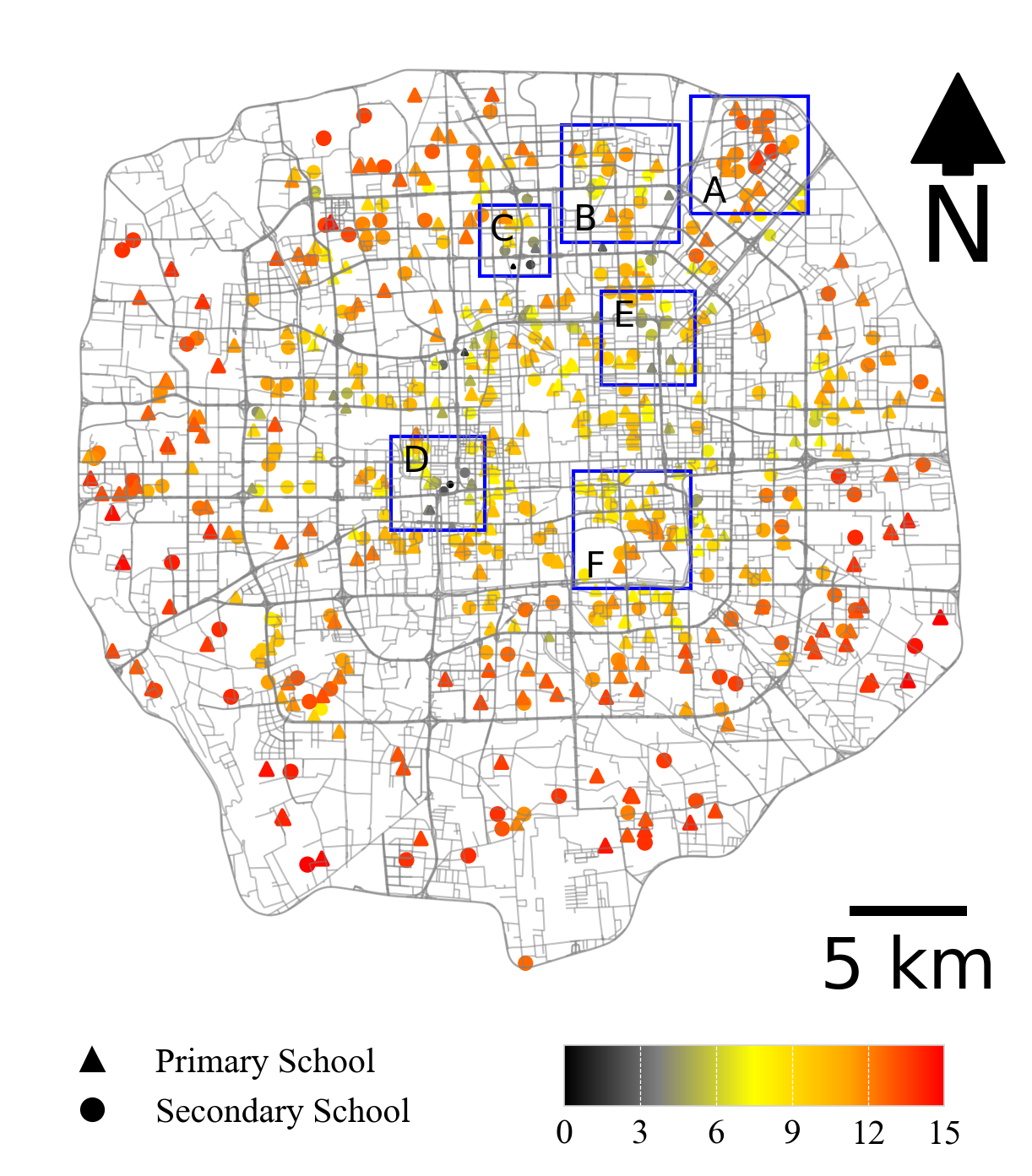}
         \caption{Distribution of the score with regard to the overall quality of the built environment around schools}
         \label{fig:five over x}
     \end{subfigure}
\caption{The overall quality of the built environment around schools}
        \label{fig7:score}
\end{figure}
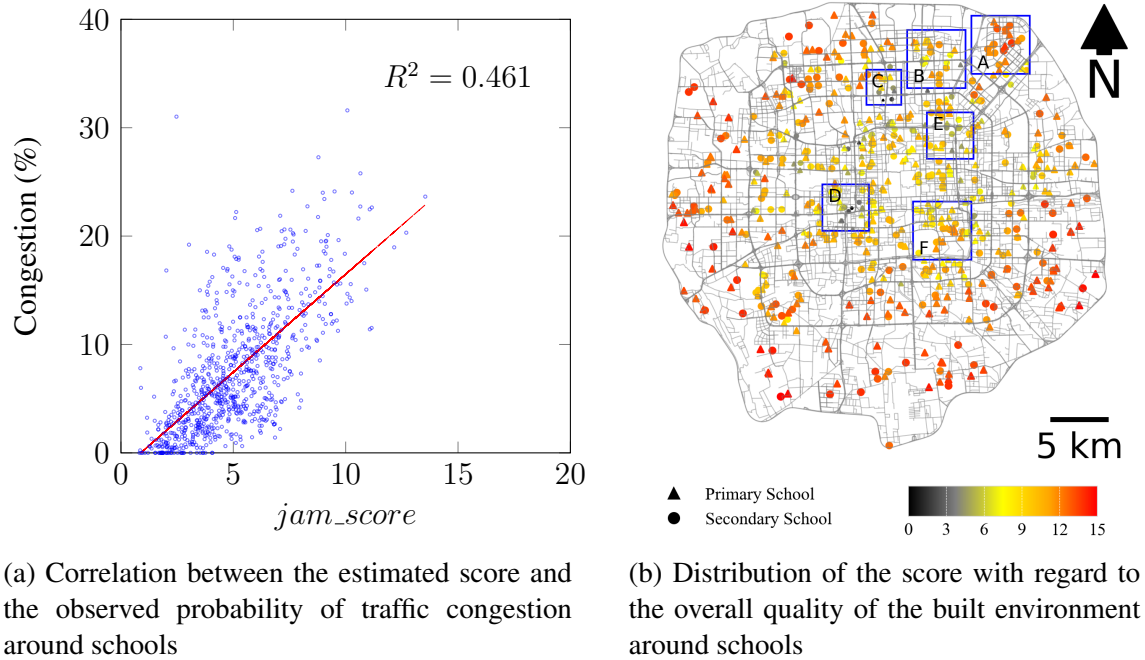
}

In the case studied area, we find that the $R^{2}$ value between the estimated score of the risk of traffic congestion and the observed traffic congestion frequency around schools is 0.461 based on the least-square regression (see Fig.~\ref{fig7:score}(a)), indicating that the scoring function (i.e., Eq.~\ref{eq:score}) has the ability to distinguish school neighborhoods with a high risk of congestion based on well-choosen built environment characteristics. From the spatial distribution of the estimated scores of the overall quality of the built environment around schools (as shown in Fig.~\ref{fig7:score}(b)), we find that schools with better built environment quality in surrounding areas are generally located on the outskirts of the case studied area, specifically within and around the Fourth Ring Road of Beijing. Moreover, two parts of the suburban areas are particularly noteworthy: The first one includes the areas covering the north to northeast part of the case studied area (e.g., Region ``B" and ``C"), where the built environment scores in the vicinity of schools are slightly lower than those of other regions. The second one is the northeastern area laying close to the Fifth Ring Road (e.g., Region ``A"), which is a hub for urban financial and commercial activities, yet the score of built environment quality around schools in this area remains relatively high. 

On the one hand, we believe that the regression models provided by this study will be useful for evaluating the current situation with regard to the risk of traffic congestion around schools. On the other hand, the derived relationships between built environment characteristics and traffic congestion on the traffic situation around exsiting schools will provide guidance for new schools that are still in the planning stage and have not yet been constructed. We believe that the following considerations should be taken into account when determing a site for new schools: (1) It should avoid either the city center or the northeastern direction within the Fifth Ring Road of Beijing; (2) It should not fall within the 500-meter radius of other schools; (3) It should not be near high-traffic roads in order to avoid the interaction between private car trips for student pick-up/drop-off and other types of travel, such as commuting and leisure trips; (4) It should avoid being too close to bus stops to reduce the disruption of buses on crowded roads during the school attending/leaving period; (5) In the surrounding area, building height should be relatively low, and new and old buildings should be highly mixed, in order to avoid attracting too much traffic due to high urban vitality; (6) In the surrounding area, land uses should be residential dominated, while avoiding commercial areas as much as possible.

\subsection{Improving the built environment around schools}

Considering that the case studied area has been developed for a long history, there might already exist enough schools within the area. In other words, it might be more urgent to improve the quality of the built environment around these existing schools rather than to build new schools at an optimal location. Under this scenario, we apply an analysis of the SHAP interaction effect between built environment characteristics to inform us the direction for udgrading the built environment in order to alleviate the traffic congestion around existing schools.  In particular, we concentrate on the interaction effects between scenescapes since we notice that areas with similar geographical location and road/building layouts can suffer from different level of traffic congestion (refer to Section 4.2). It is also noteworthy that, for the sake of concise, the SHAP interaction plots for pair-wised built environment features are reported in Appendix A.1.

\afterpage{%
  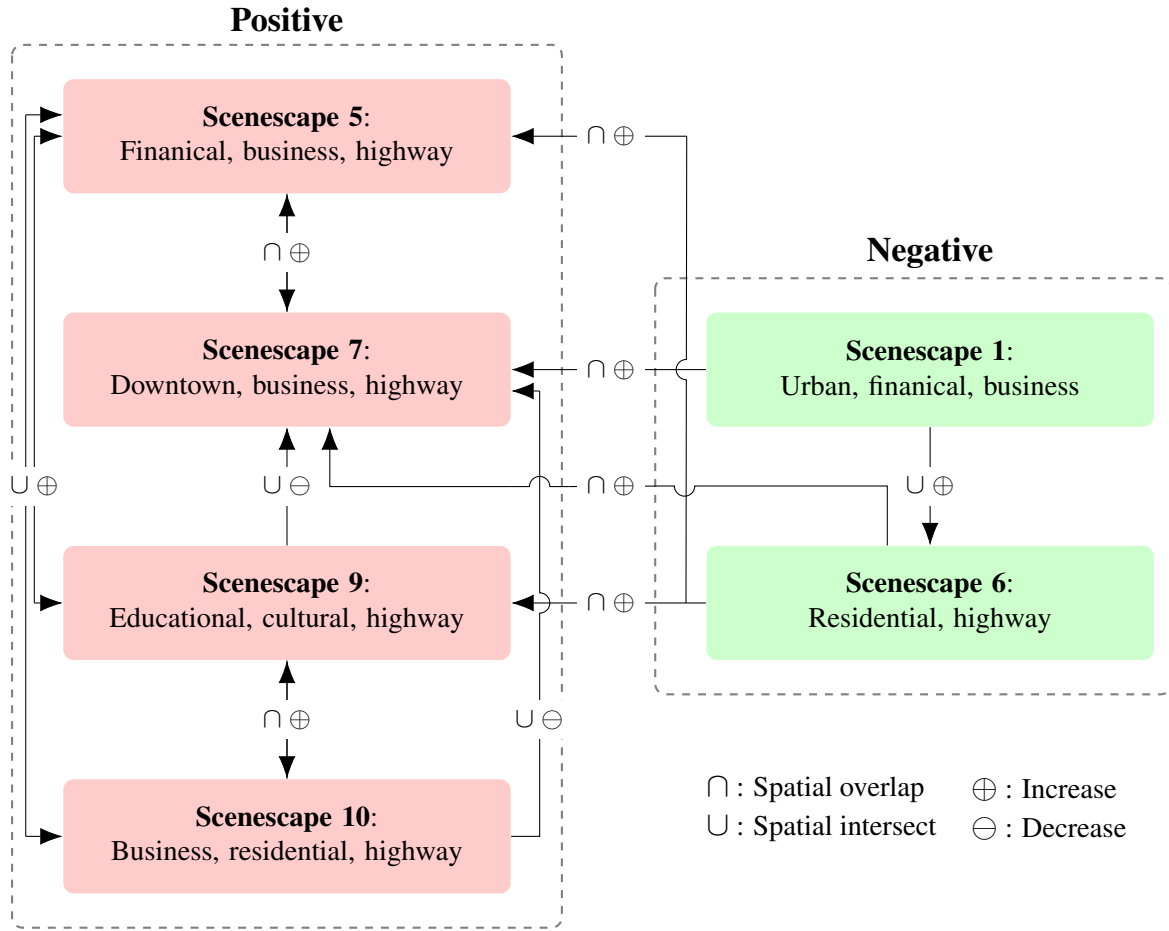
\begin{figure}[H]
   \centering
\pgfdeclarelayer{background}
\pgfdeclarelayer{foreground}
\pgfsetlayers{background,main,foreground}

\tikzstyle{scene} = [sensor, text width=6em, fill=red!20, 
    minimum height=8em, rounded corners]
    
\tikzstyle{dvs}=[fill=green!20, text width=5em, 
    text centered, minimum height=2.5em, rounded corners, anchor=center]
\tikzstyle{idvs}=[fill=orange!20, text width=5em, 
    text centered, minimum height=2.5em, rounded corners, anchor=center]  

\tikzstyle{ops1} = [dvs, text width=15em, text centered, fill=red!20, 
    minimum height=4em, rounded corners, anchor=center]
\tikzstyle{ops2} = [dvs, text width=15em, text centered, fill=green!20, 
    minimum height=4em, rounded corners, anchor=center]
        
\def\blockdist{2.3}
\def\edgedist{2.5}

\tikzset{global scale/.style={
    scale=##1,
    every node/.append style={scale=##1}
  },
  jump/.style={
     to path={
         let \p1=(\tikztostart),\p2=(\tikztotarget),\n1={atan2(\y2-\y1,\x2-\x1)} in
         (\tikztostart) -- ($($(\tikztostart)!##1!(\tikztotarget)$)!0.15cm!(\tikztostart)$)
         arc[start angle=\n1+180,end angle=\n1,radius=0.15cm] -- (\tikztotarget)}
},
jump/.default={0.5}
}

\begin{tikzpicture}[global scale = 0.9]

  
    \node (scene) [ops1] {\textbf{Scenescape 5}:\break Finanical, business, highway};
    \path (scene)+(0,-1.5*\blockdist) node (scene7) [ops1] {\textbf{Scenescape 7}:\break Downtown, business, highway};
    \path (scene)+(0,-3.0*\blockdist) node (scene9) [ops1] {\textbf{Scenescape 9}:\break Educational, cultural, highway};
    \path (scene)+(0,-4.5*\blockdist) node (scene10) [ops1] {\textbf{Scenescape 10}:\break Business, residential, highway};
    
    
    \path (scene7)+(+9.5, 0) node (scene1) [ops2] {\textbf{Scenescape 1}:\break Urban, finanical, business};
    \path (scene1)+(0,-1.5*\blockdist) node (scene6) [ops2] {\textbf{Scenescape 6}:\break Residential, highway};

  
  \draw [-{Latex[length=3mm]}] (scene) edge (scene7);
  \draw [-{Latex[length=3mm]}] (scene7) edge (scene);
  \path (scene)+(0, -0.75*\blockdist) node[fill=white] (interaction) {$\cap$~$\oplus$};
  
  \draw [-{Latex[length=3mm]}] (scene9) edge (scene7);
  \path (scene7)+(0, -0.75*\blockdist) node[fill=white] (interaction) {$\cup$~$\ominus$};
  
   \draw [-{Latex[length=3mm]}] (scene9) edge (scene10);
    \draw [-{Latex[length=3mm]}] (scene10) edge (scene9);
   \path (scene9)+(0, -0.75*\blockdist) node[fill=white] (interaction) {$\cap$~$\oplus$};

  \draw [-{Latex[length=3mm]}] (scene1) edge (scene6);
  \path (scene1)+(0, -0.75*\blockdist) node[fill=white] (interaction) {$\cup$~$\oplus$};
  
    \draw [-{Latex[length=3mm]}] (scene1) edge (scene7);
  \path (scene1)+(-2.05*\blockdist, 0) node[fill=white] (interaction) {$\cap$~$\oplus$};
  
  \draw [-{Latex[length=3mm]}] (scene6) edge (scene9);
  \path (scene6)+(-2.05*\blockdist, 0) node[fill=white] (interaction) {$\cap$~$\oplus$};

 \draw ([xshift=-0.7em, yshift=0em] scene6.west) to[jump] ([xshift=6.15em, yshift=0em] scene.east);
 \draw [-{Latex[length=3mm]}] ([xshift=6.15em, yshift=0em] scene.east) to (scene.east);
 \path (scene6)+(-2.05*\blockdist, 3.0*\blockdist) node[fill=white] (interaction) {$\cap$~$\oplus$};
 
  \draw ([xshift=-1.5em, yshift=0em] scene6.north) |- ([xshift=0.0em, yshift=4.1em] scene6.west);
  \draw [-{Latex[length=3mm]}] ([xshift=0.0em, yshift=-4.1em] scene7.east) -| ([xshift=1.5em, yshift=0em] scene7.south);
 \draw ([xshift=0.0em, yshift=-4.1em] scene7.east) to[jump] ([xshift=2.0em, yshift=-4.1em] scene7.east);
  \draw ([xshift=0.0em, yshift=4.1em] scene6.west) to[jump] ([xshift=-1.5em, yshift=4.1em] scene6.west);
  \draw ([xshift=2.0em, yshift=-4.1em] scene7.east) to ([xshift=-1.5em, yshift=4.1em] scene6.west);
 \path (scene6)+(-2.05*\blockdist, 0.75*\blockdist) node[fill=white] (interaction) {$\cap$~$\oplus$};

 \draw ([xshift=-1.0em, yshift=0em] scene9.west) to ([xshift=-1.0em, yshift=0em] scene.west);
 \draw [-{Latex[length=3mm]}] ([xshift=-1.0em, yshift=0em] scene.west) to (scene.west);
  \draw [-{Latex[length=3mm]}] ([xshift=-1.0em, yshift=0em] scene9.west) to (scene9.west);
 
 \draw ([xshift=-1.3em, yshift=0em] scene10.west) to ([xshift=-1.3em, yshift=0.75em] scene.west);
 \draw [-{Latex[length=3mm]}] ([xshift=-1.3em, yshift=0.75em] scene.west) to ([xshift=0em, yshift=0.75em] scene.west);
  \draw [-{Latex[length=3mm]}] ([xshift=-1.3em, yshift=0em] scene10.west) to (scene10.west);
 \path (scene7.west)+(-1.0em, -0.75*\blockdist) node[fill=white] (interaction) {$\cup$~$\oplus$}; 
 
  \draw ([xshift=+1.0em, yshift=-0.75em] scene7.east) to[jump] ([xshift=+1.0em, yshift=0.75em] scene10.east);
 \draw [-{Latex[length=3mm]}] ([xshift=+1.0em, yshift=-0.75em] scene7.east) to ([xshift=0em, yshift=-0.75em] scene7.east);
  \draw ([xshift=+1.0em, yshift=+0.75em] scene10.east) |- ([xshift=0em, yshift=0.0em] scene10.east);
 \path (scene9)+(1.62*\blockdist, -0.75*\blockdist) node[fill=white] (interaction) {$\cup$~$\ominus$};

   
    \path (scene)+(0.0, +0.75*\blockdist) node (positive) [text centered] {\large\textbf{Positive}};
     \path (scene1)+(0.0, +0.75*\blockdist) node (negative) [text centered] {\large\textbf{Negative}};
  

    \begin{pgfonlayer}{background}
        \path (scene.west |- scene.north)+(-0.75,0.5) node (a) {};
        \path (scene10.south -| scene10.east)+(+0.75,-0.5) node (b) {};
        \path[fill=white,rounded corners, draw=black!50, dashed, thick]
            (a) rectangle (b);

        \path (scene1.west |- scene1.north)+(-0.75,0.5) node (a) {};
        \path (scene6.south -| scene6.east)+(+0.25,-0.5) node (b) {};
        \path[fill=white,rounded corners, draw=black!50, dashed, thick]
            (a) rectangle (b);
                    
  
  \path (scene6)+(-1.5*\blockdist, -1.2*\blockdist) node (spo) [anchor=west] {{\large $\cap$} : Spatial overlap};  
  \path (scene6)+(-1.5*\blockdist, -1.45*\blockdist) node (spi) [anchor=west] {{\large $\cup$} : Spatial intersect};    
   \path (scene6)+(0.2*\blockdist, -1.2*\blockdist) node (inc) [anchor=west] {{\large $\oplus$} : Increase};    
   \path (scene6)+(0.2*\blockdist, -1.45*\blockdist) node (dec) [anchor=west] {{\large $\ominus$} : Decrease};       
                    
    \end{pgfonlayer}
        
\end{tikzpicture}
   \caption{The interaction effect of built environment characteristics on school run traffic congestion}
   \label{fig8:interaction}
\end{figure}
}

As shown in Fig.~\ref{fig8:interaction}, we summarize the interaction effects between all the important scenescapes identified by our proposed models. Interestingly, we find several contradictory effects for certain scenescapes when comparing the SHAP feature contribution without interaction with other features and the SHAP feature contribution with interaction with other features. Recall that scenescapes 1 and 6 have a negative impact on traffic congestion around schools, whereas scenescapes 5, 7, 9 and 10 have a positive impact on traffic congestion around schools. However, for a certain scenescape $A$, its impact on school run traffic congestion varies when there is another scenescape $B$ in the surrounding area; In extreme case, the impact of this scenescape could convert from positive (or negative) to negative (positive). For example, based on the interaction effects, we find that: (1) The negative impact of scenescape 6 on traffic congestion is enhanced if the area associated with scenescape 1 increases around schools; (2) The positive impact of scenscape 7 on traffic congestion is weakened if the area associated with scenescape 1 increases around schools, whereas its positive impact is enhanced if there are more scenescapes 9 and 10 in the surrounding area; (3) As the proportion of scenescape 6 increases, the positive impacts of scenescapes 5, 7 and 9 on school run traffic congestion are all enhanced.   

Based on the interaction effects, we believe that the following considerations should be taken into account when improving the built environment around schools to mitigate traffic congestion: (1) For the school neighborhood dominated by scenescape 5, we can gradually upgrade the streets and buildings to convert it into scenescape 1 rather than scenescape 6. Although both scenescapes 1 and 6 have a negative impact on traffic congestion, the convertion from scenescape 5 to scenescape 6 will increase the positive impact of the remaining scenescape 5 on traffic congestion. Instead, by converting scenescape 5 to scenescape 1, this effect will be canceled out. Moreover, scenescape 1 also increases the negative impact of scenescape 6 on traffic congestion; (2) For the school neighborhood dominated by scenescape 7, we can gradually convert the street appearance into scenescape 10 by adding the residential function. Although either scenescape 7 or scenescape 10 has a positive impact on traffic congestion, the impacts of scenescape 10 are less significant according to the model coefficients (Please refer to Fig.~\ref{fig6:mlr}). Besides, the impose of scenescape 10 will reduce the positive impact of scenescape 7 on traffic congestion around schools; (3) For the school neighborhood characterized by scenescape 10, we might remove the business function in order to convert it into scenescape 6, which will have a negative impact on school run traffic congestion.
 
\section{Conclusion and discussion}
 
This study examines whether and if so to what extent the spatial quality of the built environment influences the likelihood of traffic congestion around schools, particularly focusing on five key dimensions: geographic location, transport facility, road structure, spatial richness, and scenescapes. The empirical results in Beijing, China derived from the proposed regression models as well as the SHAP model explanations indicate that:

\begin{itemize}
\item[(1)] From a temporal perspective, traffic congestion around schools during the school attending/leaving period on the weekday is significantly higher (by 8.34\%) than during regular commute hours on the weekday. We have found a better DID setting during the national college entrance exam (``Gaokao" in Chinese) periods than previous studies to uncover the impact of school-escorted trips on traffic congestion in a more rigorous manner.
\item[(2)] From a spatial perspective, the built environment varies across school neighborhoods, which in turn leads to varying levels of school run traffic congestion. By establishing an evaluation system to assess the overall quality of the built environment around schools, we have identified the key influencing factors to school run traffic congestion, concerning the complexity of road traffic (including bus stops, road centrality), the concentration of urban population (including distance to city center,  school mix, building height), and the scenescapes (including the spatial arrangements of commercial, residential and transport scenes).
\end{itemize}

Based on these findings, we have also proposed several practical criteria to help urban planners to select optimal locations for new schools as well as to improve the built environment around existing schools in order to prevent potential traffic congestion during school runs. While the study could be further refined through improved data sources and methods for characterizing the built environment more comprehensively, its results provide valuable insights for urban planners. These findings can serve as a foundation for developing urban upgrade strategies aimed at alleviating traffic congestion in school neighborhoods and other specific urban areas.

\newpage

\bibliography{references}

\newpage

\section*{Appendix}
\subsection*{A.1 Analysis of the interaction effects between typical scenescapes}
\renewcommand{\thefigure}{A-\arabic{figure}}
\setcounter{figure}{0}

\begin{figure}[htbp]
     \centering
     \begin{subfigure}[b]{0.515\textwidth}
         \centering
         \includegraphics[width=\textwidth, trim={0.5cm 0 1.2cm 0},clip]{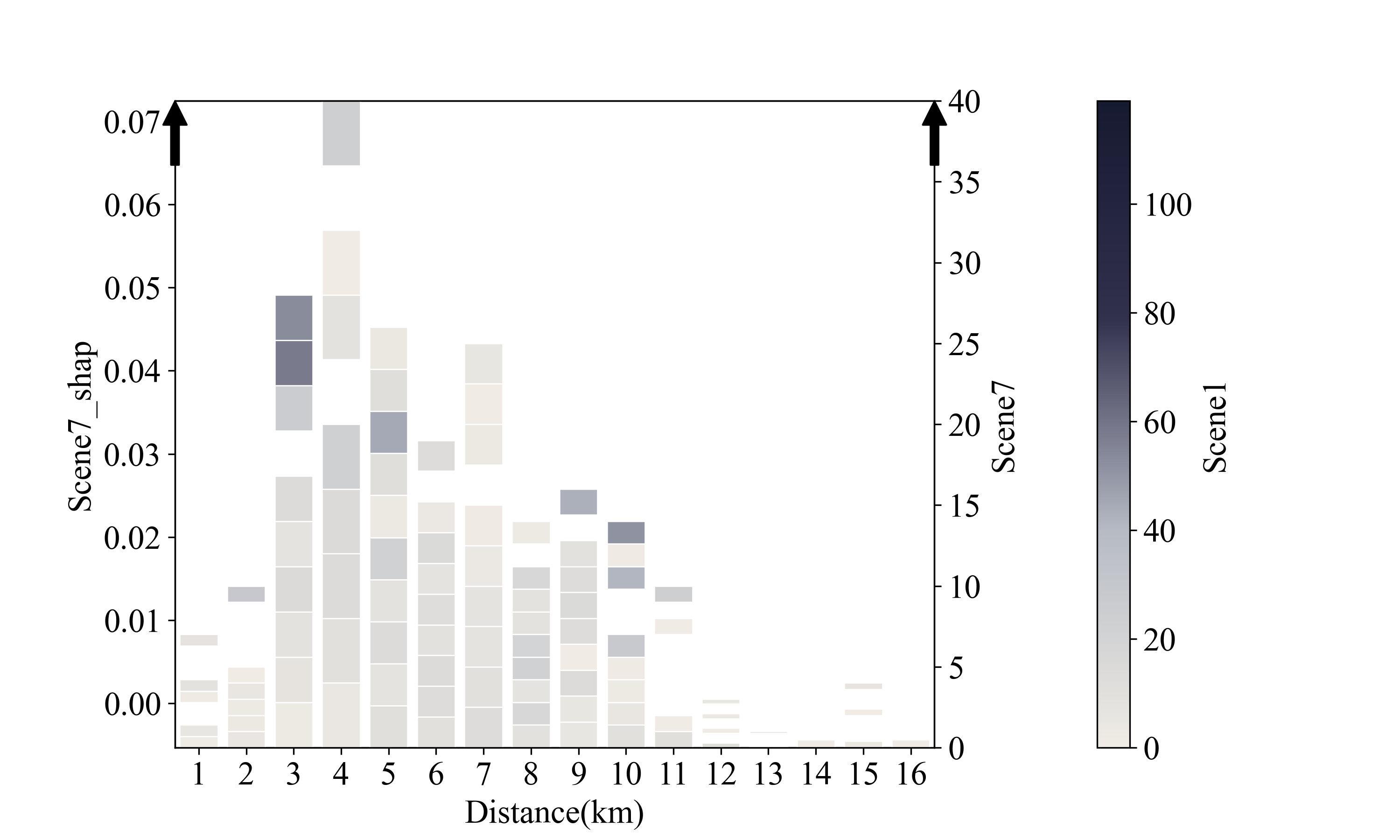}
         \caption{Interaction effect by distance}
         \label{fig:five over x}
     \end{subfigure}
     \hfill
     \begin{subfigure}[b]{0.475\textwidth}
         \centering
         \includegraphics[width=\textwidth]{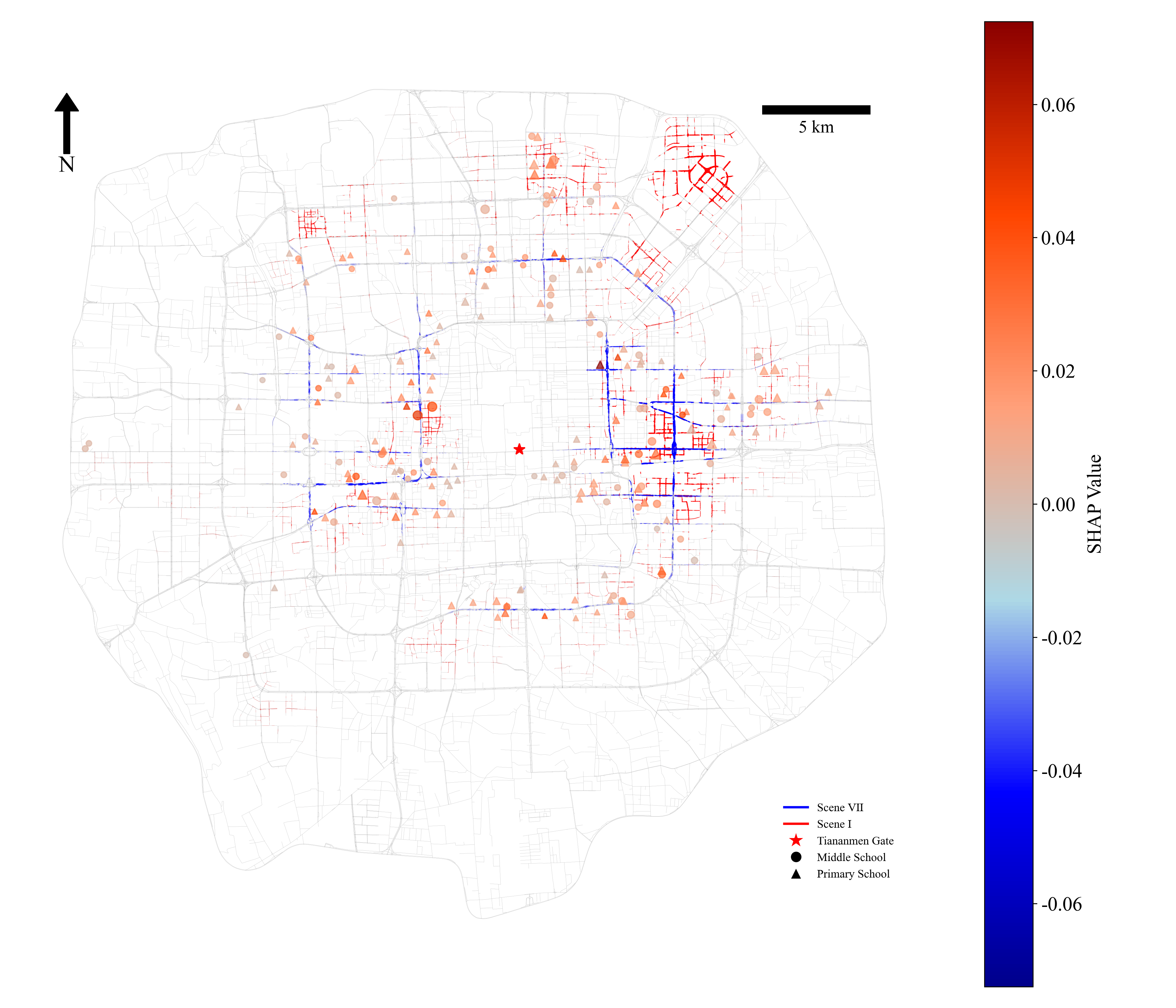}
         \caption{Interaction effect by geographical location}
         \label{fig:five over x}
     \end{subfigure}
\caption{The distribution of SHAP interaction values of scenescape 1 on scenescape 7}
        \label{figs1:interaction}
\end{figure}

\begin{figure}[htbp]
     \centering
     \begin{subfigure}[b]{0.515\textwidth}
         \centering
         \includegraphics[width=\textwidth, trim={0.5cm 0 1.2cm 0},clip]{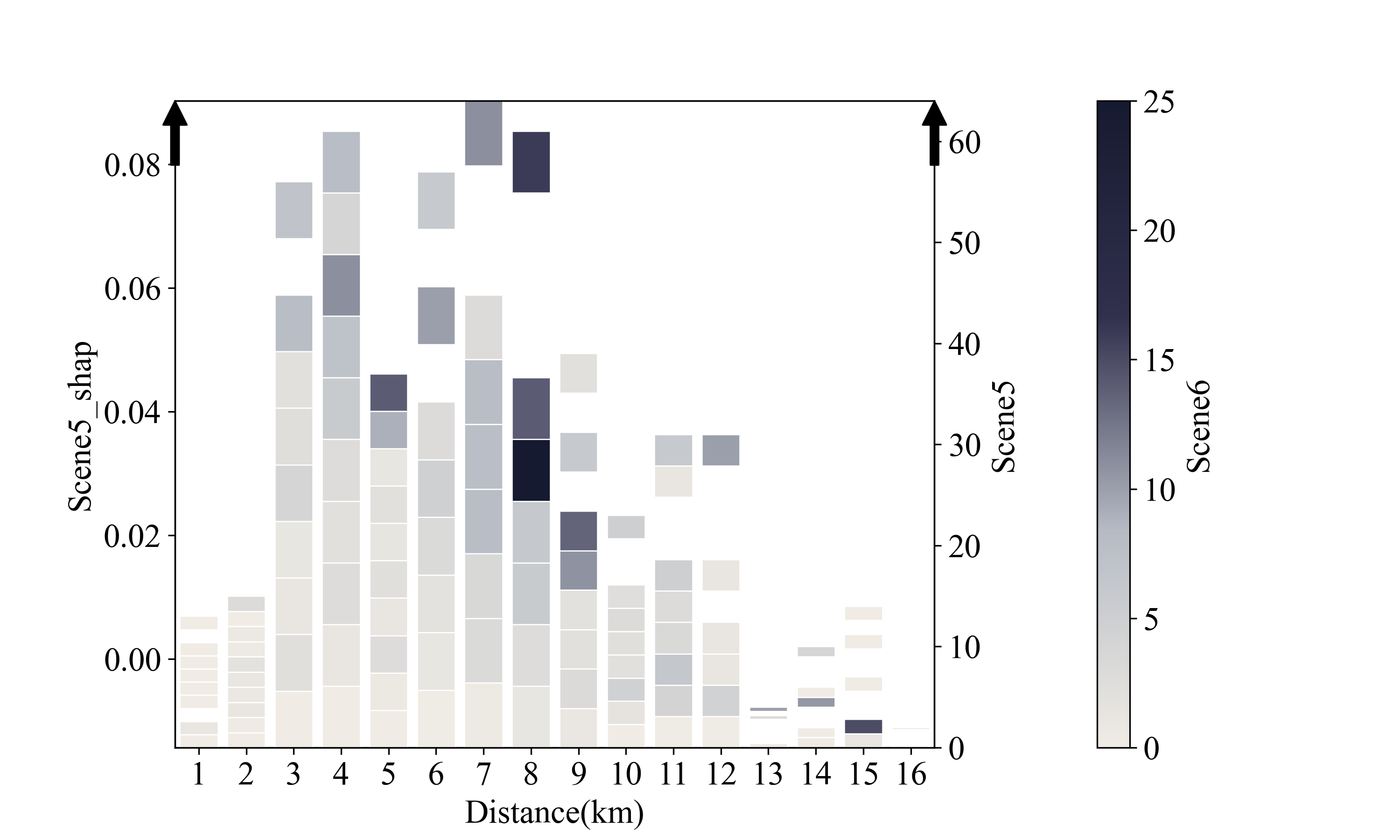}
         \caption{Interaction effect by distance}
         \label{fig:five over x}
     \end{subfigure}
     \hfill
     \begin{subfigure}[b]{0.475\textwidth}
         \centering
         \includegraphics[width=\textwidth]{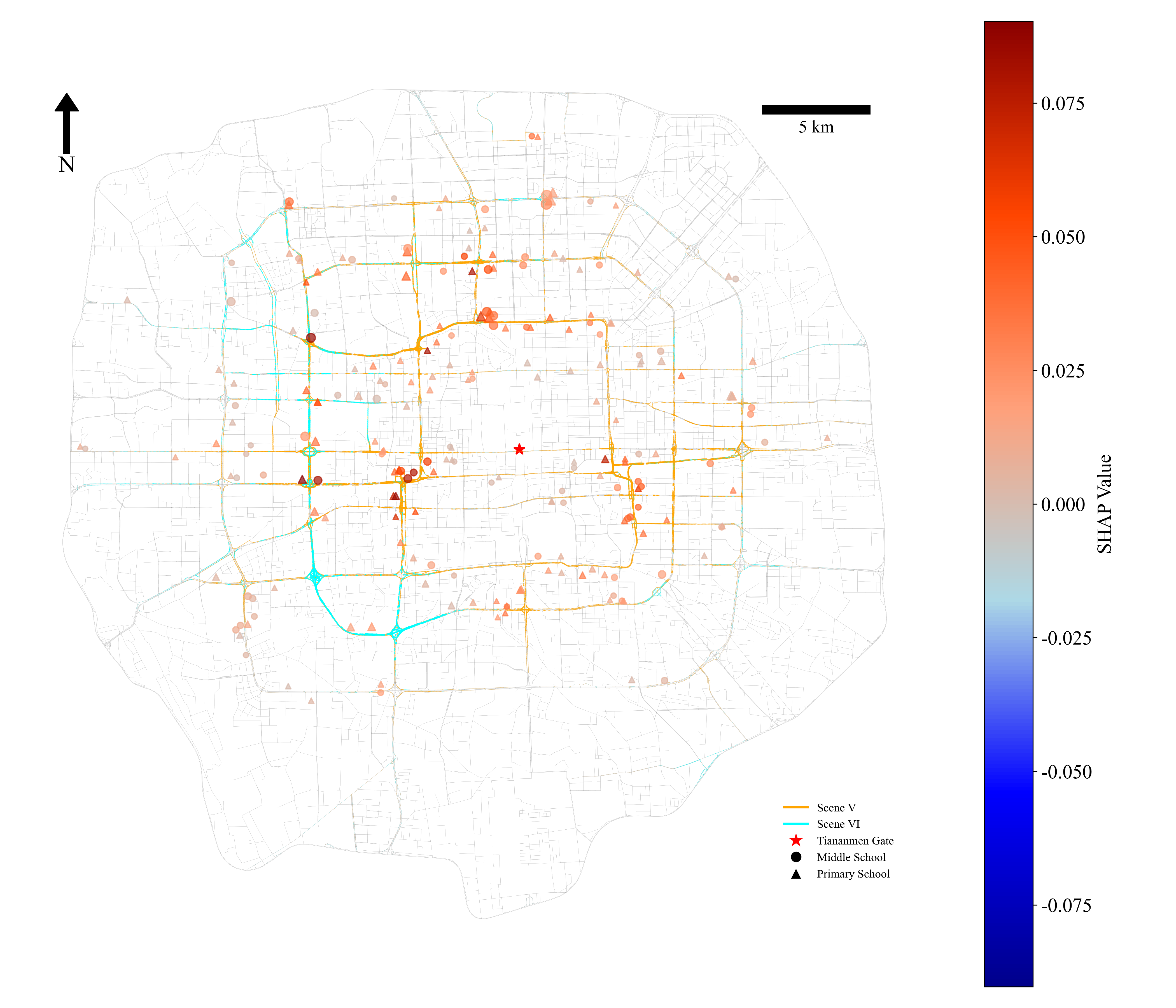}
         \caption{Interaction effect by geographical location}
         \label{fig:five over x}
     \end{subfigure}
\caption{The distribution of SHAP interaction values of scenescape 6 on scenescape 5}
        \label{figs2:interaction}
\end{figure}

\begin{figure}[htbp]
     \centering
     \begin{subfigure}[b]{0.515\textwidth}
         \centering
         \includegraphics[width=\textwidth, trim={0.5cm 0 1.2cm 0},clip]{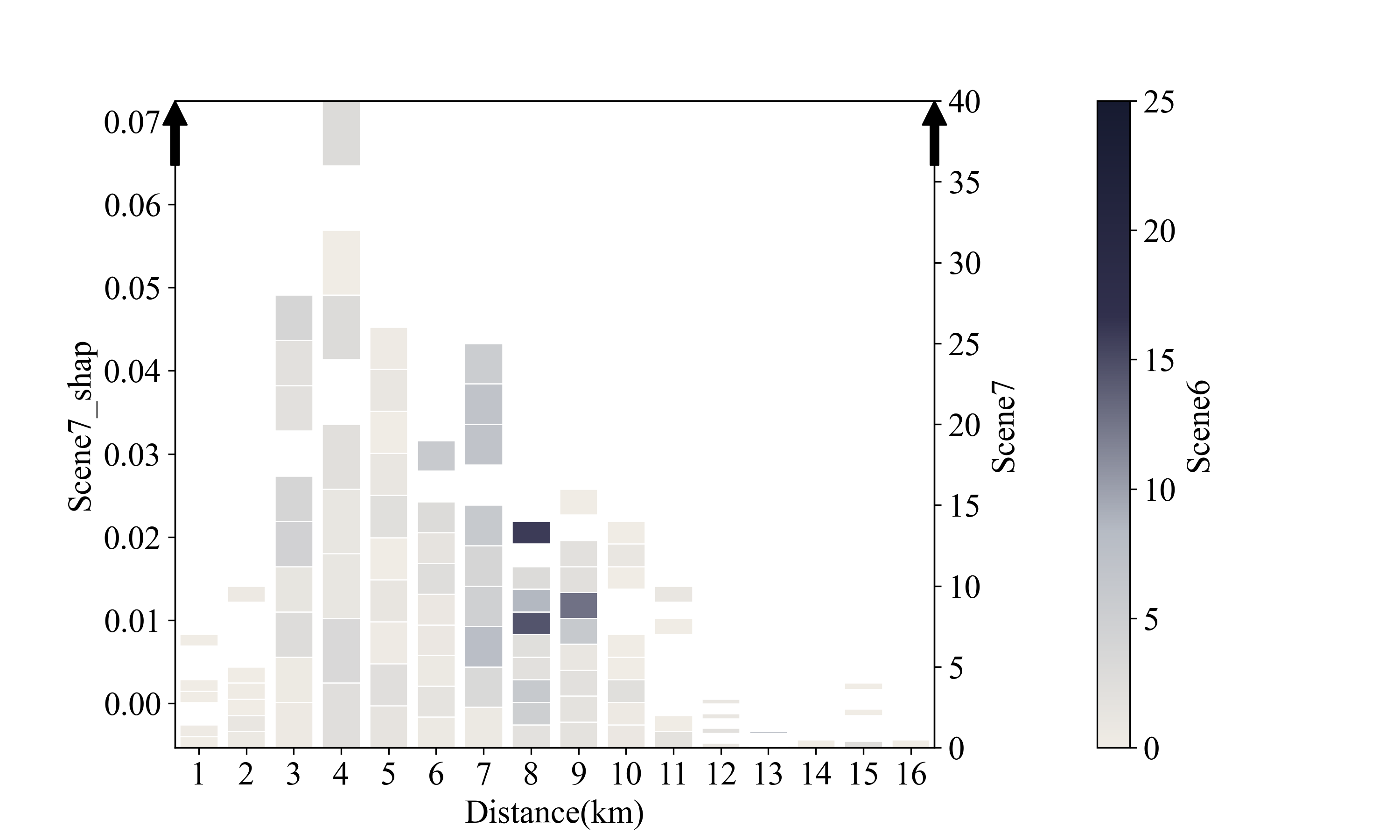}
         \caption{Interaction effect by distance}
         \label{fig:five over x}
     \end{subfigure}
     \hfill
     \begin{subfigure}[b]{0.475\textwidth}
         \centering
         \includegraphics[width=\textwidth]{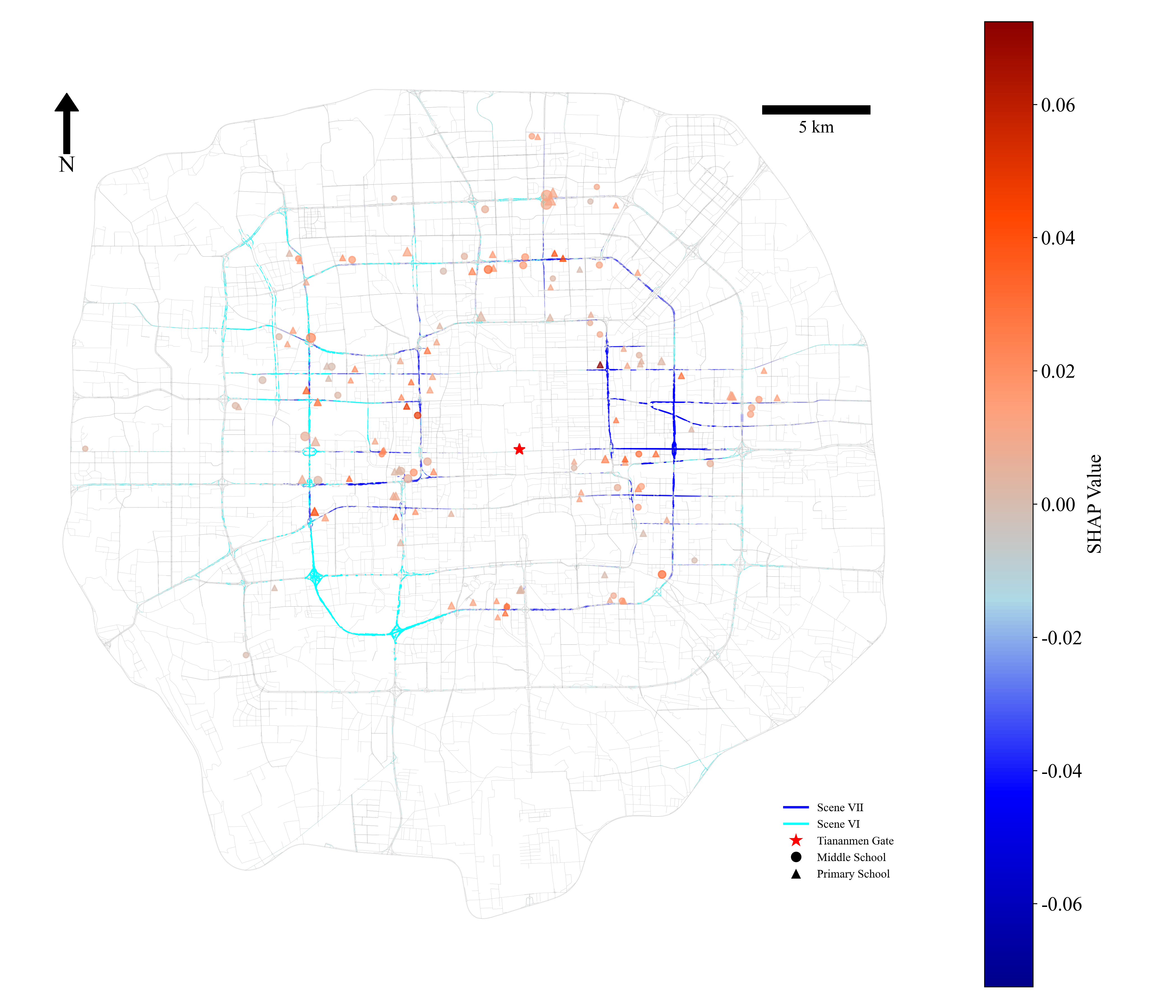}
         \caption{Interaction effect by geographical location}
         \label{fig:five over x}
     \end{subfigure}
\caption{The distribution of SHAP interaction values of scenescape 6 on scenescape 7}
        \label{figs3:interaction}
\end{figure}

\begin{figure}[htbp]
     \centering
     \begin{subfigure}[b]{0.515\textwidth}
         \centering
         \includegraphics[width=\textwidth, trim={0.5cm 0 1.2cm 0},clip]{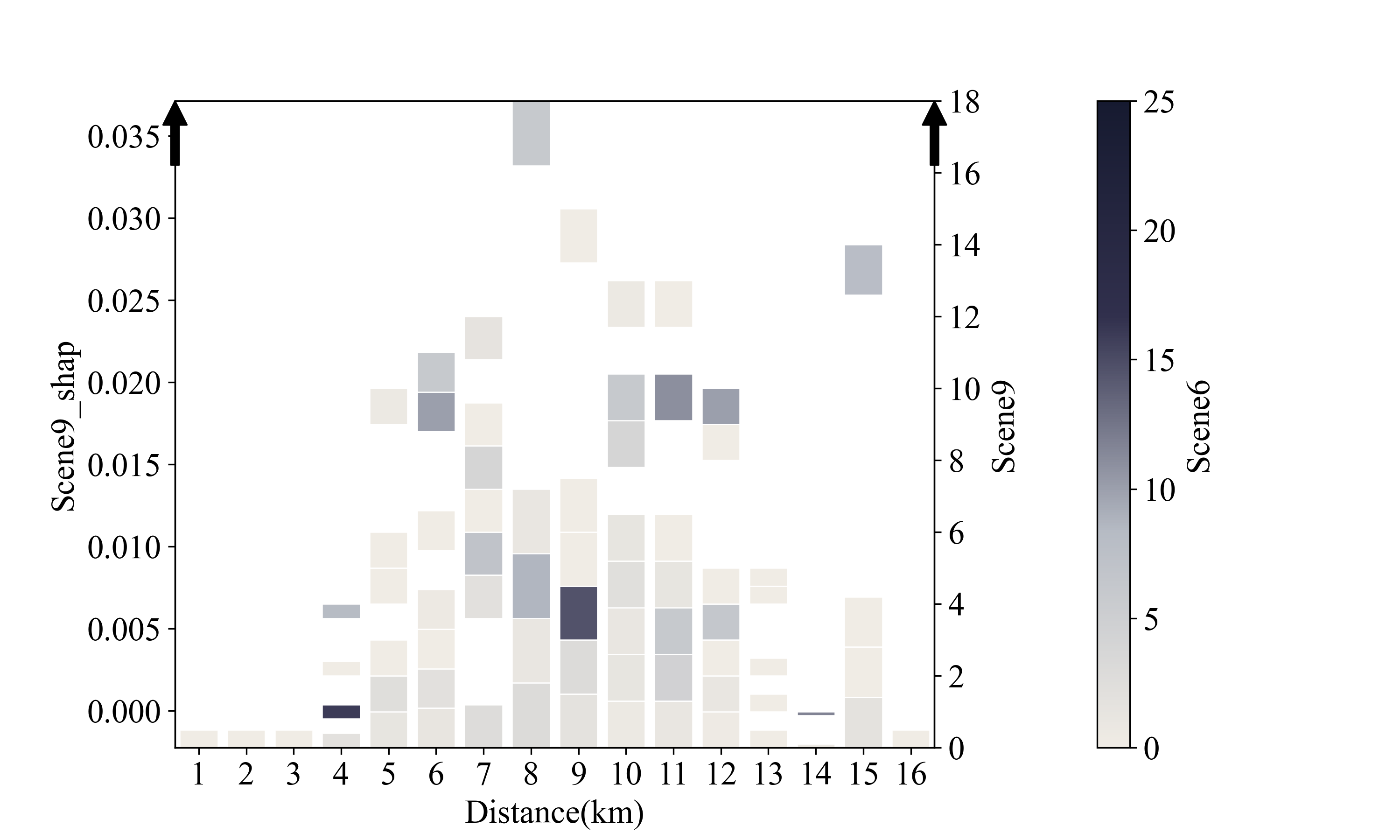}
         \caption{Interaction effect by distance}
         \label{fig:five over x}
     \end{subfigure}
     \hfill
     \begin{subfigure}[b]{0.475\textwidth}
         \centering
         \includegraphics[width=\textwidth]{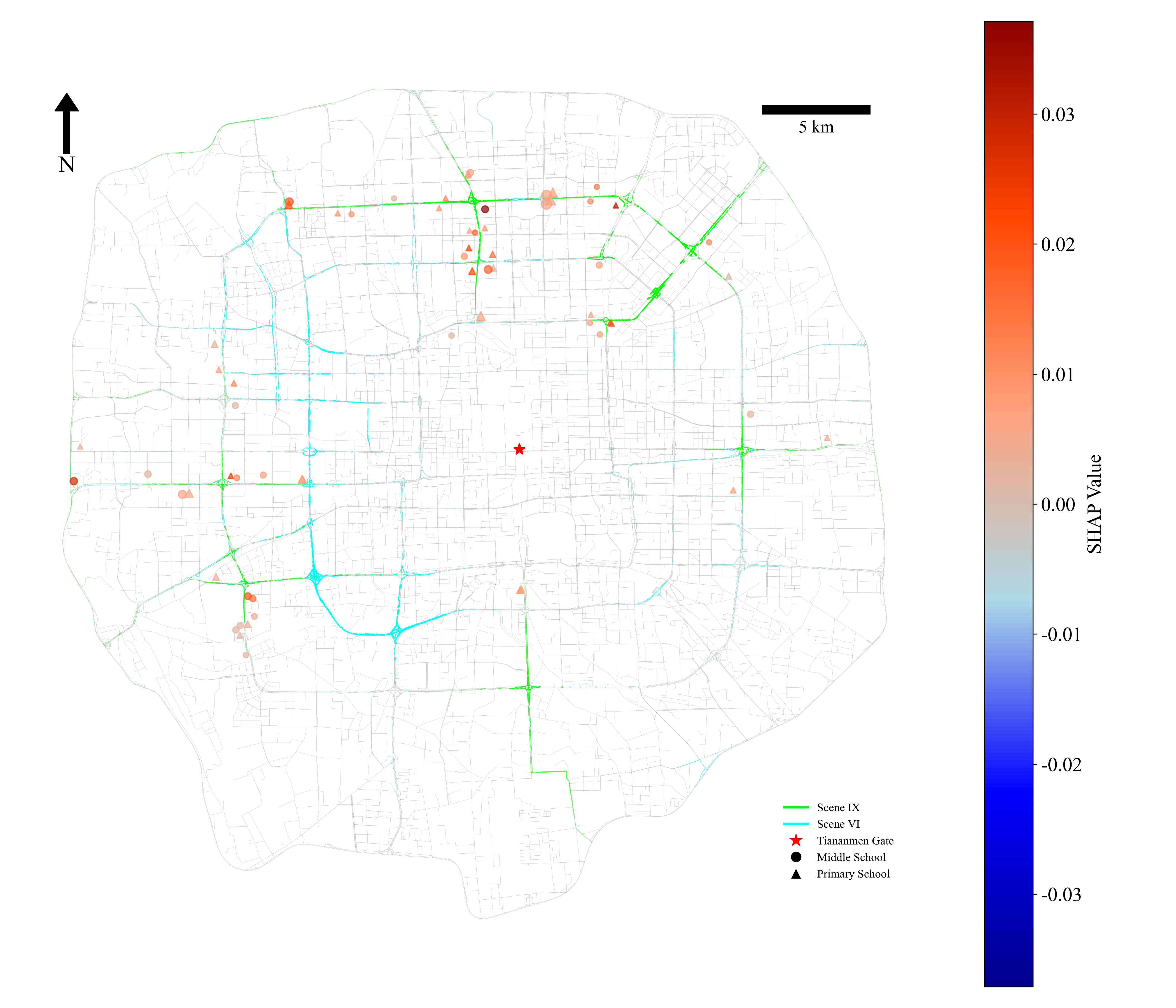}
         \caption{Interaction effect by geographical location}
         \label{fig:five over x}
     \end{subfigure}
\caption{The distribution of SHAP interaction values of scenescape 6 on scenescape 9}
        \label{figs4:interaction}
\end{figure}

\begin{figure}[htbp]
     \centering
     \begin{subfigure}[b]{0.515\textwidth}
         \centering
         \includegraphics[width=\textwidth, trim={0.5cm 0 1.2cm 0},clip]{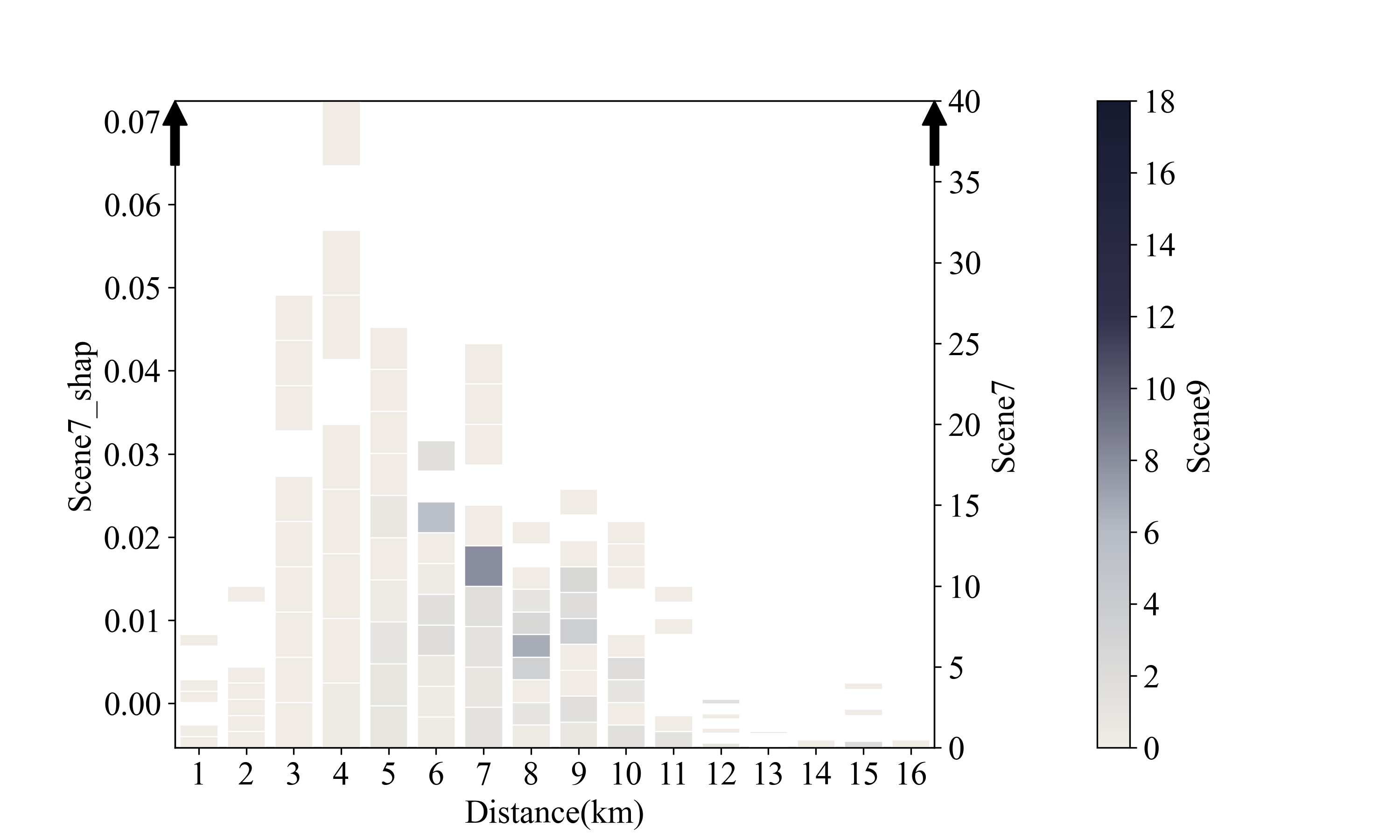}
         \caption{Interaction effect by distance}
         \label{fig:five over x}
     \end{subfigure}
     \hfill
     \begin{subfigure}[b]{0.475\textwidth}
         \centering
         \includegraphics[width=\textwidth]{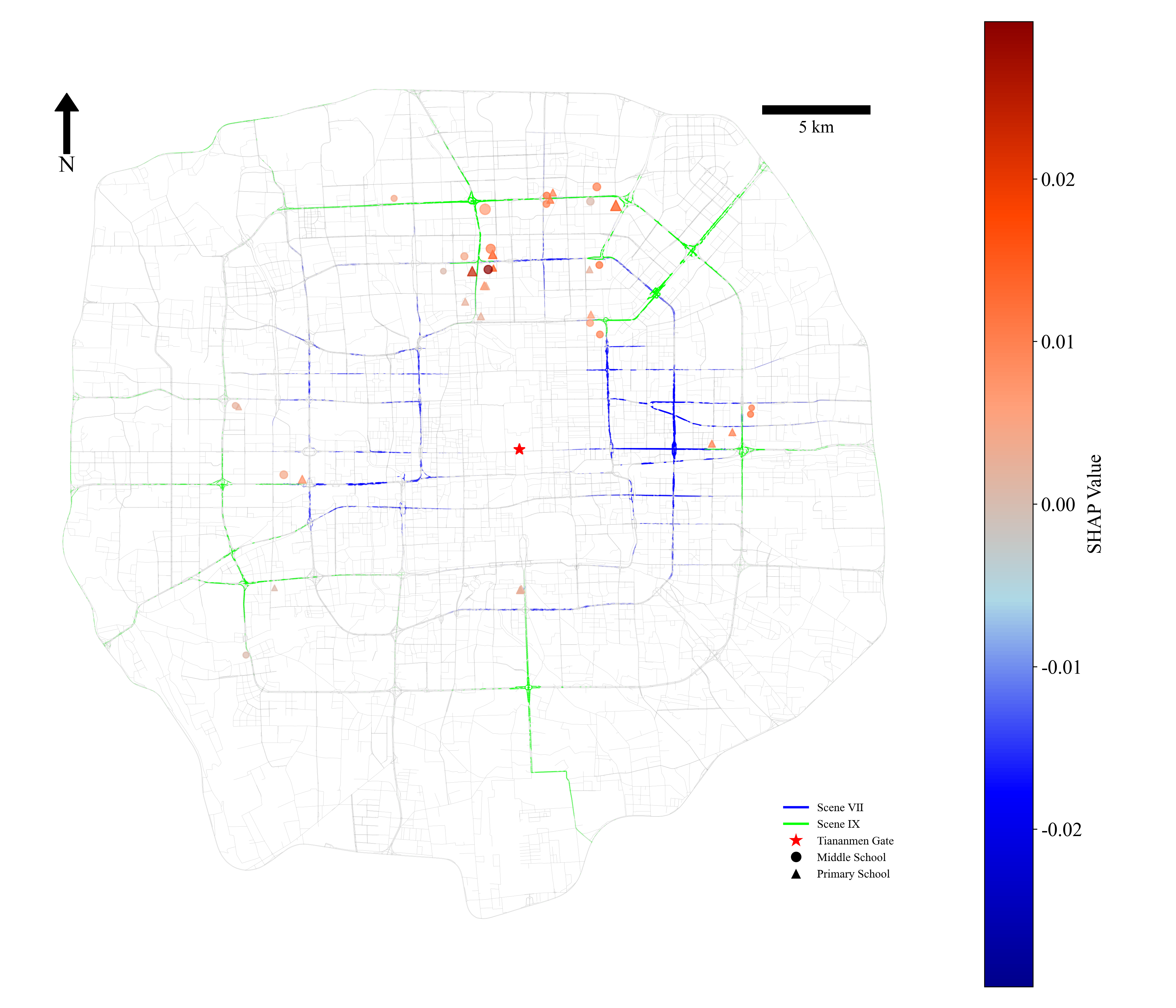}
         \caption{Interaction effect by geographical location}
         \label{fig:five over x}
     \end{subfigure}
\caption{The distribution of SHAP interaction values of scenescape 9 on scenescape 7}
        \label{figs5:interaction}
\end{figure}

\begin{figure}[htbp]
     \centering
     \begin{subfigure}[b]{0.515\textwidth}
         \centering
         \includegraphics[width=\textwidth, trim={0.5cm 0 1.2cm 0},clip]{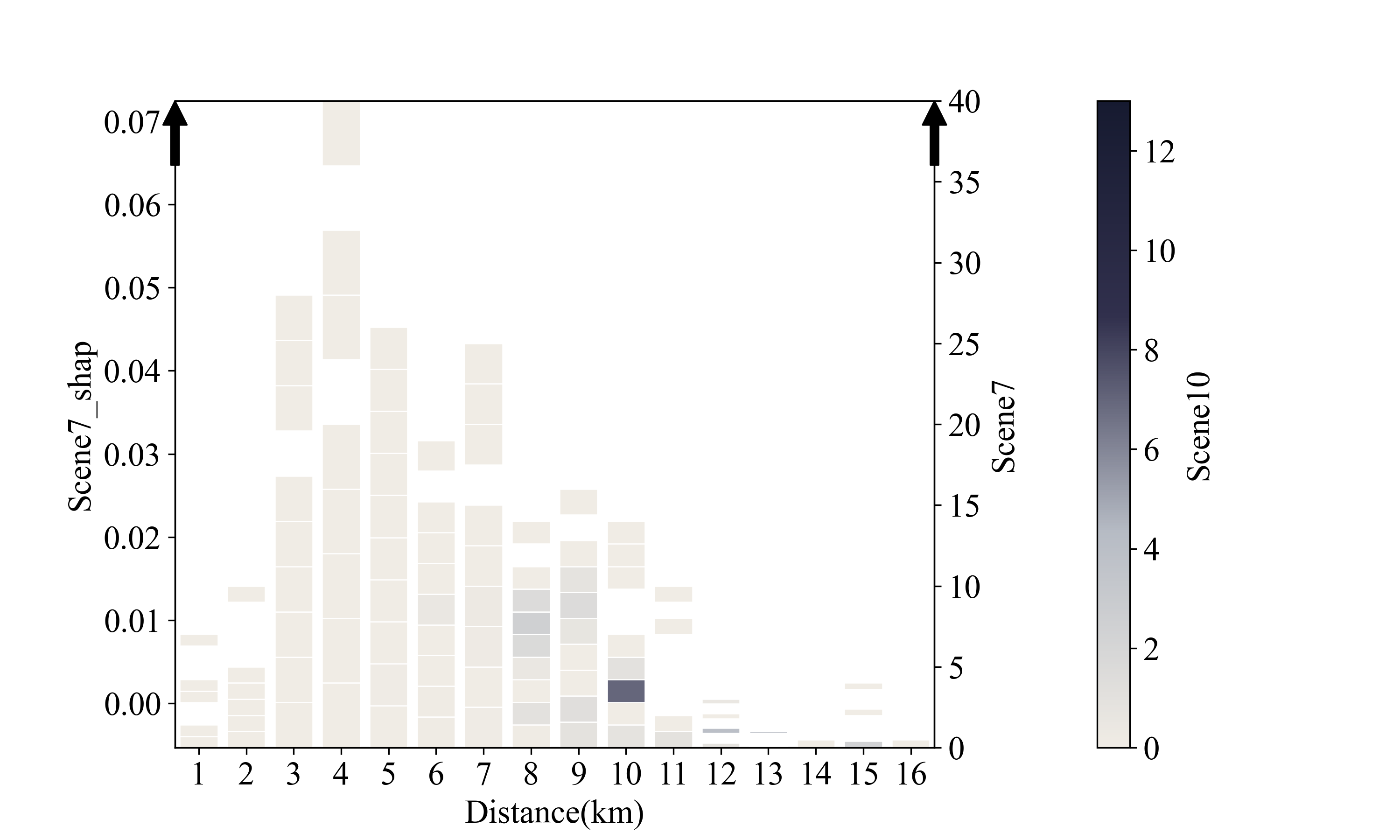}
         \caption{Interaction effect by distance}
         \label{fig:five over x}
     \end{subfigure}
     \hfill
     \begin{subfigure}[b]{0.475\textwidth}
         \centering
         \includegraphics[width=\textwidth]{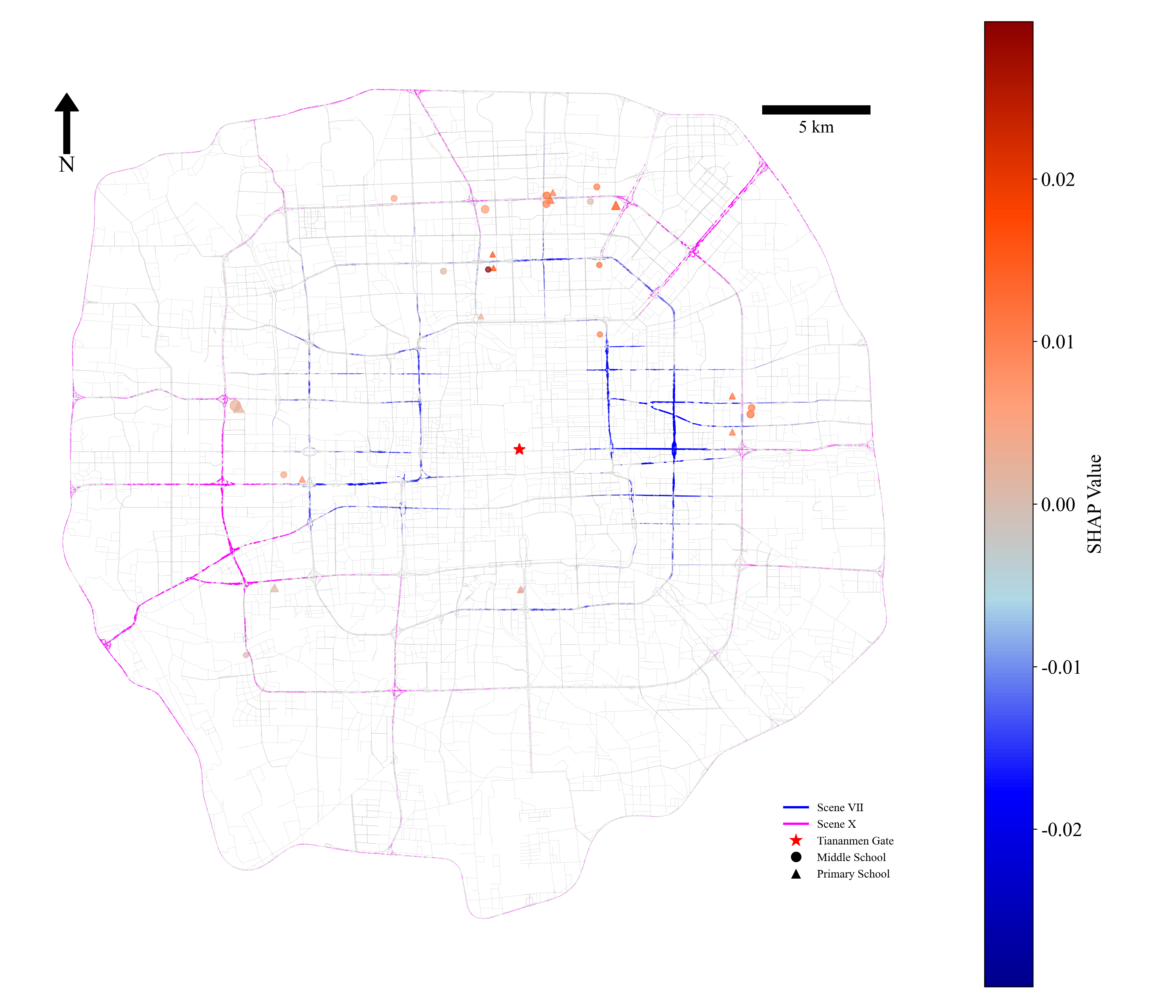}
         \caption{Interaction effect by geographical location}
         \label{fig:five over x}
     \end{subfigure}
\caption{The distribution of SHAP interaction values of scenescape 10 on scenescape 7}
        \label{figs6:interaction}
\end{figure}

\begin{figure}[htbp]
     \centering
     \begin{subfigure}[b]{0.515\textwidth}
         \centering
         \includegraphics[width=\textwidth, trim={0.5cm 0 1.2cm 0},clip]{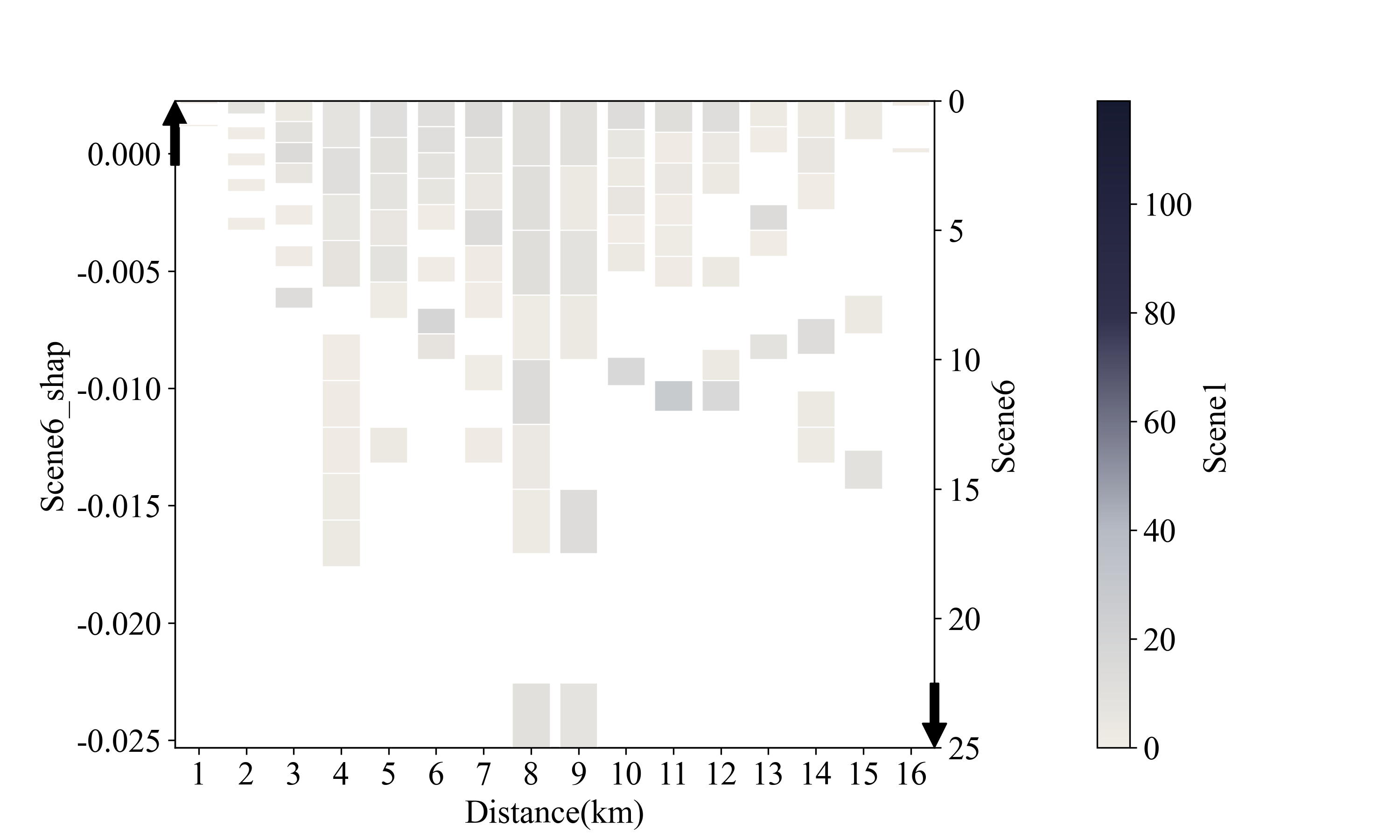}
         \caption{Interaction effect by distance}
         \label{fig:five over x}
     \end{subfigure}
     \hfill
     \begin{subfigure}[b]{0.475\textwidth}
         \centering
         \includegraphics[width=\textwidth]{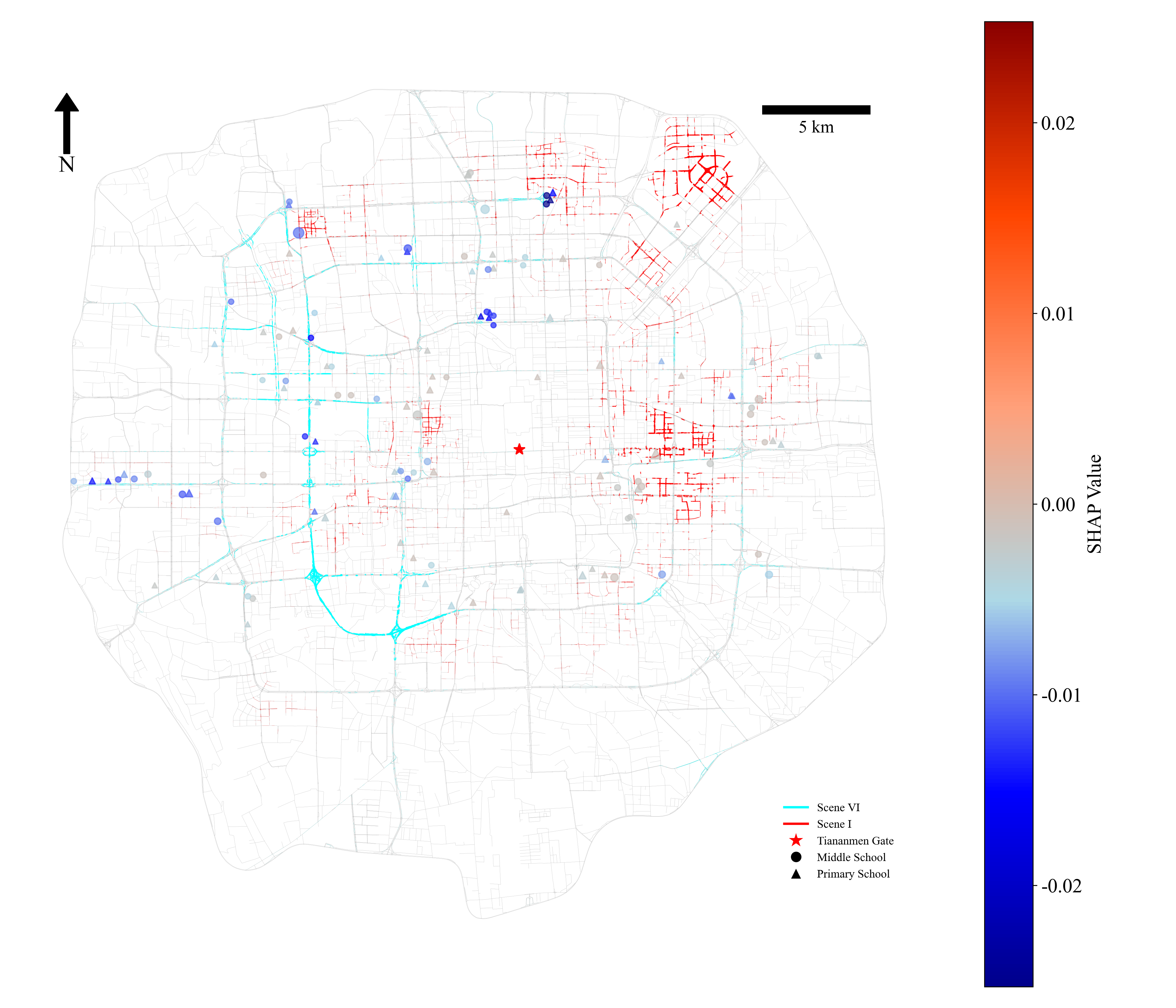}
         \caption{Interaction effect by geographical location}
         \label{fig:five over x}
     \end{subfigure}
\caption{The distribution of SHAP interaction values of scenescape 1 on scenescape 6}
        \label{figs7:interaction}
\end{figure}

Fig.~\ref{figs1:interaction} shows that, in school districts at the same distance from the center of Tiananmen Square, Scenescape 1, which initially helped reduce congestion, ended up amplifying the congestion-promoting effect of Scenescape 7 as its quantity increased. Figs.~\ref{figs2:interaction}, \ref{figs3:interaction}, and \ref{figs4:interaction} show that in these same districts, Scenescape 6, which originally reduced congestion, began to strengthen the congestion-promoting effects of Scenescapes 5, 7, and 9 as its frequency increased.  Figs.~\ref{figs5:interaction} and \ref{figs6:interaction}, illustrate that in these districts, Scenescapes 9 and 10, which initially promoted congestion, started to counteract the congestion-promoting effect of Scenescape 7 as their quantities increased. Finally, Fig.~\ref{figs7:interaction} shows that in the same districts, Scenescape 1, which originally reduced congestion, became more effective at enhancing the congestion-reducing impact of Scenescape 6 as its quantity increased.

\end{document}